\documentclass[sigconf, natbib=True, screen=False]{acmart}

\usepackage{multirow}

\usepackage{tikz}
\usepackage{pgfplots}
\pgfplotsset{compat=1.18}
\usetikzlibrary{shapes.misc}

\usepackage{algorithm2e}

\usepackage{cleveref}
\Crefname{equation}{Eq.}{Eqs.}
\Crefname{figure}{Fig.}{Figs.}
\Crefname{table}{Tab.}{Tabs.} 
\Crefname{appendix}{App.}{Apps.} 
\crefformat{section}{\S#2#1#3}
\crefformat{subsection}{\S#2#1#3}

\theoremstyle{definition}
\newtheorem{definition}{Def.}

\makeatletter

\setcopyright{cc} 
\copyrightyear{2025} 
\acmYear{2025} 
\setcctype{by-sa} 
\acmConference[WWW '25]{Proceedings of the ACM Web Conference 2025}{April 28-May 2, 2025}{Sydney, NSW, Australia} 
\acmBooktitle{Proceedings of the ACM Web Conference 2025 (WWW '25), April 28-May 2, 2025, Sydney, NSW, Australia} 
\acmDOI{10.1145/3696410.3714589} 
\acmISBN{979-8-4007-1274-6/25/04}

\DeclareMathOperator*{\argminA}{arg\,min}
\begin{document}

\title{Joint Evaluation of Fairness and Relevance in Recommender Systems with Pareto Frontier}

\author{Theresia Veronika Rampisela}

\orcid{0000-0003-1233-7690}
\affiliation{%
  \institution{University of Copenhagen}
  \city{Copenhagen}
  \country{Denmark}}
\email{thra@di.ku.dk}

\author{Tuukka Ruotsalo}
\orcid{0000-0002-2203-4928}
\affiliation{%
 \institution{University of Copenhagen}
 \city{Copenhagen}
 \country{Denmark}
}
\affiliation{%
 \institution{LUT University}
 \city{Lahti}
 \country{Finland}
}
\email{tr@di.ku.dk}

\author{Maria Maistro}
\orcid{0000-0002-7001-4817}
\affiliation{%
 \institution{University of Copenhagen}
 \city{Copenhagen}
 \country{Denmark}
}
\email{mm@di.ku.dk}

\author{Christina Lioma}
\orcid{0000-0003-2600-2701}
\affiliation{%
 \institution{University of Copenhagen}
 \city{Copenhagen}
 \country{Denmark}
}
\email{c.lioma@di.ku.dk}

\begin{abstract}
Fairness and relevance are two important aspects of recommender systems (RSs). Typically, they are evaluated either (i) separately by individual measures of fairness and relevance, or (ii) jointly using a single measure that accounts for fairness with respect to relevance. 
However, approach (i) often does not provide a reliable joint estimate of the goodness of the models, as it has two different best models: one for fairness and another for relevance. 
Approach (ii) is also problematic because these measures tend to be ad-hoc and do not relate well to traditional relevance measures, like NDCG. 
Motivated by this, we present a new approach for jointly evaluating fairness and relevance in RSs: Distance to Pareto Frontier (DPFR). 
Given some user-item interaction data, we compute their Pareto frontier for a pair of existing relevance and fairness measures, and then use the distance from the frontier as a measure of the jointly achievable fairness and relevance. Our approach is modular and intuitive as it can be computed with existing measures. 
Experiments with 4 RS models, 3 re-ranking strategies, and 6 datasets show that existing metrics have inconsistent associations with our Pareto-optimal solution, making DPFR a more robust and theoretically well-founded joint measure for assessing fairness and relevance.
Our code: \href{https://github.com/theresiavr/DPFR-recsys-evaluation}{github.com/theresiavr/DPFR-recsys-evaluation}.
\end{abstract}



\maketitle

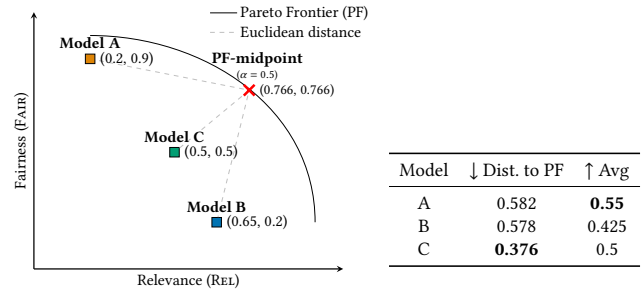
\begin{figure}[!h]
    \centering
    \begin{minipage}{.6\columnwidth}
        \centering
        \scalebox{.6}{
        \begin{tikzpicture}
            \tikzstyle{blue rectangle}=[fill={rgb,255: red,1; green,115; blue,178}, draw=black, shape=rectangle]
            \tikzstyle{green circle pareto}=[fill={rgb,255: red,2; green,158; blue,115}, draw=black, shape=rectangle]
            \tikzstyle{orange rectangle}=[fill={rgb,255: red,222; green,132; blue,5}, draw=black, shape=rectangle]
            \tikzstyle{pink rectangle}=[fill={rgb,255: red,204; green,120; blue,188}, draw=black, shape=rectangle]
            \tikzset{cross/.style={cross out, draw=black, minimum size=2*(#1-\pgflinewidth), inner sep=0pt, outer sep=0pt}, 
            cross/.default={5pt}}
            \begin{axis}
            [
                xmin = 0,
                ymin = 0,
                xmax = 1+0.1,
                ymax = 1+0.1,
                xlabel={\large Relevance (\textsc{Rel})}, 
                ylabel={\large Fairness (\textsc{Fair})}, 
                ticks=none,
                clip=false,
                axis lines=left,
                axis line style=thick,
                x label style={at={(axis description cs:0.5,0)},anchor=north},
                y label style={at={(axis description cs:0,.5)},anchor=south},
                legend style={at={(0.83,0.87)}, anchor=south,legend columns=1},
                legend style={font=\small, draw=none},
                legend cell align={left}
            ]

            \addlegendimage{}
            \addlegendimage{dashed,gray!50}

            \node (0) at (0.2, 1) {};
            \node (1) at (1, 0.2) {};
            \node [style=green circle pareto] (9) at (0.5, 0.5) {};
            \node [style=orange rectangle] (bestF) at (0.2, 0.9) {};
            \node [style=blue rectangle] (bestR) at (0.65, 0.2) {};
            \draw [in=90, out=0] (0.center) to (1.center);
            \addlegendentry{\large Pareto Frontier (PF)};
            \draw (0.7656854249492, 0.7656854249492) 
            node[cross,red, ultra thick,
            label={[align=center]0:{\normalsize(0.766, 0.766)}}, 
            label={[align=center, xshift=5pt]88:{\large\textbf{PF-midpoint}}\\[-2pt]{\footnotesize($\alpha=0.5$)}}] (10) {};
            \draw[dashed, color=gray!50] (9) -- (10);
            \addlegendentry{\large Euclidean distance};
            \draw[dashed, color=gray!50] (10) -- (bestF);
            \draw[dashed, color=gray!50] (10) -- (bestR);
            \node[label={\large \textbf{Model C}}, label={0:{\large(0.5, 0.5)}}]  at (9) {};
            \node[label={\large \textbf{Model A}}, label={0:{\large(0.2, 0.9)}}]  at (bestF) {};
            \node[label={\large \textbf{Model B}}, label={0:{\large(0.65, 0.2)}}]  at (bestR) {};
            \end{axis}%
        \end{tikzpicture}
        }
    \end{minipage}%
    \begin{minipage}[t][][b]{0.4\columnwidth}

     \scalebox{.8}{
        \begin{tabular}{ccc}
            \toprule
            Model & $\downarrow$ Dist.~to PF & $\uparrow$ Avg\\
            \midrule
            A & 0.582 & \textbf{0.55} \\
            B & 0.578 & 0.425 \\
            C & \textbf{0.376} & 0.5 \\
            \bottomrule
        \end{tabular}}
    \end{minipage}
    
    \caption{
    $(x, y)$ denotes the pair of relevance and fairness score. 
    Example: 
    Model A is best for fairness, 
    Model B is best for relevance, and Model C is the closest 
    to the Pareto Frontier (PF) midpoint, when relevance and fairness are equally weighted ($\alpha=0.5$). 
    Averaging relevance and fairness (Avg) leads to falsely concluding that Model A is best for both aspects. Note that distance to PF also beats other existing measures of fairness and relevance (see $\S$\ref{ss:corr}).
    }
    \label{fig:pareto_teaser}

\end{figure}

\section{Introduction
}
\label{s:intro}

Relevance and fairness are important aspects of recommender systems (RSs). Relevance is typically evaluated using common ranking measures (e.g., NDCG), while various fairness measures for RSs exist \cite{Wang2023ASystems,Amigo2023ASystems}. 
Some fairness measures integrate relevance, so that they evaluate fairness w.r.t.~relevance. 
The problem with these joint measures is that they tend to be ad-hoc, unstable, and they do not account very well for both aspects simultaneously \cite{Rampisela2024CanRelevance}. 
Another way of evaluating relevance and fairness is to use a different measure for each aspect. However, this does not always provide a reliable joint estimate of the goodness of the models, as it may have two different best models: one for fairness and another for relevance.   
This can be avoided by aggregating the scores of the two measures into a single score, or by aggregating the resulting model rankings into one using rank fusion. These approaches are also problematic because: 
(i) the scores of the two measures may 
have different distributions and different scales, making them hard to combine; 
(ii) the two measures may not even be computed with the same input, making their combination hard to interpret (relevance scores are computed for individual users and then averaged, while fairness measures for individual items are typically based on individual item recommendation frequency); and 
(iii) the resulting scores are less understandable as it is unknown how close the models are to an ideal balance of fairness and relevance, e.g., an acceptable trade-off between fairness and relevance scores. 

To address the above limitations, we contribute an approach that builds on the set of all Pareto-optimal solutions \cite{Censor1977ParetoProblems}. Our approach addresses issue (i) and (ii) above by avoiding direct combination of measures. We directly address (iii) by computing the distance of the model scores to a desired fairness-relevance balance. 
Our approach uses Pareto-optimality, a popular concept in multi-objective optimization problems across domains, including RSs \cite{Ribeiro2015MultiobjectiveSystems}. 
A recommendation is Pareto-optimal if there are no other possible recommendations with the same \textsc{Rel} score that achieve better fairness.\footnote{The opposite is also true, but in RS scenario the \textsc{Rel} score is usually the primary objective, 
not the \textsc{Fair} score.} 
In other words, given Pareto-optimal solutions, we cannot get other recommendations that empirically perform better, 
unless relevance is sacrificed. 
In our approach, we combine existing \textsc{Fair} measures and \textsc{Rel} measures as follows. We build a Pareto Frontier (PF) that first maximises relevance, finds the best fairness achievable under the relevance constraint, and then jointly quantifies fairness and relevance as the distance from an optimal solution, see Fig.~\ref{fig:pareto_teaser}. 

Our approach, \emph{Distance to PF of Fairness and Relevance} (DPFR) has several strengths. First, DPFR is \emph{modular}; it can be used with well-known existing measures of relevance and fairness. DPFR is also \emph{tractable} as one can control the weight ($\alpha$) of fairness w.r.t.~relevance. As the resulting score is the distance to the scores of a traditional relevance measure and a well-known fairness measure, DPFR is also \emph{intuitive} in its interpretation. Most importantly, DPFR is a principled way of jointly evaluating relevance and fairness based on an empirical best solution that uses Pareto-optimality. Experiments with different RS models, re-ranking approaches and datasets show that there exists a noticeable gap between using current measures of relevance and fairness and our Pareto-optimal joint evaluation of relevance and fairness. This gap is bigger in larger datasets and when using rank-based relevance measures (i.e., MAP, NDCG), as opposed to set-based relevance measures (i.e., 
Precision, Recall). 

In this work, we focus on \textbf{individual item fairness}. This type of fairness is commonly defined as all items having equal exposure, where exposure typically refers to the frequency of item appearance in the recommendation list across all users~\cite{Patro2020FairRec:Platforms, Mansoury2020FairMatch:Systems, Rampisela2024EvaluationStudy}. Individual item fairness is important in ensuring that each item/product in the system has a chance to be recommended to any user \cite{Lazovich2022MeasuringMetrics}.

\section{Related work}\label{s:prev_work}

Evaluating fairness and relevance together is a type of multi-aspect evaluation. 
However, none of the existing multi-aspect evaluation methods \cite{Maistro2021PrincipledRankings, Lioma2017EvaluationLists,Palotti2018MM:Engines} can be used in this case as 
these methods require separate labels that are unavailable in RS scenarios. 
Specifically, it is not possible to label an item as `fair', because item fairness depends on other recommended items. The same item can be a fair recommendation in one ranking, but unfair in another. In RSs, fairness is typically defined as treating users or items without discrimination \cite{Biega2018EquityRankings}. This is often quantified as the opportunity for 
having equal relevance (for users) or exposure (for items) 
\cite{Biega2018EquityRankings, Wang2022ProvidingSystems}, computed either individually or for
groups of items/users \cite{Raj2022MeasuringResults, Zehlike2022FairnessSystemsc}. 

The problem of jointly evaluating RS relevance and fairness is further aggravated by the fact that improved fairness is often achieved at the expense of relevance to users \cite{Mehrotra2018TowardsSystems}. 
We posit that this trade-off makes multi-objective optimization a suitable solution. Pareto optimality is a well-known objective for such optimization, 
and it has been previously used in RS but only to recommend items to users \cite{Ribeiro2015MultiobjectiveSystems,  Zheng2022AOptimization, Ge2022TowardLearning, Xu2023P-MMF:System}. 
Because the true PF is often unknown due to the problem complexity \cite{Laszczyk2019SurveyMeasures,Audet2020PerformanceOptimization}, prior work has used the model's training loss w.r.t.~two different aspects \cite{Lin2019ARecommendation} or scores from different models \cite{Nia2022RethinkingNetworks,Paparella2023Post-hocRecommendation} to generate the PF. 
Our work differs from this in terms of both the purpose of using Pareto-optimal solutions, and the nature of the PF. Specifically, we exploit Pareto-optimality through PF as a robust \textit{evaluation} method, instead of as a recommendation method. 
In addition, our generated PF is based on the ground truth (i.e., the test set), a common RS evaluation approach, instead of the recommender models' empirical performance, which may not be optimal. Thus, our PF is also model-agnostic, as opposed to the PF in \cite{Xu2023P-MMF:System}. 
Our approach differs also from FAIR~\cite{Gao2022FAIR:Evaluation} since the PF considers the empirically achievable optimal solution based on the dataset, while FAIR compares against the desired fairness distribution which might not be achievable. Lastly, \cite{Paparella2023Post-hocRecommendation} selects the optimal solution based on its distance to the utopia point (the theoretical ideal scores), whereas the utopia point may not be realistic due to dataset or measure characteristics~\cite{Rampisela2024EvaluationStudy,Moffat2013SevenMetrics}. Since our PF is generated based on test data, any of its solutions is empirically achievable.

\section{Distance to Pareto Frontier(DPFR)}
\label{s:our_method}

We present definitions 
($\S$\ref{ss:motivation}),  
and then explain DPFR in different steps: 
given a \textsc{Fair} and a \textsc{Rel} measure, how to 
generate PF based on the ground truth data in the test set ($\S$\ref{ss:generation}); how to choose a reference point in the PF based on $\alpha$ (e.g., the midpoint for $\alpha=0.5$) ($\S$\ref{ss:pareto-for-eval}); and how to compute the distance of the \textsc{Fair} and \textsc{Rel} scores to the reference point with a distance measure $d$ ($\S$\ref{ss:pareto-for-eval}). 
Additionally, we present a computationally efficient adaptation of DPFR ($\S$\ref{ss:compute-eff}). 

\subsection{Definitions}
\label{ss:motivation}

We adapt the Pareto-optimality definition \cite{vanVeldhuizen1999MultiobjectiveInnovations}: the multi-objective problem is finding the optimum \textsc{Fair} score $s_f$, and \textsc{Rel} score, $s_r$ from a list of possible recommendations across all users. We define the tuple $s = (s_r, s_f) \in S$, where $S$ is the Cartesian product of all possible \textsc{Rel} and \textsc{Fair} scores. The relation $\geq_A$ ($>_A$) means `better or equal to' (`better to') according to an aspect $A \in \{\textsc{Rel}, \textsc{Fair}\}$. 

\begin{definition}[Pareto Dominance]
A tuple $s = (s_r, s_f)$ dominates $s'=(s_r', s_f')$ iff $s$ is partially better than $s'$, i.e., $s_r \geq_\textsc{Rel} s'_r$ and $s_f \geq_\textsc{Fair} s'_f$, in addition to $s_r >_\textsc{Rel} s'_r$ or $s_f >_\textsc{Fair} s'_f$. 
\end{definition}

\begin{definition}[Pareto Optimality]
A solution (recommendation list) that has \textsc{Rel} and \textsc{Fair} scores of $x = (x_r, x_f) \in S$ is Pareto-optimal iff there is no other solution with $x'=(x_r', x_f') \in S$ that dominates $x$.
\end{definition}

\begin{definition}[Pareto Frontier]
The set of all Pareto-optimal tuples.
\end{definition}

\subsection{Pareto Frontier generation}
\label{ss:generation}
Given user-item preference data (e.g., test set), the aim is to explore the empirical, maximum feasible fairness towards individual items, such that the recommendation satisfies Pareto-optimality w.r.t.~fairness scores across all items and an average relevance score across users, e.g., MAP@$10=0.9$.\footnote{This is how \textsc{Fair} and \textsc{Rel} measures are usually computed.} This is done to measure how far a model performance is, from these Pareto-optimal solutions.
Enumerating all possible recommendations for users and items to find the complete set of Pareto-optimal solutions is computationally infeasible, and there is no analytical solution either. Instead, we contribute an algorithm that iteratively builds upon a maximally relevant initial recommendation list. Our algorithm iteratively finds Pareto-optimal recommendations by prioritising relevance over fairness, as recommendations are usually optimised for relevance (with or without fairness). This prioritisation is known as lexicographic optimization \cite{Ryu2018Multi-objectiveWeight}. 
We call our algorithm {\sc Oracle2Fair} (full technical description in App.~\ref{app:algo}). 
Our algorithm generates the PF of fairness and relevance in two steps: 
 \textbf{(1) initialisation} of the recommendations with an \textit{Oracle} (App.\ref{app:algo}, Algorithm~\ref{alg:oracle}).  The \textit{Oracle} generates a recommendation with the highest empirical score for relevance, based on user interactions that are part of the test set. This step is followed by \textbf{(2) replacements} to make the recommendations as \textit{Fair} as possible; at the end of this algorithm, the \textsc{Fair} scores should reach the empirically fairest score while maintaining as much relevance as possible. 
Throughout the PF generation, items in a user's train/val split are not recommended to the same user. Henceforth, \emph{relevant items} refers to the items in a user's test split.

\noindent \textbf{(1) Initialisation}. 
The Oracle recommends at most $k=10$ relevant items, from the $n$ items in the dataset, to each of the $m$ users in the test split, one user at a time. 
The recommendation begins with users having exactly $k$ items in the test split; only these items can be recommended to those users to gain the maximum relevance. Recommendations to other users are made maximally relevant and fair as follows: if a user has $>k$ relevant items, we pick $k$ items with the least exposure among them. Item exposure is computed based on what has been recommended to other users who already have exactly $k$ items. Note that this process is not trivial (see App.~\ref{app:algo}, 
Algo~\ref{alg:oracle}, ll.~\ref{ln:startgtk}--\ref{ln:endgtk}). If a user has $<k$ relevant items, we recommend those items at the top (to maximise top-weighted \textsc{Rel} measures) and fill the rest of their recommendation slots with the least exposed items in the dataset (Algo~\ref{alg:oracle}, ll.~\ref{ln:startltk}--\ref{ln:endltk}). This least-exposure prioritisation strategy ensures that the solutions are Pareto-optimal. 

\noindent \textbf{(2) Replacements}. 
The algorithm iteratively replaces the recommended items to achieve maximum fairness, such that each replacement results in a fairer recommendation than the previous. 
We compute the \textsc{Fair} and \textsc{Rel} measures after each replacement as follows. The most popular item, which is recommended most often, is replaced with one of these item types, 
in decreasing order of priority: an unexposed item, then the least popular item in the recommendation; this increases fairness from the previous recommendations. We do this one item and one user at a time, starting with the users that have the most popular item at the bottom of their recommendation list, to ensure that the decrease in relevance is minimum as the replacement item is mostly not relevant to that user. Nonetheless, the \textsc{Oracle2Fair} prioritises replacing the recommendations of users for whom the replacement item is relevant (if any). As fairness increases and relevance decreases/stays the same from the previous recommendation, the new recommendation is also Pareto-optimal. 
We continue the replacement until the maximum times any item is recommended is $\left\lceil km/n \right\rceil$, i.e., the upper bound of how many times an item can be recommended, if all items in the dataset must appear in the recommendation as uniformly as possible. We explicitly used $\left\lceil km/n \right\rceil$ as a stopping condition for \textsc{Oracle2Fair}. 
To ensure maximum \textsc{Rel} scores (especially in top-weighted measures), each time a replacement takes place, we rerank the recommendations based on descending relevance.

The resulting (\textsc{Rel, Fair}) scores reflecting Pareto-optimal recommendations from this process make up the PF. If there are duplicates in the \textsc{Rel} value, we keep the best \textsc{Fair} score for a single value of \textsc{Rel}. While it cannot be reasonably verified that the resulting PF matches the theoretical PF, this is a close way to build the full PF, as opposed to building the PF from trained models scores ($\S$\ref{s:prev_work}).

\subsection{Distance computation} 
\label{ss:pareto-for-eval}

For each pair of \textsc{Fair} and \textsc{Rel} measures, we find a reference point using a tunable parameter $\alpha \in [0,1]$; $\alpha=0$ means only relevance is accounted for, and $\alpha=1$ means only fairness is accounted for. Next, we explain how to compute the reference point. We first use the following equation to find the length of a subset $T$ of the PF: 
$
    lenPF(T) = \sum_{t=1}^{|T|-1} d_E(x^t, x^{t+1})
$.
 Given that $P$ is the set of all Pareto-optimal solutions, $x^t = (x_r^{t}, x_f^{t})$ is the pair of Pareto-optimal solutions $(x_r,x_f)$ with the $t$-th highest $x_r$ in $P$, and $d_E$ is the Euclidean distance. The overall PF length is $lenPF(P)$ or simply $lenPF$. 
 
 The reference point is $s_{\alpha} = x^{t'}$, where $t'$ is computed as follows:
 $
    t' = \argminA_{j \in \left[1,\dots,|P|-1\right]}\left|lenPF(T^j)-\ \alpha \cdot lenPF \right|
$. $T^{j}$ is a subset of $P$ containing the $j$ highest $x_r$ scores. So, the reference point is a point in the PF whose cumulative traversal distance is closest to the $\alpha$-weighted PF distance travelled from the first point in the PF. The reference point $s_\alpha$ is how far the PF is traversed, from the pair with the best \textsc{Rel} score to the one with the best \textsc{Fair} score, multiplied by $\alpha$. As the PFs 
may have different density of points along the frontiers, the reference point is not computed based on a percentile (e.g., median) to avoid bias towards the denser part. 
Next, the distance between each model's ($x_r, x_f$) scores and the reference point $s_\alpha$ is computed with a distance measure $d$ that accommodates 2d-vectors. The model with the closest distance is the best model in terms of both relevance and fairness, given the weight $\alpha$. 
We call this \emph{Distance to Pareto frontier of Fairness and Relevance} (DPFR). 

\subsection{Efficient computation of Pareto Frontier}
\label{ss:compute-eff}

Generating the PF as in $\S$\ref{ss:generation} is costly. An efficient alternative is to compute a subset of the PF. We pick a fixed amount of Pareto-optimal solutions to compute, $p$ (e.g., 10). However, to reliably approximate the PF, these solutions should be spread according to the PF distribution, as opposed to e.g., only computing the first $p$ points of the PF. The spread of the points is important, as the reference point in DPFR is computed based on the overall estimated PF. 
In the estimated PF, the first point corresponds to the measure scores of the initial recommendation given by the Oracle, and the rest are spread evenly throughout the PF generation. 
To select at which point of the \textsc{Oracle2Fair} algorithm the measures should be computed, we first estimate the total number of replacements needed by examining the distribution of recommended items frequency. This is done by getting the individual frequency count of all items in the recommendation, and subtracting the ideal upper bound of item count $\lceil km/n \rceil$ ($\S$\ref{ss:generation}) from each count. 
The number of expected replacements is the sum of the difference between the item frequency count and the ideal upper bound of item count in $\S$\ref{ss:generation}. Items with recommendation frequency counts less than the upper bound are excluded. With the estimated total number of replacement $numRep$, we compute the measures every $numRep\ \text{div} (p-1)$ replacements done by \textsc{Oracle2Fair}, such that the measures are computed a total of $p-1$ times during the replacement process + 1 time before the replacement starts. These $p$ points are spread evenly in terms of distance in the PF, which is important as DPFR is a distance-based measurement. 

\section{Experimental setup}
\label{s:experiments}

We study how our joint evaluation approach, DPFR, compares to existing single- and multi-aspect evaluation measures of relevance and fairness. 
Next, we present our experimental setup. 
The experiments are run in a cluster of CPUs and GPUs (e.g., Intel(R) Xeon(R) Silver 4214R CPU @ 2.40GHz, AMD EPYC 7413 and 7443, Titan X/Xp/V, Titan RTX, Quadro RTX 6000, A40, A100, and H100).

\noindent \textbf{Datasets.} We use six real-world datasets of various sizes and domains: 
e-commerce (A\-ma\-zon Luxury Beauty, i.e., Amazon-lb \cite{Ni2019JustifyingAspects}),
movies (ML-10M and ML-20M \cite{Harper2015TheContext}), 
music (Lastfm \cite{Cantador20112nd2011},  
videos (QK-video \cite{Yuan2022Tenrec:Systems}), 
and jokes (Jester \cite{Goldberg2001Eigentaste:Algorithm}). 
The datasets are as provided by \cite{Zhao2021RecBole:Algorithms}, except for QK-video, which we obtain from \cite{Yuan2022Tenrec:Systems}. 
Statistics of the datasets are in Tab.~\ref{tab:stats} and extended statistics are in App.~\ref{app:stat}.

\begin{table}[tb]
\caption{Statistics of the preprocessed datasets.}

\resizebox{0.98\columnwidth}{!}{
\begin{tabular}{lrrrr}
\toprule
\textbf{dataset} & \multicolumn{1}{l}{\textbf{\#users ($m$)}} & \multicolumn{1}{l}{\textbf{\#items ($n$)}} & \multicolumn{1}{l}{\textbf{\#interactions}} & \multicolumn{1}{l}{\textbf{sparsity (\%)}} \\ 
\midrule                                                                   
Lastfm \cite{Cantador20112nd2011}& 1,859 & 2,823 & 71,355 & 98.64\% \\
Amazon-lb \cite{Ni2019JustifyingAspects} & 1,054 & 791 & 12,397 & 98.51\% \\
QK-video \cite{Yuan2022Tenrec:Systems} & 4,656 & 6,423 & 51,777 & 99.83\% \\
Jester \cite{Goldberg2001Eigentaste:Algorithm} & 63,724 & 100 & 2,150,060 & 66.26\% \\ 
ML-10M \cite{Harper2015TheContext} & 49,378 & 9,821 & 5,362,685 & 98.89\% \\
ML-20M \cite{Harper2015TheContext} & 89,917 & 16,404 & 10,588,141 & 99.28\% \\
\bottomrule
\end{tabular}
}
\label{tab:stats}
\end{table}

\noindent \textbf{Preprocessing.} We remove duplicate interactions (we keep the most recent). We keep only users and items with $\geq5$ interactions. 
We convert ratings equal/above the following threshold to $1$ and discard the rest: for Amazon-lb and ML-*, the threshold is 3, as their ratings range between $[1,5]$ and $[0.5, 5]$ resp.; the threshold for Jester is $0$, as the ratings range in $[-10, 10]$. Lastfm and QK-video have no ratings. QK-video has several interaction types; we use the `sharing' interactions. 
For Jester, we remove users with $>80$ interactions, to provide a large enough number of item candidates for recommendation during testing.\footnote{\label{fn:jester}Some users in Jester have interacted with almost all 100 items. If a user has 80 items in the train/val set, there would only be 20 candidate items to recommend during test, which makes it easier to achieve higher relevance. 
} 

\noindent \textbf{Data splits.} To obtain the train/val/test sets for Amazon-lb and ML-*, we use global temporal splits \cite{Meng2020ExploringModels} with a ratio of 
6:2:2 on the preprocessed datasets. Global random splits with the same ratio are used for the other datasets that have no timestamps. 
From all splits, we remove users with $<5$ interactions in the train set. 

\noindent\textbf{Measures}.
We measure relevance (\textsc{Rel}) with Hit Rate (HR), MRR, Precision (P), Recall (R), MAP, and NDCG. We measure individual item fairness (\textsc{Fair}), as per  \cite{Rampisela2024EvaluationStudy}, with Jain Index (Jain) \cite{jain1984quantitative, Zhu2020FARM:APPs}, Qualification Fairness (QF) \cite{Zhu2020FARM:APPs}, Gini Index (Gini) \cite{Gini1912VariabilitaMutabilita, Mansoury2020FairMatch:Systems}, Fraction of Satisfied Items (FSat) \cite{Patro2020FairRec:Platforms}, and Entropy (Ent) \cite{Shannon1948ACommunication, Patro2020FairRec:Platforms}.\footnote{
These are set-based measures, but we do not expect the conclusions to differ for rank-based measures (App.~\ref{app:discussion}).
} 
We also use joint measures (\textsc{Fair+Rel}) as per \cite{Rampisela2024CanRelevance}: 
Item Better-Off (IBO) \cite{Saito2022FairRanking},\footnote{
The measure Item Worse-Off is not used as its formulation is highly similar to IBO.}
Mean Max Envy (MME) \cite{Saito2022FairRanking},
Inequity of Amortized Attention (IAA) \cite{Biega2018EquityRankings, Borges2019EnhancingAutoencoders}, Individual-user-to-individual-item fairness (II-F) \cite{Diaz2020EvaluatingExposure, Wu2022JointRecommendation}, and 
All-users-to-individual-item fairness (AI-F) \cite{Diaz2020EvaluatingExposure}. 
We denote by $\uparrow$/$\downarrow$ measures where higher/lower is better.
DPFR is computed with Euclidean distance and $\alpha=0.5$ (PF midpoint) for all datasets, so the midpoint is based on the empirically achievable scores, per dataset and measure pairs. 
For all runs, we use $k=10$. 

\noindent\textbf{Recommenders}. We use 4 
common collaborative filtering-based recommenders: ItemKNN \cite{Deshpande2004Item-basedAlgorithms},
BPR \cite{Rendle2009BPR:Feedback},
MultiVAE \cite{Liang2018VariationalFiltering}, and 
NCL \cite{Lin2022ImprovingLearning}, with RecBole \cite{Zhao2021RecBole:Algorithms} and tune their hyperparameters. We train for 300 epochs with early stopping, and keep the configuration with the best NDCG@10 during validation. Each user's train/val items are excluded from their recommendations during testing. 

\noindent\textbf{Fair Re-rankers.} 
To have fairer recommendations, we reorder the top $k'$ items that are pre-optimised for relevance. Ideally $k'>k$ to allow exposing items that are not in the top $k$. As there are few relevant items per user in RS data,\footnote{The median number of relevant items per user across all datasets is 2--53, see App.~\ref{app:stat}.}
$k'$ should not be too big (e.g., 100). So, we re-rank the top $k'=25$ items per dataset and model using three methods: GS, CM, and BC (explained below). 
We re-rank separately per user for CM and BC, or altogether for GS, for all $k'm$ recommended items, where $m$ is the number of users. 
Other fair ranking methods exist but cannot be used as they apply to group or two-sided fairness only (e.g., \cite{Zehlike2017FAIR:Algorithm, Zehlike2020ReducingApproach, Patro2020FairRec:Platforms}), or to stochastic rankings only (e.g., \cite{Wu2022JointRecommendation, Oosterhuis2021ComputationallyFairness}), or do not scale to larger datasets (e.g., \cite{Biega2018EquityRankings, Saito2022FairRanking}).
    
   \noindent \textit{1. Greedy Substitution (GS) \cite{Wang2022ProvidingSystems}} is a re-ranker for individual item fairness. We modify the GS algorithm, to replace the most popular items with the least popular ones, both considering how many times an item is at the top $k'$ recommendations for all users (App.~\ref{app:gs}). As such, items can be swapped across users. To determine which items are most popular (i.e., to be replaced) and least popular (replacement items), the parameter $\beta=0.05$ is used. We pair these two item types, and for each pair, we calculate the loss of (predicted) relevance if the items are swapped. We then replace up to 25\% of the initial recommendations, starting from item pairs with the least loss. 
    
    \noindent \textit{2. COMBMNZ (CM) \cite{Lee1997AnalysesCombination}} is a common rank fusion method. Two rankings are fused for each user: one based on the (min-max) normalised predicted relevance score and another based on the coverage of each top $k'$ item (to approximate fairness). We calculate item coverage only based on their appearance in the top $k$ across all users and min-max normalise the score across all users. 
    As favouring items with higher coverage would boost unfairness, we generate the ranking using $1$ minus the normalised coverage. CM uses a multiplier based on the item appearance count in the two rankings above; this count is also only based on the top $k$. 
    The resulting ranking is a fused ranking of fairness and relevance.
    
    \noindent \textit{3. Borda Count (BC)} is a common rank fusion method. For each user, we combine the original recommendation list and the rankings based on increasing item coverage, as in CM. Unlike CM, BC uses points. Higher points are given to items placed at the top. The result is a fused ranking of fairness and relevance. 

\section{Experimental results}
\label{ss:result-pareto}
We now present the evaluation scores of 16 runs (4 recommenders $\times$ 3 re-rankers, including no reranking) ($\S$\ref{sss:results-diversity}). The relevance and fairness scores of these runs are the input to our DPFR approach. 
Not all combinations of evaluation measures are suited for PF. We explain this in $\S$\ref{sss:no-fit}. We present the generated PF ($\S$\ref{sss:result-pf}) and compare existing measures to DPFR ($\S$\ref{ss:corr}). We compare the results of efficient DPFR to other joint evaluation approaches ($\S$\ref{ss:compare-eff}).

\subsection{Groundwork runs} \label{sss:results-diversity}

The scores of \textsc{Rel}, \textsc{Fair}, and \textsc{Fair+Rel} measures for our 16 runs are shown in the appendix (Tab.~\ref{tab:base-rerank-all-1}--\ref{tab:base-rerank-all-2}). 
Two main findings emerge from Tab.~\ref{tab:base-rerank-all-1}--\ref{tab:base-rerank-all-2}. First, for all six datasets, \textbf{none of the best models according to \textsc{Rel} are also the best according to \textsc{Fair} measures}. This is similar to our toy example (Fig.~\ref{fig:pareto_teaser}), where one model ranks highest for fairness and another for relevance. 
Second, \textbf{the five \textsc{Fair+Rel} measures have no unanimous agreement on the best model}. 
IBO has a different best model from the others in 4/6 times, but sometimes agrees with one or more \textsc{Fair} measures. 
MME and AI-F agree on the best model 5/6 times, and sometimes agree on the best model with \textsc{Fair} measures.
The best model according to IAA and II-F is always the same, and 4/6 times the same as the best model based on the \textsc{Rel} measures. 
The overall picture is inconclusive, with some \textsc{Fair+Rel} measures aligning more with \textsc{Fair} measures, and others aligning more with \textsc{Rel} measures. 

\begin{figure}[htbp]
    \centering
    \includegraphics[width=\columnwidth]{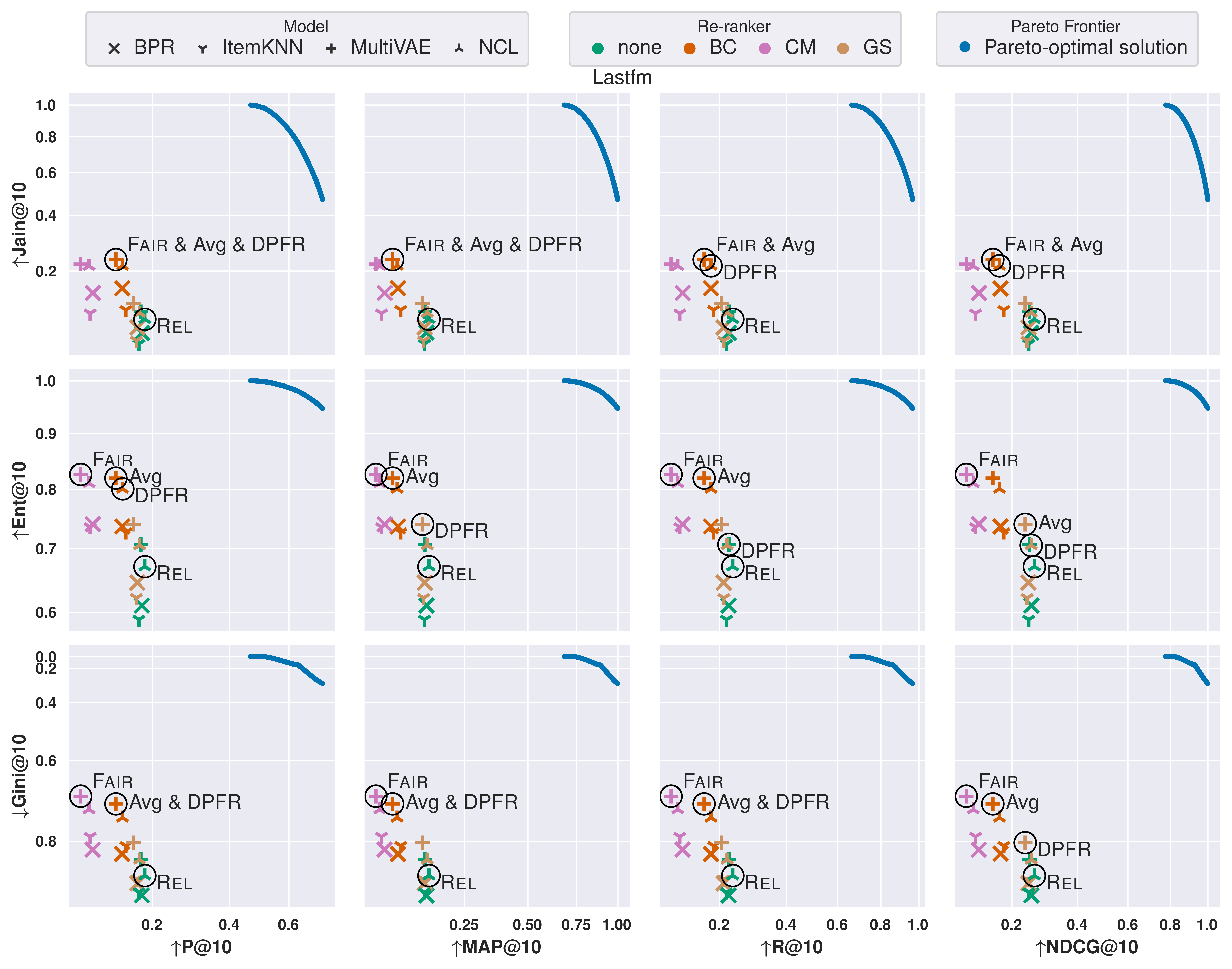}
    \includegraphics[width=\columnwidth]{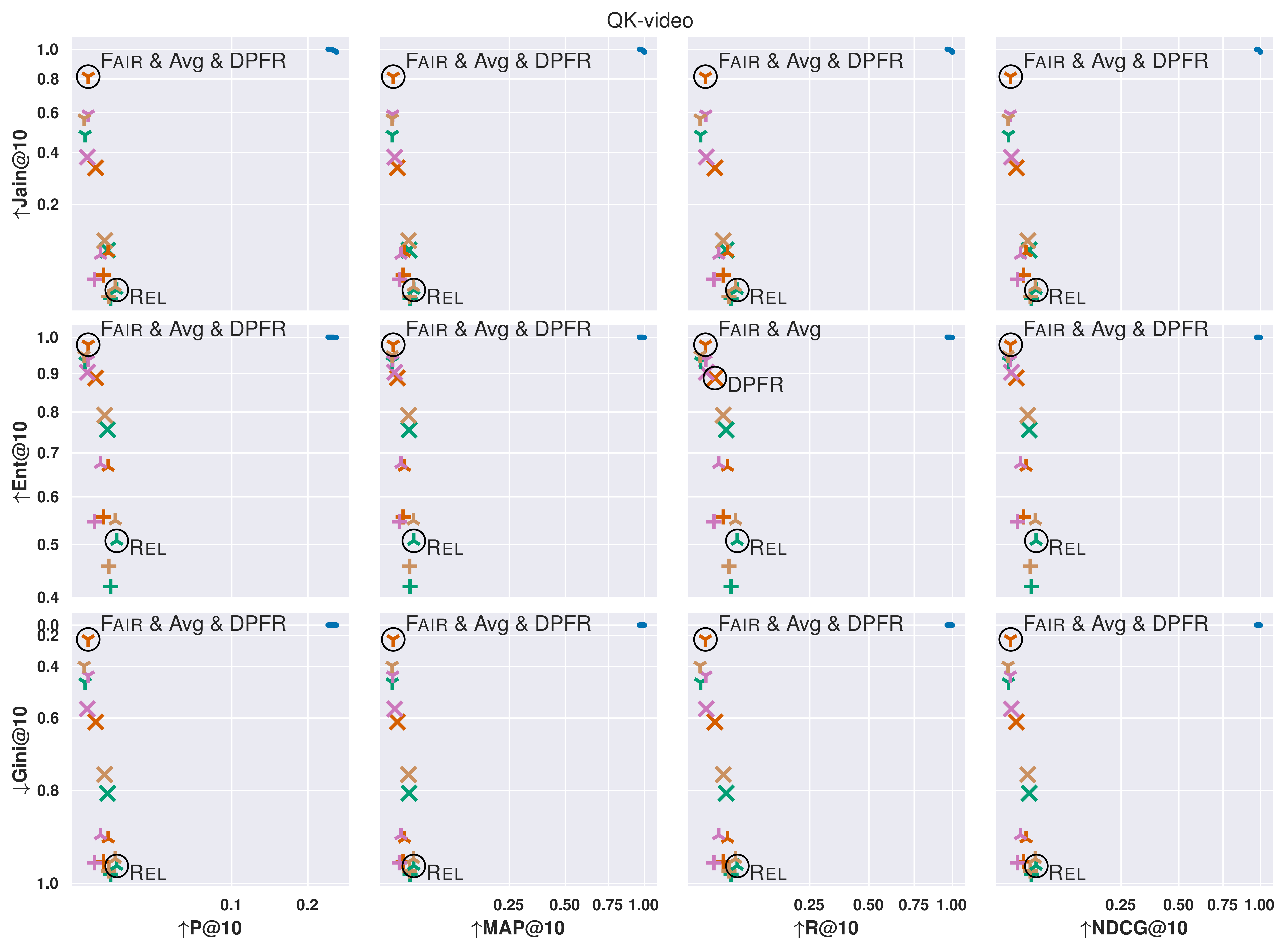}
    \caption{Pareto Frontier of fairness and relevance (in blue) and recommender scores for Lastfm and QK-video on exponential-like scales. 
    \textsc{Rel, Fair}, Avg (mean of \textsc{Rel, Fair}), and DPFR are the best model per evaluation approach.}
    \label{fig:pareto-pair-plot}
\end{figure}

\subsection{Measure compatibility with DPFR}
\label{sss:no-fit}

 Which pairs of \textsc{Rel} and \textsc{Fair} measures are suitable to generate the PF? We answer this based on the PF slope. The slope is calculated using the two endpoints of the PF, i.e., the start and end of the  \textsc{Oracle2Fair} algorithm. A slope of zero means the \textsc{Rel} scores of the PF vary, but the \textsc{Fair} scores do not. As we compute the PF for multiple measures simultaneously, we expect a zero gradient for cases where the initial recommendation according to a \textsc{Fair} measure is already the fairest, even if other \textsc{Fair} scores are not. An undefined gradient value occurs when the initial recommendation is already the fairest and the most relevant according to a pair of \textsc{Fair} and \textsc{Rel} measures. Thus, we posit that a PF with a gradient value other than zero or undefined makes the corresponding pair of measures fit for PF generation (it allows for trade-offs in both aspects). 
The \textsc{Rel}-\textsc{Fair} measure pairs that are fit for DPFR based on their gradient are: \{P, MAP, R, NDCG\} $\times$ \{Jain, Ent, Gini\} (App.~\ref{app:actual-scores}, Tab.~\ref{tab:app-gradient}). Only results from these pairs are shown henceforth. Next, we explain what causes an undefined or zero gradient. 

\noindent \textbf{Causes of zero/undefined gradient.}
Generating the PF requires a ranking of items. Any score that is based on a single relevant item, e.g., HR and MRR, is by design not suitable. 
Out of the \textsc{Fair} measures, QF and FSat sometimes behave inconsistently depending on the dataset properties, as follows. 
A dataset with relatively few relevant items can already be made maximally fair at the start of the PF generation, as QF quantifies fairness with ignorance to frequency of exposure; the score does not change as long as the same set of items appears in the top $k$ recommendations of all users, no matter how many times each. 
When all items in the dataset already occur in the initial recommendations of our Oracle, nothing can be done to improve QF. 
For FSat, in few cases, the score is already maximum at the start of the PF generation. A maximum FSat score is achieved when all items in the dataset have at least the maximum possible exposure, if the available recommendation slots are shared equally across all items.\footnote{$\left\lfloor km/n\right\rfloor$ times (the total number of recommendation slots across users divided by the number of items).} 
In principle, QF and FSat can still be used for DPFR when the initial recommendation by Oracle is not the maximum yet. Otherwise, the interpretation would be less meaningful in joint evaluation, as there is no trade-off between different aspects.

\subsection{The generated PF} \label{sss:result-pf}

Fig.~\ref{fig:pareto-pair-plot} shows the PF plots of the pairs of \textsc{Fair} and \textsc{Rel} measures that are suited for DPFR, only for Lastfm and QK-video, which  are representative of the overall trends in all our datasets (see Fig.~\ref{fig:app-pairplot} in the Appendix). The scores plotted are those computed in $\S$\ref{ss:generation}. The corresponding scores of our recommendation models are in App.~\ref{app:actual-scores}. 
We see that, as the recommendations are made fairer, the generated PF for all datasets is a series of monotonic scores of \textsc{Fair}, specifically monotonic increasing \textsc{Fair} scores (except $\downarrow$Gini), and the remaining measures are monotonic decreasing. 
The monotonic property  theoretically and empirically holds for the \textsc{Fair} measures, as we replace an item with the most exposure by another item with the least exposure, thereby making the recommendation fairer. 
Note that some users do not have exactly $k$ items in the test set, so the perfect relevance score cannot be reached for Precision@$k$ and Recall@$k$ \cite{Moffat2013SevenMetrics}. NDCG and MAP are implemented with normalisation\footnote{Only the first $\min{(|R^*_u|, k)}$ items in a user $u$'s recommendations are considered, where $R^*_u$ is the set of relevant items for user $u$.} so that they can still achieve a score of 1 in this situation. 

The datasets which were randomly split as they have no timestamps (QK-video, Jester) have relatively short, compact PF. This happens because the random split results in a uniform distribution of items in each split, which means that items in the test split are quite diverse (64--100\% of all unique items in the dataset). Considering that the \textsc{Oracle2Fair} algorithm starts by recommending items in the test split and stops when the recommendation reaches the fairest, there is not much room for change in \textsc{Fair} scores, as the initial recommendation is already rather fair. Additionally, there are not many relevant items per user in these datasets (i.e., the median for both datasets is 6 or less); random non-relevant items were chosen to make up for the remaining recommendations.\footnote{The randomly-split Lastfm does not have a short PF because on average it has more relevant items per user compared to QK-video and Jester (see App.~\ref{app:stat}).} 
Thus, the PF generation decreases relevance only marginally in 2/6 cases. 
Correspondingly, we find that in QK-video and Jester, there exist Pareto-optimal recommendations, that are close to maximally fair and maximally relevant, with the exception of P@10. These can be seen in the measure pairs of \{MAP, R, NDCG\} $\times$ \{Jain, Ent, Gini\}, where the PF is close to the coordinates of $(1,1)$, or $(1,0)$ for 
the measure pairs with Gini. Thus, in theory, a fair recommendation does not necessarily have to sacrifice relevance.

\subsection{Agreement between measures} \label{ss:corr}

We study the agreement between DPFR and other evaluation approaches in ranking our 16 runs from best to worst. Low agreement means that the other approaches have few ties to the Pareto-optimal solutions that DPFR uses, and vice versa. 
We compare DPFR to (a) existing \textsc{Rel} and \textsc{Fair} measures, (b) existing joint \textsc{Fair+Rel} measures ($\S$\ref{s:experiments}), and also (c) the average (arithmetic mean) of \textsc{Fair} and \textsc{Rel} scores from the selected measure pairs that are used to generate the PFs. 
To compute the average for a measure where lower values are better (i.e., Gini), we compute $1-$the Gini score instead.

\subsubsection{Comparison of existing measures to DPFR}
We find that for all datasets and all measure pairs, \textbf{the best model as per  DPFR is always different from the best model as per \textsc{Rel} measures}. Moreover, \textbf{half the time, the best model as per DPFR is different from the best model as per a \textsc{Fair} measure}. 
Existing \textsc{Fair+Rel} measures tend to have the same best model as either \textsc{Fair} or \textsc{Rel} measures (73.3\% of the time), instead of having a more balanced evaluation of both aspects. These findings are expected as existing joint evaluation measures use relevance in their formulation differently than the \textsc{Rel} measures. 
Overall, the best model found with DPFR is less skewed towards relevance or fairness.

\subsubsection{Correlation of measures}
\begin{figure}[htbp]
    \centering   \includegraphics[width=0.8\columnwidth, trim = 2.5mm 2.5mm 2mm 0mm, clip,]{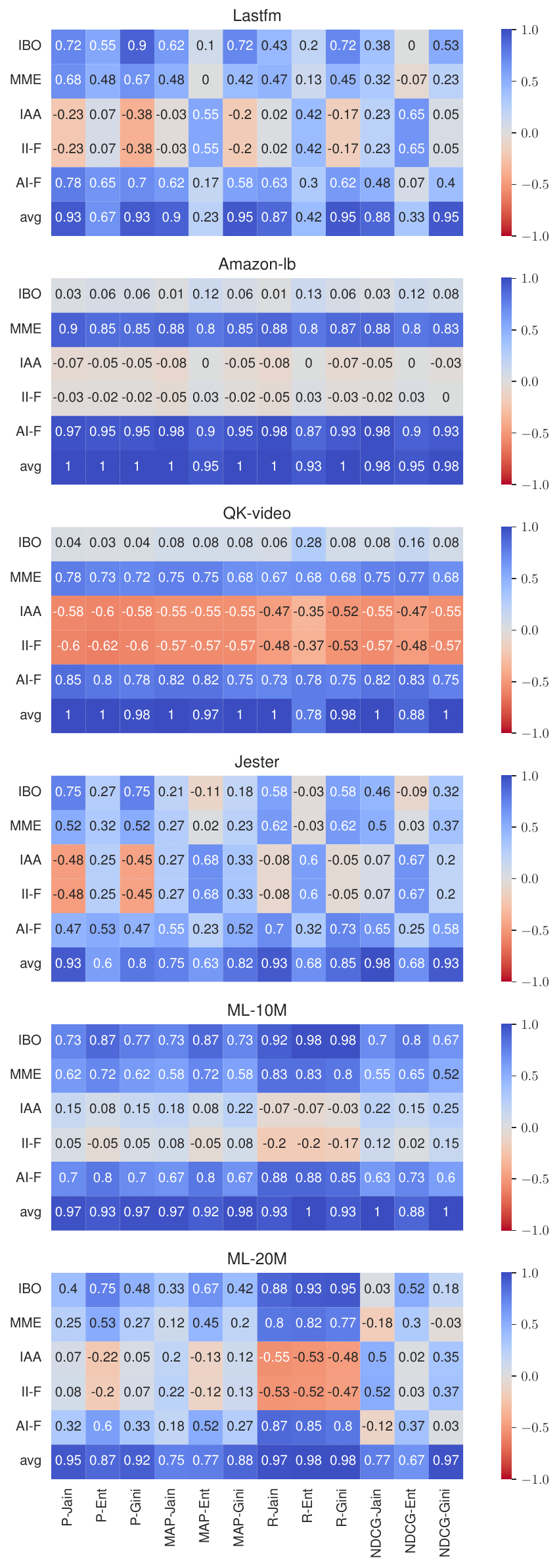}
    \caption{ 
    Kendall's $\tau$ correlation heatmap between the rank ordering of existing joint evaluation measures (including the average of \textsc{Fair} and \textsc{Rel} scores, avg), and DPFR.}
    \label{fig:corrplot}
\end{figure}

For each dataset, we compute the Kendall's\footnote{Ties are handled, unlike in Spearman's $\rho$.} $\tau$ \cite{KENDALL1945THEPROBLEMS} correlations between the ranking given by DPFR and by the joint evaluation baselines (see Fig.~\ref{fig:corrplot}). 
Rankings are considered equivalent if $\tau \geq 0.9$ \cite{Maistro2021PrincipledRankings, Voorhees2001EvaluationDocuments}. 
We see similar agreement trends in datasets where recommenders have higher \textsc{Rel} scores (Lastfm and Jester) or lower (Amazon-lb and QK-video). 

Overall, most times DPFR orders models differently ($\tau<0.9$) than all \textsc{Fair+Rel} measures except AI-F. We see similar trends between IAA and II-F, and between MME and AI-F. IBO can have similar trends as MME (except for Amazon-lb and QK-video). For all datasets, IAA and II-F have overall either weak or negative $\tau$ with DPFR (e.g., $[-0.2, 0.25]$ for ML-10M and $[-0.62, -0.35]$ for QK-video). A notable exception is DPFR with \{MAP, R, NDCG\} $\times$ {Ent} for Lastfm and Jester, where we see moderate correlations, $\tau \in [0.42,0.68]$.

Ent differs from this trend because DPFR with \{MAP, R, NDCG\} $\times$ \{Jain, Gini\} has PF gradients of greater magnitude. This only affects Lastfm and Jester (they have higher \textsc{Rel} scores than the other datasets). DPFR with P has different patterns from other \textsc{Rel} measures: the raw DPFR scores of pairs involving P are lower on average, as the scores from Oracle do not start from $1$, but much less, and therefore closer to the models' scores (Fig.~\ref{fig:pareto-pair-plot}). 
Meanwhile, IBO has varying $\tau$ across datasets: a huge range of $\tau$, i.e., $[0.00,0.9]$ for Lastfm, weak correlations $[0.01, 0.13]$ for Amazon-lb, and moderate to strong correlations $[0.67, 0.98]$ for ML-10M. These variations might be because IBO is based on the number of items satisfying a certain criterion, rather than an average of scores across users and/or items, i.e., how other \textsc{Fair+Rel} measures are defined. 

Among the joint measures, AI-F correlates the strongest with DPFR, as both AI-F and DPFR, indirectly or directly, consider 
the recommendation frequency of each item and compare it with that of other items. 
However, the rank orderings given by AI-F are not equivalent to DPFR, as $\tau < 0.9$ for 5/6 datasets (excl. Amazon-lb). 
For the same measure-pair and between datasets, the $\tau$ of AI-F and DPFR also varies a lot. E.g., $\tau=0.07$ for NDCG-Ent for Lastfm, but $\tau=0.9$ 
for Amazon-lb. We thereby do not recommend using any of the \textsc{Fair+Rel} measures (none correlates with Pareto optimality). 

Taking the mean of \textsc{Fair} and \textsc{Rel} scores (avg) at a glance seems to correlate highly with DPFR. However, while it gives equivalent rankings ($\tau\geq 0.9$) in some cases (e.g., for Amazon-lb, most of ML-10M and QK-video, and half of ML-20M), it only does so for (1) datasets with lower \textsc{Rel} scores (Amazon-lb, QK-video), i.e., in cases where all models perform poorly, we have low variance in \textsc{Rel}, which leads to fairness dominating both avg and DPFR; (2) datasets with low variance in \textsc{Fair} scores (ML-*).  
In such cases, quantifying the evaluation jointly is challenging as one aspect dominates over the other. 
In the other datasets, the rank ordering given by the average is inconsistent: sometimes $\tau \geq 0.9$ for one dataset, but not for the others. This inconsistency between datasets holds for all measure pairs, except for P-Jain and NDCG-Gini. Due to these inconsistencies, we discourage using the arithmetic mean. 

Overall, our correlation analysis shows that existing joint \textsc{Fair+Rel} evaluation measures cannot be used as a reliable proxy for DPFR. 

\subsubsection{Best model disagreement} 

We take a closer look at how DPFR relates to computing averages, as they are similar approaches in terms of combining scores from a measure pair. As comparing the raw scores of DPFR and the average is invalid, we instead count the disagreement between the best model based on DPFR and the mean of \textsc{Fair} and \textsc{Rel} scores  (Tab.~\ref{tab:best_model_disagreement}). 
The aim is to study whether one would come to the same conclusion regarding the best model, using the two different joint evaluation approaches.

Among the 12 measure pairs that are fit for DPFR, we find that \textbf{the best model according to DPFR is not always the same according to the average of \textsc{Fair} and \textsc{Rel} scores of the same pair}; in one case the disagreement is up to 58.33\% (for Lastfm). The disagreement is generally much higher in the more complex rank-based measures (0--83.33\%) compared to simpler set-based \textsc{Rel} measures (0--50.00\%). 
Therefore, there are many cases where the mean of \textsc{Fair} and \textsc{Rel} scores is not the best case, especially for Lastfm and Jester where \textsc{Rel} scores are higher and vary more. 
In these two datasets, more often than not, DPFR leads to different conclusions than a simple average. Yet, sometimes the average agrees with DPFR in the best model: for QK-video, disagreement is low (8.33\%), and there is a perfect agreement on the best model for Amazon-lb; we posit that these are due to equally poor and low variance in the \textsc{Rel} scores. This is in line with our correlation analysis. 
As there is a huge range of variability across datasets (0--58.33\%), we do not recommend using a simple average to get the same result as DPFR, as it is unreliable and inconsistent. 
Generally, averaging fails to reach the same conclusion as DPFR almost half the time, especially when the \textsc{Rel} and \textsc{Fair} scores vary highly.

\begin{table}[tbp]
\centering
\caption{The percentage of best model disagreement when taking the mean of \textsc{Fair} and \textsc{Rel} scores as opposed to using DPFR, separated by the \textsc{Rel} measure type.  P@$k$ and R@$k$ are set-based, NDCG and MAP are rank-based. We only consider the 12 measure pairs with a nonzero, defined gradient ($\S$\ref{sss:no-fit}).}
\label{tab:best_model_disagreement}
\resizebox{0.70\columnwidth}{!}{
\begin{tabular}{lrrr}
\toprule
{} &                    Set-based & Rank-based &    All \\
\hline
\midrule
Lastfm    &          50.00 &      66.67 &  58.33 \\
Amazon-lb &          0.00 &       0.00 &   0.00 \\
QK-video  &          16.67 &       0.00 &   8.33 \\
Jester    &          16.67 &      83.33 &  50.00 \\
ML-10M    &           0.00 &      66.67 &  33.33 \\
ML-20M       &       0.00 &       50.00 &  25.00 \\
\hline
\midrule
All datasets &      13.89 &       44.44 &  29.17 \\
\bottomrule
\end{tabular}
}
\end{table}

\subsection{Efficient DPFR} 
\label{ss:compare-eff}

\subsubsection{The efficiency of the PF generation} 
We study the efficiency of DPFR by comparing the PF, an estimated version of PF on a subset of points, and the \textsc{Fair+Rel} measures. The estimated version of PF uses 3--12 points as per $\S$\ref{ss:compute-eff}. 
We compute the amount of points in the estimated PF as \% of those in the PF, and the resulting computation times. One point in the PF translates to one round of computing all \textsc{Fair} and \textsc{Rel} measures, so fewer points mean faster. For brevity, we report the estimated PF with only 3, 6, and 12 points in Tab.~\ref{tab:big-est-time} 
(see App.~\ref{app:actual-scores}, Tab.~\ref{tab:corr_est} for extended results).
 \balance

\begin{table}[tbp]
    \centering
    \caption{Efficiency and effectiveness of PF, estimated PF (Est.~PF), and \textsc{Fair+Rel} measures: \% of data points in the Est.~PF (\% pts), computation time. Dist is the average distance between midpoints in the Est.~PF and PF over 12 measure pairs. Minimum agreement (Min $\tau$) is the Kendall's $\tau$ correlation between DPFR with PF and Est.~PF. 
    Both PFs compute 11 \textsc{Rel} and \textsc{Fair} measures simultaneously. 
    The times for other evaluation measures are averaged (Avg/model) and summed (All models) over 16 model combinations.}
    \label{tab:big-est-time}
\scalebox{0.65}{
     \begin{tabular}{lllrrrrrr}
\toprule
& & {} &  Lastfm &  Amazon-lb &  QK-video &  Jester &  ML-10M &  ML-20M \\
\midrule
\multicolumn{2}{c}{\#pts}& PF& 4882 & 847 & 499 & 16202 & 2781 & 3783 \\
\cline{1-9}
\multicolumn{2}{c}{\multirow[c]{3}{*}{\%pts}} & Est.~PF (12 pts) & 0.25 &      1.42 &     2.40 &   0.07 &   0.43 &   0.32 \\
& & Est.~PF (6 pts) &  0.12 &      0.71 &     1.20 &   0.04 &   0.22 &   0.16 \\
& & Est.~PF (3 pts) & 0.06 &      0.35 &     0.60 &   0.02 &   0.11 &   0.08\\
\hline
\midrule
\multirow[c]{14}{*}{\rotatebox[origin=c]{90}{$\downarrow$ Computation time (mins.)}}& & PF           &   19.18 &       0.56 &     10.49 &  847.42 &   28.99 &   75.77 \\
& & Est.~PF (12 pts) &    2.02 &       0.19 &      4.23 &  552.16 &    1.90 &    2.60 \\
& & Est.~PF (6 pts)  &    2.00 &       0.19 &      4.12 &  551.72 &    1.84 &    2.54 \\
& & Est.~PF (3 pts)  &    2.01 &      0.19 &     4.07 &  552.26 &   1.82 &   2.52\\
\cline{2-9}
& \multirow[c]{5}{*}{\rotatebox[origin=c]{90}{Avg/model}} & IBO    &    <0.3s &     <0.3s &      0.01 &    0.01 &   <0.3s &     0.01  \\
& & MME    &    2.04 &       0.03 &     19.51 &    0.09 &   15.25 &    89.13 \\
& & IAA    &    <0.3s &     <0.3s &      0.01 &    0.02 &    0.01 &     0.02 \\
& & II-F   &    <0.3s &     <0.3s &      0.01 &    0.01 &    0.01 &     0.02 \\
& & AI-F   &    <0.3s &     <0.3s &      0.01 &    0.01 &   <0.3s &     0.01 \\
\cline{2-9}
& \multirow[c]{5}{*}{\rotatebox[origin=c]{90}{All models}} & IBO  &    0.02 &     <0.3s &      0.10 &    0.12 &    0.06 &     0.15 \\
& & MME  &   32.63 &       0.49 &    312.14 &    1.38 &  244.03 &  1426.10 \\
& & IAA  &    0.03 &     <0.3s &      0.10 &    0.36 &    0.13 &     0.30  \\
& & II-F &    0.04 &     <0.3s &      0.16 &    0.14 &     0.1 &     0.25 \\
& & AI-F &   0.03 &     <0.3s &      0.11 &    0.13 &    0.07 &     0.17 \\
\hline
\midrule
\multicolumn{2}{c}{\multirow[c]{3}{*}{$\uparrow$ 
Min $\tau$}} & Est.~PF (12 pts)& 0.95 &       1.00 &      1.00 &    0.98 &    0.98 &    0.97\\
\multicolumn{2}{c}{} &  Est.~PF (6 pts)&  0.90 &       0.97 &      1.00 &    0.98 &    0.95 &    0.92\\
\multicolumn{2}{c}{} &  Est.~PF (3 pts)&  0.78 &       0.98 &      1.00 &    1.00 &    0.97 &    0.75\\
\midrule
\multicolumn{2}{c}{\multirow[c]{3}{*}{$\downarrow$ Dist.}} & Est.~PF (12 pts)& 0.01 &      0.02 &     0.00 &   0.00 &   0.01 &   0.01 \\
\multicolumn{2}{c}{} &  Est.~PF (6 pts)& 0.03 &      0.05 &     0.00 &   0.00 &   0.02 &   0.02 \\
\multicolumn{2}{c}{} &  Est.~PF (3 pts)& 0.03 &      0.05 &     0.00 &   0.00 &   0.03 &   0.05 \\
\bottomrule
\end{tabular}
}
\end{table}

The PF (Fig.~\ref{fig:pareto-pair-plot}) has hundreds to tens of thousands of points each, while the estimated PF only contains 0.02--2.40\% of the points, which means reduced computational complexity for the PF generation. In terms of actual computation time (Tab.~\ref{tab:big-est-time}), computing the PF with \textsc{Oracle2Fair} take 0.56--75.77 mins to compute, but only 0.19--4.07 mins for the PFs estimated with 3 points for all datasets except Jester. For Jester, it takes $\sim$14 hours and the estimation only takes $\sim$9 hours. 
However, this is expected for Jester as it has 62K users in the test split, as opposed to the 3.5K or fewer in the other datasets (i.e., see Tab.~\ref{tab:data_split} in App.~\ref{app:stat}). 
While computing the estimated PFs is on average slower than computing the joint measures IBO, IAA, II-F, and AI-F, it is expected as the (estimated) PFs compute 11 measures simultaneously. Yet, in most cases (except Amazon-lb and Jester), the estimated PFs are still faster to compute than the time to compute MME for one model per dataset, let alone to compute MME for all models. For ML-20M, computing the estimated PF is even up to 35 times faster than computing MME of one model.

\subsubsection{The effectiveness of efficient DPFR} 
To what extent is the DPFR from the efficiently generated PF (estimated PF) a reasonable proxy for fairness-relevance joint evaluation using the PF, in terms of giving a similar ordering of models? 
We compare the DPFR from the PF and estimated PF using Kendall's $\tau$ correlations. Further, as DPFR is computed based on a $\alpha$-based reference point lying on the PF, to quantify possible accuracy loss of the estimated PF, in Tab.~\ref{tab:big-est-time} we show the error of the midpoint estimation. This error is the Euclidean distance between the reference point in the PF and estimated PF (i.e., the midpoint in our case), as per \cite{Wang2016AEngineering}.

We first analyse the error of the midpoint estimation. For the 12 measure pairs and 6 datasets, the midpoint coordinates on average do not move much: the distance is 0.00--0.05, even when the PF is only estimated with 3 points. 
Ergo, the correlations between the rank ordering of models given by the DPFR of PF and its estimation, are still equivalent ($\tau \geq 0.9$) when estimated with 6 or 12 points \cite{Maistro2021PrincipledRankings,Voorhees2001EvaluationDocuments}. Even the 3-point estimation maintains high agreement ($\tau \in [0.75,1]$), with only 5 cases having $\tau<0.9$ across 6 datasets and 12 measure-pairs. 
Therefore, it is possible to only compute a small number of points in the PF, e.g., 6 or 12 points, and still make a reliable PF estimation, evidenced by the small shift of the PF midpoint and the rank ordering of the models remaining equivalent ($\tau \geq 0.9$), if not identical ($\tau=1$) for all measure pairs and datasets. 

\section{Discussion and conclusions}
Recommendation evaluation has long used measures that quantify only relevance, but has recently shifted to include fairness. However, there exists no de-facto way to robustly quantify these two aspects. We propose a novel approach (DPFR) that uses fairness and relevance measures under a joint evaluation scheme for RSs. DPFR computes the empirical best possible recommendation, jointly accounting for a given pair of relevance and fairness measures, in a principled way according to Pareto-optimality. DPFR is modular, tractable, and intuitively understandable. It can be used with existing measures for relevance and fairness, and allows for different trade-offs of relevance and fairness. We empirically show that existing evaluation measures of fairness w.r.t.~relevance \cite{Biega2018EquityRankings,Diaz2020EvaluatingExposure,Wu2022JointRecommendation, Saito2022FairRanking, Borges2019EnhancingAutoencoders} behave inconsistently: they disagree with optimal solutions based on DPFR computed on more robust and well-understood measures of relevance, such as NDCG, and fairness, such as Gini. 
We uncover some weaknesses of these measures, but more research is warranted to study their behaviour. Admittedly, existing joint measures are not originally defined to be aligned with existing relevance and fairness measures \cite{jain1984quantitative,Zhu2020FARM:APPs,Gini1912VariabilitaMutabilita,Mansoury2021ASystems,Mansoury2020FairMatch:Systems,Patro2020FairRec:Platforms}, so it is not surprising that they have different results from DPFR. However, existing measures show varying performance also from each other and from well-understood relevance and fairness measures. Thus, DPFR can be a viable alternative for robust, interpretable, and provenly optimal evaluation in offline scenarios. We also show that DPFR can be computed fast while reaching equivalent conclusions. 
Overall, DPFR demonstrates distinct benefits in mitigating false conclusions by up to 50\% compared to basic aggregation methods like averaging. Surprisingly, simple averaging aligns more with our Pareto-optimal based DPFR, than existing joint measures. 
We recommend combining either MAP-Ent or NDCG-Ent: the conclusions are distinguishable from simply averaging, or taking the best model 
based on fairness or relevance measures. 

\clearpage
\begin{acks}
The work is supported by the Algorithms, Data, and Democracy project (ADD-project), funded by Villum Foundation and Velux Foundation. Qiuchi Li contributed to the idea of computing the reference point. We thank the DIKU IR Lab and the anonymous reviewers, who have provided helpful feedback to improve earlier versions of the manuscript.
\end{acks}

\bibliographystyle{ACM-Reference-Format}
\bibliography{references}


\begin{thebibliography}{56}


\ifx \showCODEN    \undefined \def \showCODEN     #1{\unskip}     \fi
\ifx \showDOI      \undefined \def \showDOI       #1{#1}\fi
\ifx \showISBNx    \undefined \def \showISBNx     #1{\unskip}     \fi
\ifx \showISBNxiii \undefined \def \showISBNxiii  #1{\unskip}     \fi
\ifx \showISSN     \undefined \def \showISSN      #1{\unskip}     \fi
\ifx \showLCCN     \undefined \def \showLCCN      #1{\unskip}     \fi
\ifx \shownote     \undefined \def \shownote      #1{#1}          \fi
\ifx \showarticletitle \undefined \def \showarticletitle #1{#1}   \fi
\ifx \showURL      \undefined \def \showURL       {\relax}        \fi
\providecommand\bibfield[2]{#2}
\providecommand\bibinfo[2]{#2}
\providecommand\natexlab[1]{#1}
\providecommand\showeprint[2][]{arXiv:#2}

\bibitem[Amig{\'{o}} et~al\mbox{.}(2023)]%
        {Amigo2023ASystems}
\bibfield{author}{\bibinfo{person}{Enrique Amig{\'{o}}}, \bibinfo{person}{Yashar Deldjoo}, \bibinfo{person}{Stefano Mizzaro}, {and} \bibinfo{person}{Alejandro Bellog{\'{i}}n}.} \bibinfo{year}{2023}\natexlab{}.
\newblock \showarticletitle{{A unifying and general account of fairness measurement in recommender systems}}.
\newblock \bibinfo{journal}{\emph{Information Processing {\&} Management}} \bibinfo{volume}{60}, \bibinfo{number}{1} (\bibinfo{date}{1} \bibinfo{year}{2023}), \bibinfo{pages}{103115}.
\newblock
\showISSN{0306-4573}
\urldef\tempurl%
\url{https://doi.org/10.1016/J.IPM.2022.103115}
\showDOI{\tempurl}


\bibitem[Audet et~al\mbox{.}(2020)]%
        {Audet2020PerformanceOptimization}
\bibfield{author}{\bibinfo{person}{Charles Audet}, \bibinfo{person}{Jean Bigeon}, \bibinfo{person}{Dominique Cartier}, \bibinfo{person}{Sébastien Le~Digabel}, {and} \bibinfo{person}{Ludovic Salomon}.} \bibinfo{year}{2020}\natexlab{}.
\newblock \showarticletitle{{Performance indicators in multiobjective optimization}}.
\newblock \bibinfo{journal}{\emph{European Journal of Operational Research}} \bibinfo{volume}{292}, \bibinfo{number}{2} (\bibinfo{year}{2020}), \bibinfo{pages}{397--422}.
\newblock
\urldef\tempurl%
\url{https://doi.org/10.1016/j.ejor.2020.11.016}
\showDOI{\tempurl}


\bibitem[Biega et~al\mbox{.}(2018)]%
        {Biega2018EquityRankings}
\bibfield{author}{\bibinfo{person}{Asia~J. Biega}, \bibinfo{person}{Krishna~P. Gummadi}, {and} \bibinfo{person}{Gerhard Weikum}.} \bibinfo{year}{2018}\natexlab{}.
\newblock \showarticletitle{{Equity of attention: Amortizing individual fairness in rankings}}.
\newblock \bibinfo{journal}{\emph{41st International ACM SIGIR Conference on Research and Development in Information Retrieval, SIGIR 2018}}  \bibinfo{volume}{18} (\bibinfo{date}{6} \bibinfo{year}{2018}), \bibinfo{pages}{405--414}.
\newblock
\showISBNx{9781450356572}
\urldef\tempurl%
\url{https://doi.org/10.1145/3209978.3210063}
\showDOI{\tempurl}


\bibitem[Borges and Stefanidis(2019)]%
        {Borges2019EnhancingAutoencoders}
\bibfield{author}{\bibinfo{person}{Rodrigo Borges} {and} \bibinfo{person}{Kostas Stefanidis}.} \bibinfo{year}{2019}\natexlab{}.
\newblock \showarticletitle{{Enhancing Long Term Fairness in Recommendations with Variational Autoencoders}}. In \bibinfo{booktitle}{\emph{Proceedings of the 11th International Conference on Management of Digital EcoSystems}}. \bibinfo{publisher}{ACM}, \bibinfo{address}{New York, NY, USA}, \bibinfo{pages}{95--102}.
\newblock
\showISBNx{9781450362382}
\urldef\tempurl%
\url{https://doi.org/10.1145/3297662}
\showDOI{\tempurl}


\bibitem[Cantador et~al\mbox{.}(2011)]%
        {Cantador20112nd2011}
\bibfield{author}{\bibinfo{person}{Iván Cantador}, \bibinfo{person}{Peter Brusilovsky}, {and} \bibinfo{person}{Tsvi Kuflik}.} \bibinfo{year}{2011}\natexlab{}.
\newblock \showarticletitle{{2nd Workshop on Information Heterogeneity and Fusion in Recommender Systems (HetRec 2011)}}. In \bibinfo{booktitle}{\emph{Proceedings of the 5th ACM conference on Recommender systems}} \emph{(\bibinfo{series}{RecSys 2011})}. \bibinfo{publisher}{ACM}, \bibinfo{address}{New York, NY, USA}.
\newblock


\bibitem[Censor(1977)]%
        {Censor1977ParetoProblems}
\bibfield{author}{\bibinfo{person}{Yair Censor}.} \bibinfo{year}{1977}\natexlab{}.
\newblock \showarticletitle{{Pareto optimality in multiobjective problems}}.
\newblock \bibinfo{journal}{\emph{Applied Mathematics {\&} Optimization}} \bibinfo{volume}{4}, \bibinfo{number}{1} (\bibinfo{date}{3} \bibinfo{year}{1977}), \bibinfo{pages}{41--59}.
\newblock
\showISSN{14320606}
\urldef\tempurl%
\url{https://doi.org/10.1007/BF01442131/METRICS}
\showDOI{\tempurl}


\bibitem[Deshpande and Karypis(2004)]%
        {Deshpande2004Item-basedAlgorithms}
\bibfield{author}{\bibinfo{person}{Mukund Deshpande} {and} \bibinfo{person}{George Karypis}.} \bibinfo{year}{2004}\natexlab{}.
\newblock \showarticletitle{{Item-based top-N recommendation algorithms}}.
\newblock \bibinfo{journal}{\emph{ACM Transactions on Information Systems}} \bibinfo{volume}{22}, \bibinfo{number}{1} (\bibinfo{date}{1} \bibinfo{year}{2004}), \bibinfo{pages}{143--177}.
\newblock
\showISSN{10468188}
\urldef\tempurl%
\url{https://doi.org/10.1145/963770.963776}
\showDOI{\tempurl}


\bibitem[Diaz et~al\mbox{.}(2020)]%
        {Diaz2020EvaluatingExposure}
\bibfield{author}{\bibinfo{person}{Fernando Diaz}, \bibinfo{person}{Bhaskar Mitra}, \bibinfo{person}{Michael~D Ekstrand}, \bibinfo{person}{Asia~J Biega}, {and} \bibinfo{person}{Ben Carterette}.} \bibinfo{year}{2020}\natexlab{}.
\newblock \showarticletitle{{Evaluating Stochastic Rankings with Expected Exposure}}. In \bibinfo{booktitle}{\emph{Proceedings of the 29th ACM International Conference on Information {\&} Knowledge Management}}. \bibinfo{publisher}{ACM}, \bibinfo{address}{New York, NY, USA}, \bibinfo{pages}{275--284}.
\newblock
\showISBNx{9781450368599}
\urldef\tempurl%
\url{https://doi.org/10.1145/3340531}
\showDOI{\tempurl}


\bibitem[Gao et~al\mbox{.}(2022)]%
        {Gao2022FAIR:Evaluation}
\bibfield{author}{\bibinfo{person}{Ruoyuan Gao}, \bibinfo{person}{Yingqiang Ge}, {and} \bibinfo{person}{Chirag Shah}.} \bibinfo{year}{2022}\natexlab{}.
\newblock \showarticletitle{{FAIR: Fairness-aware information retrieval evaluation}}.
\newblock \bibinfo{journal}{\emph{Journal of the Association for Information Science and Technology}} \bibinfo{volume}{73}, \bibinfo{number}{10} (\bibinfo{date}{10} \bibinfo{year}{2022}), \bibinfo{pages}{1461--1473}.
\newblock
\showISSN{2330-1643}
\urldef\tempurl%
\url{https://doi.org/10.1002/ASI.24648}
\showDOI{\tempurl}


\bibitem[Ge et~al\mbox{.}(2022)]%
        {Ge2022TowardLearning}
\bibfield{author}{\bibinfo{person}{Yingqiang Ge}, \bibinfo{person}{Xiaoting Zhao}, \bibinfo{person}{Lucia Yu}, \bibinfo{person}{Saurabh Paul}, \bibinfo{person}{Diane Hu}, \bibinfo{person}{Chu~Cheng Hsieh}, {and} \bibinfo{person}{Yongfeng Zhang}.} \bibinfo{year}{2022}\natexlab{}.
\newblock \showarticletitle{{Toward pareto efficient fairness-utility trade-off in recommendation through reinforcement learning}}. In \bibinfo{booktitle}{\emph{WSDM 2022 - Proceedings of the 15th ACM International Conference on Web Search and Data Mining}}. \bibinfo{publisher}{Association for Computing Machinery, Inc}, \bibinfo{address}{Virtual Event, AZ, USA}, \bibinfo{pages}{316--324}.
\newblock
\showISBNx{9781450391320}
\urldef\tempurl%
\url{https://doi.org/10.1145/3488560.3498487}
\showDOI{\tempurl}


\bibitem[Gini(1912)]%
        {Gini1912VariabilitaMutabilita}
\bibfield{author}{\bibinfo{person}{C. Gini}.} \bibinfo{year}{1912}\natexlab{}.
\newblock \bibinfo{booktitle}{\emph{{Variabilit{\`{a}} e mutabilit{\`{a}}}}}.
\newblock \bibinfo{publisher}{Tipogr. di P. Cuppini}, \bibinfo{address}{Rome}.
\newblock


\bibitem[Goldberg et~al\mbox{.}(2001)]%
        {Goldberg2001Eigentaste:Algorithm}
\bibfield{author}{\bibinfo{person}{Ken Goldberg}, \bibinfo{person}{Theresa Roeder}, \bibinfo{person}{Dhruv Gupta}, {and} \bibinfo{person}{Chris Perkins}.} \bibinfo{year}{2001}\natexlab{}.
\newblock \showarticletitle{{Eigentaste: A Constant Time Collaborative Filtering Algorithm}}.
\newblock \bibinfo{journal}{\emph{Information Retrieval}} \bibinfo{volume}{4}, \bibinfo{number}{2} (\bibinfo{date}{7} \bibinfo{year}{2001}), \bibinfo{pages}{133--151}.
\newblock
\showISSN{13864564}
\urldef\tempurl%
\url{https://doi.org/10.1023/A:1011419012209/METRICS}
\showDOI{\tempurl}


\bibitem[Harper and Konstan(2015)]%
        {Harper2015TheContext}
\bibfield{author}{\bibinfo{person}{F~Maxwell Harper} {and} \bibinfo{person}{Joseph~A Konstan}.} \bibinfo{year}{2015}\natexlab{}.
\newblock \showarticletitle{{The MovieLens datasets: History and context}}.
\newblock \bibinfo{journal}{\emph{ACM Trans. Interact. Intell. Syst.}} \bibinfo{volume}{5}, \bibinfo{number}{4} (\bibinfo{year}{2015}), \bibinfo{pages}{1--19}.
\newblock
\urldef\tempurl%
\url{https://doi.org/10.1145/2827872}
\showDOI{\tempurl}


\bibitem[Jain et~al\mbox{.}(1998)]%
        {jain1984quantitative}
\bibfield{author}{\bibinfo{person}{Rajendra~K Jain}, \bibinfo{person}{Dah-Ming~W Chiu}, \bibinfo{person}{William~R Hawe}, {and} \bibinfo{person}{{others}}.} \bibinfo{year}{1998}\natexlab{}.
\newblock \bibinfo{title}{{A Quantitative Measure Of Fairness And Discrimination For Resource Allocation In Shared Computer Systems}}.
\newblock
\newblock
\urldef\tempurl%
\url{http://arxiv.org/abs/cs/9809099}
\showURL{%
\tempurl}


\bibitem[Kendall(1945)]%
        {KENDALL1945THEPROBLEMS}
\bibfield{author}{\bibinfo{person}{M.~G. Kendall}.} \bibinfo{year}{1945}\natexlab{}.
\newblock \showarticletitle{{The Treatment of Ties in Ranking Problems}}.
\newblock \bibinfo{journal}{\emph{Biometrika}} \bibinfo{volume}{33}, \bibinfo{number}{3} (\bibinfo{date}{11} \bibinfo{year}{1945}), \bibinfo{pages}{239--251}.
\newblock
\showISSN{0006-3444}
\urldef\tempurl%
\url{https://doi.org/10.1093/BIOMET/33.3.239}
\showDOI{\tempurl}


\bibitem[Laszczyk and Myszkowski(2019)]%
        {Laszczyk2019SurveyMeasures}
\bibfield{author}{\bibinfo{person}{Maciej Laszczyk} {and} \bibinfo{person}{Paweł~B. Myszkowski}.} \bibinfo{year}{2019}\natexlab{}.
\newblock \showarticletitle{{Survey of quality measures for multi-objective optimization: Construction of complementary set of multi-objective quality measures}}.
\newblock \bibinfo{journal}{\emph{Swarm and Evolutionary Computation}}  \bibinfo{volume}{48} (\bibinfo{date}{8} \bibinfo{year}{2019}), \bibinfo{pages}{109--133}.
\newblock
\showISSN{2210-6502}
\urldef\tempurl%
\url{https://doi.org/10.1016/J.SWEVO.2019.04.001}
\showDOI{\tempurl}


\bibitem[Lazovich et~al\mbox{.}(2022)]%
        {Lazovich2022MeasuringMetrics}
\bibfield{author}{\bibinfo{person}{Tomo Lazovich}, \bibinfo{person}{Luca Belli}, \bibinfo{person}{Aaron Gonzales}, \bibinfo{person}{Amanda Bower}, \bibinfo{person}{Uthaipon Tantipongpipat}, \bibinfo{person}{Kristian Lum}, \bibinfo{person}{Ferenc Husz{\'{a}}r}, {and} \bibinfo{person}{Rumman Chowdhury}.} \bibinfo{year}{2022}\natexlab{}.
\newblock \showarticletitle{{Measuring disparate outcomes of content recommendation algorithms with distributional inequality metrics}}.
\newblock \bibinfo{journal}{\emph{Patterns}} \bibinfo{volume}{3}, \bibinfo{number}{8} (\bibinfo{date}{8} \bibinfo{year}{2022}).
\newblock
\showISSN{2666-3899}
\urldef\tempurl%
\url{https://doi.org/10.1016/j.patter.2022.100568}
\showDOI{\tempurl}


\bibitem[Lee(1997)]%
        {Lee1997AnalysesCombination}
\bibfield{author}{\bibinfo{person}{Joon~Ho Lee}.} \bibinfo{year}{1997}\natexlab{}.
\newblock \showarticletitle{{Analyses of multiple evidence combination}}. In \bibinfo{booktitle}{\emph{SIGIR '97: Proceedings of the 20th annual international ACM SIGIR conference on Research and development in information retrieval}}. \bibinfo{publisher}{Association for Computing Machinery (ACM)}, \bibinfo{address}{Philadelphia}, \bibinfo{pages}{267--276}.
\newblock
\urldef\tempurl%
\url{https://doi.org/10.1145/258525.258587}
\showDOI{\tempurl}


\bibitem[Liang et~al\mbox{.}(2018)]%
        {Liang2018VariationalFiltering}
\bibfield{author}{\bibinfo{person}{Dawen Liang}, \bibinfo{person}{Rahul~G. Krishnan}, \bibinfo{person}{Matthew~D. Hoffman}, {and} \bibinfo{person}{Tony Jebara}.} \bibinfo{year}{2018}\natexlab{}.
\newblock \showarticletitle{{Variational autoencoders for collaborative filtering}}.
\newblock \bibinfo{journal}{\emph{The Web Conference 2018 - Proceedings of the World Wide Web Conference, WWW 2018}}  \bibinfo{volume}{10} (\bibinfo{date}{4} \bibinfo{year}{2018}), \bibinfo{pages}{689--698}.
\newblock
\showISBNx{9781450356398}
\urldef\tempurl%
\url{https://doi.org/10.1145/3178876.3186150}
\showDOI{\tempurl}


\bibitem[Lin et~al\mbox{.}(2019)]%
        {Lin2019ARecommendation}
\bibfield{author}{\bibinfo{person}{Xiao Lin}, \bibinfo{person}{Hongjie Chen}, \bibinfo{person}{Changhua Pei}, \bibinfo{person}{Fei Sun}, \bibinfo{person}{Xuanji Xiao}, \bibinfo{person}{Hanxiao Sun}, \bibinfo{person}{Yongfeng Zhang}, \bibinfo{person}{Wenwu Ou}, {and} \bibinfo{person}{Peng Jiang}.} \bibinfo{year}{2019}\natexlab{}.
\newblock \showarticletitle{{A pareto-eficient algorithm for multiple objective optimization in e-commerce recommendation}}. In \bibinfo{booktitle}{\emph{RecSys 2019 - 13th ACM Conference on Recommender Systems}}. \bibinfo{publisher}{Association for Computing Machinery, Inc}, \bibinfo{address}{Copenhagen, Denmark}, \bibinfo{pages}{20--28}.
\newblock
\showISBNx{9781450362436}
\urldef\tempurl%
\url{https://doi.org/10.1145/3298689.3346998}
\showDOI{\tempurl}


\bibitem[Lin et~al\mbox{.}(2022)]%
        {Lin2022ImprovingLearning}
\bibfield{author}{\bibinfo{person}{Zihan Lin}, \bibinfo{person}{Changxin Tian}, \bibinfo{person}{Yupeng Hou}, {and} \bibinfo{person}{Wayne~Xin Zhao}.} \bibinfo{year}{2022}\natexlab{}.
\newblock \showarticletitle{{Improving Graph Collaborative Filtering with Neighborhood-enriched Contrastive Learning}}. In \bibinfo{booktitle}{\emph{WWW 2022 - Proceedings of the ACM Web Conference 2022}}. \bibinfo{publisher}{Association for Computing Machinery, Inc}, \bibinfo{address}{Virtual Event, Lyon, France}, \bibinfo{pages}{2320--2329}.
\newblock
\showISBNx{9781450390965}
\urldef\tempurl%
\url{https://doi.org/10.1145/3485447.3512104}
\showDOI{\tempurl}


\bibitem[Lioma et~al\mbox{.}(2017)]%
        {Lioma2017EvaluationLists}
\bibfield{author}{\bibinfo{person}{Christina Lioma}, \bibinfo{person}{Jakob~Grue Simonsen}, {and} \bibinfo{person}{Birger Larsen}.} \bibinfo{year}{2017}\natexlab{}.
\newblock \showarticletitle{{Evaluation measures for relevance and credibility in ranked lists}}. In \bibinfo{booktitle}{\emph{ICTIR 2017 - Proceedings of the 2017 ACM SIGIR International Conference on the Theory of Information Retrieval}}. \bibinfo{publisher}{Association for Computing Machinery, Inc}, \bibinfo{address}{Amsterdam, The Netherlands}, \bibinfo{pages}{91--98}.
\newblock
\showISBNx{9781450344906}
\urldef\tempurl%
\url{https://doi.org/10.1145/3121050.3121072}
\showDOI{\tempurl}


\bibitem[Maistro et~al\mbox{.}(2021)]%
        {Maistro2021PrincipledRankings}
\bibfield{author}{\bibinfo{person}{Maria Maistro}, \bibinfo{person}{Lucas~Chaves Lima}, \bibinfo{person}{Jakob~Grue Simonsen}, {and} \bibinfo{person}{Christina Lioma}.} \bibinfo{year}{2021}\natexlab{}.
\newblock \showarticletitle{{Principled Multi-Aspect Evaluation Measures of Rankings}}. In \bibinfo{booktitle}{\emph{International Conference on Information and Knowledge Management, Proceedings}}. \bibinfo{publisher}{Association for Computing Machinery}, \bibinfo{address}{Virtual Event, Queensland, Australia}, \bibinfo{pages}{1232--1242}.
\newblock
\showISBNx{9781450384469}
\urldef\tempurl%
\url{https://doi.org/10.1145/3459637.3482287}
\showDOI{\tempurl}


\bibitem[Mansoury et~al\mbox{.}(2020)]%
        {Mansoury2020FairMatch:Systems}
\bibfield{author}{\bibinfo{person}{Masoud Mansoury}, \bibinfo{person}{Himan Abdollahpouri}, \bibinfo{person}{Mykola Pechenizkiy}, \bibinfo{person}{Bamshad Mobasher}, {and} \bibinfo{person}{Robin Burke}.} \bibinfo{year}{2020}\natexlab{}.
\newblock \showarticletitle{{FairMatch: A Graph-based Approach for Improving Aggregate Diversity in Recommender Systems}}.
\newblock \bibinfo{journal}{\emph{UMAP 2020 - Proceedings of the 28th ACM Conference on User Modeling, Adaptation and Personalization}}  \bibinfo{volume}{20} (\bibinfo{date}{7} \bibinfo{year}{2020}), \bibinfo{pages}{154--162}.
\newblock
\showISBNx{9781450368612}
\urldef\tempurl%
\url{https://doi.org/10.1145/3340631.3394860}
\showDOI{\tempurl}


\bibitem[Mansoury et~al\mbox{.}(2021)]%
        {Mansoury2021ASystems}
\bibfield{author}{\bibinfo{person}{Masoud Mansoury}, \bibinfo{person}{Himan Abdollahpouri}, \bibinfo{person}{Mykola Pechenizkiy}, \bibinfo{person}{Bamshad Mobasher}, {and} \bibinfo{person}{Robin Burke}.} \bibinfo{year}{2021}\natexlab{}.
\newblock \showarticletitle{{A Graph-Based Approach for Mitigating Multi-Sided Exposure Bias in Recommender Systems}}.
\newblock \bibinfo{journal}{\emph{ACM Transactions on Information Systems (TOIS)}} \bibinfo{volume}{40}, \bibinfo{number}{2} (\bibinfo{date}{11} \bibinfo{year}{2021}), \bibinfo{pages}{32}.
\newblock
\showISSN{15582868}
\urldef\tempurl%
\url{https://doi.org/10.1145/3470948}
\showDOI{\tempurl}


\bibitem[Mehrotra et~al\mbox{.}(2018)]%
        {Mehrotra2018TowardsSystems}
\bibfield{author}{\bibinfo{person}{Rishabh Mehrotra}, \bibinfo{person}{James McInerney}, \bibinfo{person}{Hugues Bouchard}, \bibinfo{person}{Mounia Lalmas}, {and} \bibinfo{person}{Fernando Diaz}.} \bibinfo{year}{2018}\natexlab{}.
\newblock \showarticletitle{{Towards a fair marketplace: Counterfactual evaluation of the trade-off between relevance, fairness {\&} satisfaction in recommendation systems}}.
\newblock \bibinfo{journal}{\emph{International Conference on Information and Knowledge Management, Proceedings}}  \bibinfo{volume}{18} (\bibinfo{date}{10} \bibinfo{year}{2018}), \bibinfo{pages}{2243--2252}.
\newblock
\showISBNx{9781450360142}
\urldef\tempurl%
\url{https://doi.org/10.1145/3269206.3272027}
\showDOI{\tempurl}


\bibitem[Meng et~al\mbox{.}(2020)]%
        {Meng2020ExploringModels}
\bibfield{author}{\bibinfo{person}{Zaiqiao Meng}, \bibinfo{person}{Richard McCreadie}, \bibinfo{person}{Craig MacDonald}, {and} \bibinfo{person}{Iadh Ounis}.} \bibinfo{year}{2020}\natexlab{}.
\newblock \showarticletitle{{Exploring Data Splitting Strategies for the Evaluation of Recommendation Models}}. In \bibinfo{booktitle}{\emph{RecSys 2020 - 14th ACM Conference on Recommender Systems}}. \bibinfo{publisher}{Association for Computing Machinery, Inc}, \bibinfo{address}{Virtual Event, Brazil}, \bibinfo{pages}{681--686}.
\newblock
\showISBNx{9781450375832}
\urldef\tempurl%
\url{https://doi.org/10.1145/3383313.3418479}
\showDOI{\tempurl}


\bibitem[Moffat(2013)]%
        {Moffat2013SevenMetrics}
\bibfield{author}{\bibinfo{person}{Alistair Moffat}.} \bibinfo{year}{2013}\natexlab{}.
\newblock \showarticletitle{{Seven Numeric Properties of Effectiveness Metrics}}. In \bibinfo{booktitle}{\emph{Information Retrieval Technology}}, \bibfield{editor}{\bibinfo{person}{Rafael~E Banchs}, \bibinfo{person}{Fabrizio Silvestri}, \bibinfo{person}{Tie-Yan Liu}, \bibinfo{person}{Min Zhang}, \bibinfo{person}{Sheng Gao}, {and} \bibinfo{person}{Jun Lang}} (Eds.). \bibinfo{publisher}{Springer Berlin Heidelberg}, \bibinfo{address}{Berlin, Heidelberg}, \bibinfo{pages}{1--12}.
\newblock
\showISBNx{978-3-642-45068-6}


\bibitem[Ni et~al\mbox{.}(2019)]%
        {Ni2019JustifyingAspects}
\bibfield{author}{\bibinfo{person}{Jianmo Ni}, \bibinfo{person}{Jiacheng Li}, {and} \bibinfo{person}{Julian McAuley}.} \bibinfo{year}{2019}\natexlab{}.
\newblock \showarticletitle{{Justifying Recommendations using Distantly-Labeled Reviews and Fine-Grained Aspects}}. In \bibinfo{booktitle}{\emph{EMNLP-IJCNLP 2019 - 2019 Conference on Empirical Methods in Natural Language Processing and 9th International Joint Conference on Natural Language Processing, Proceedings of the Conference}}. \bibinfo{publisher}{Association for Computational Linguistics}, \bibinfo{address}{Hong Kong, China}, \bibinfo{pages}{188--197}.
\newblock
\showISBNx{9781950737901}
\urldef\tempurl%
\url{https://doi.org/10.18653/V1/D19-1018}
\showDOI{\tempurl}


\bibitem[Nia et~al\mbox{.}(2022)]%
        {Nia2022RethinkingNetworks}
\bibfield{author}{\bibinfo{person}{Vahid~Partovi Nia}, \bibinfo{person}{Alireza Ghaffari}, \bibinfo{person}{Mahdi Zolnouri}, {and} \bibinfo{person}{Yvon Savaria}.} \bibinfo{year}{2022}\natexlab{}.
\newblock \bibinfo{title}{{Rethinking pareto frontier for performance evaluation of deep neural networks}}.
\newblock
\newblock


\bibitem[Oosterhuis(2021)]%
        {Oosterhuis2021ComputationallyFairness}
\bibfield{author}{\bibinfo{person}{Harrie Oosterhuis}.} \bibinfo{year}{2021}\natexlab{}.
\newblock \showarticletitle{{Computationally Efficient Optimization of Plackett-Luce Ranking Models for Relevance and Fairness}}. In \bibinfo{booktitle}{\emph{Proceedings of the 44th International ACM SIGIR Conference on Research and Development in Information Retrieval}} \emph{(\bibinfo{series}{SIGIR '21})}. \bibinfo{publisher}{Association for Computing Machinery}, \bibinfo{address}{New York, NY, USA}, \bibinfo{pages}{1023--1032}.
\newblock
\showISBNx{9781450380379}
\urldef\tempurl%
\url{https://doi.org/10.1145/3404835.3462830}
\showDOI{\tempurl}


\bibitem[Palotti et~al\mbox{.}(2018)]%
        {Palotti2018MM:Engines}
\bibfield{author}{\bibinfo{person}{Joao Palotti}, \bibinfo{person}{Guido Zuccon}, {and} \bibinfo{person}{Allan Hanbury}.} \bibinfo{year}{2018}\natexlab{}.
\newblock \showarticletitle{{MM: A new framework for multidimensional evaluation of search engines}}. In \bibinfo{booktitle}{\emph{International Conference on Information and Knowledge Management, Proceedings}}. \bibinfo{publisher}{Association for Computing Machinery}, \bibinfo{address}{Torino, Italy}, \bibinfo{pages}{1699--1702}.
\newblock
\showISBNx{9781450360142}
\urldef\tempurl%
\url{https://doi.org/10.1145/3269206.3269261}
\showDOI{\tempurl}


\bibitem[Paparella et~al\mbox{.}(2023)]%
        {Paparella2023Post-hocRecommendation}
\bibfield{author}{\bibinfo{person}{Vincenzo Paparella}, \bibinfo{person}{Vito~Walter Anelli}, \bibinfo{person}{Franco~Maria Nardini}, \bibinfo{person}{Raffaele Perego}, {and} \bibinfo{person}{Tommaso Di~Noia}.} \bibinfo{year}{2023}\natexlab{}.
\newblock \showarticletitle{{Post-hoc Selection of Pareto-Optimal Solutions in Search and Recommendation}}.
\newblock \bibinfo{journal}{\emph{International Conference on Information and Knowledge Management, Proceedings}} (\bibinfo{date}{10} \bibinfo{year}{2023}), \bibinfo{pages}{2013--2023}.
\newblock
\showISBNx{9798400701245}
\urldef\tempurl%
\url{https://doi.org/10.1145/3583780.3615010}
\showDOI{\tempurl}


\bibitem[Patro et~al\mbox{.}(2020)]%
        {Patro2020FairRec:Platforms}
\bibfield{author}{\bibinfo{person}{Gourab~K. Patro}, \bibinfo{person}{Arpita Biswas}, \bibinfo{person}{Niloy Ganguly}, \bibinfo{person}{Krishna~P. Gummadi}, {and} \bibinfo{person}{Abhijnan Chakraborty}.} \bibinfo{year}{2020}\natexlab{}.
\newblock \showarticletitle{{FairRec: Two-Sided Fairness for Personalized Recommendations in Two-Sided Platforms}}. In \bibinfo{booktitle}{\emph{The Web Conference 2020 - Proceedings of the World Wide Web Conference, WWW 2020}}. \bibinfo{publisher}{Association for Computing Machinery, Inc}, \bibinfo{address}{Taipei, Taiwan}, \bibinfo{pages}{1194--1204}.
\newblock
\showISBNx{9781450370233}
\urldef\tempurl%
\url{https://doi.org/10.1145/3366423.3380196}
\showDOI{\tempurl}


\bibitem[Raj and Ekstrand(2022)]%
        {Raj2022MeasuringResults}
\bibfield{author}{\bibinfo{person}{Amifa Raj} {and} \bibinfo{person}{Michael~D. Ekstrand}.} \bibinfo{year}{2022}\natexlab{}.
\newblock \showarticletitle{{Measuring Fairness in Ranked Results}}. In \bibinfo{booktitle}{\emph{Proceedings of the 45th International ACM SIGIR Conference on Research and Development in Information Retrieval}}. \bibinfo{publisher}{ACM}, \bibinfo{address}{New York, NY, USA}, \bibinfo{pages}{726--736}.
\newblock
\showISBNx{9781450387323}
\urldef\tempurl%
\url{https://doi.org/10.1145/3477495.3532018}
\showDOI{\tempurl}


\bibitem[Rampisela et~al\mbox{.}(2024a)]%
        {Rampisela2024EvaluationStudy}
\bibfield{author}{\bibinfo{person}{Theresia~Veronika Rampisela}, \bibinfo{person}{Maria Maistro}, \bibinfo{person}{Tuukka Ruotsalo}, {and} \bibinfo{person}{Christina Lioma}.} \bibinfo{year}{2024}\natexlab{a}.
\newblock \showarticletitle{{Evaluation Measures of Individual Item Fairness for Recommender Systems: A Critical Study}}.
\newblock \bibinfo{journal}{\emph{ACM Trans. Recomm. Syst.}} \bibinfo{volume}{3}, \bibinfo{number}{2} (\bibinfo{date}{11} \bibinfo{year}{2024}).
\newblock
\urldef\tempurl%
\url{https://doi.org/10.1145/3631943}
\showDOI{\tempurl}


\bibitem[Rampisela et~al\mbox{.}(2024b)]%
        {Rampisela2024CanRelevance}
\bibfield{author}{\bibinfo{person}{Theresia~Veronika Rampisela}, \bibinfo{person}{Tuukka Ruotsalo}, \bibinfo{person}{Maria Maistro}, {and} \bibinfo{person}{Christina Lioma}.} \bibinfo{year}{2024}\natexlab{b}.
\newblock \showarticletitle{{Can We Trust Recommender System Fairness Evaluation? The Role of Fairness and Relevance}}. In \bibinfo{booktitle}{\emph{Proceedings of the 47th International ACM SIGIR Conference on Research and Development in Information Retrieval}} \emph{(\bibinfo{series}{SIGIR '24})}. \bibinfo{publisher}{Association for Computing Machinery}, \bibinfo{address}{New York, NY, USA}, \bibinfo{pages}{271--281}.
\newblock
\showISBNx{9798400704314}
\urldef\tempurl%
\url{https://doi.org/10.1145/3626772.3657832}
\showDOI{\tempurl}


\bibitem[Rendle et~al\mbox{.}(2009)]%
        {Rendle2009BPR:Feedback}
\bibfield{author}{\bibinfo{person}{Steffen Rendle}, \bibinfo{person}{Christoph Freudenthaler}, \bibinfo{person}{Zeno Gantner}, {and} \bibinfo{person}{Lars Schmidt-Thieme}.} \bibinfo{year}{2009}\natexlab{}.
\newblock \showarticletitle{{BPR: Bayesian Personalized Ranking from Implicit Feedback}}. In \bibinfo{booktitle}{\emph{UAI '09: Proceedings of the Twenty-Fifth Conference on Uncertainty in Artificial Intelligence}}. \bibinfo{publisher}{AUAI Press}, \bibinfo{address}{Montreal, Quebec, Canada}, \bibinfo{pages}{452--461}.
\newblock
\urldef\tempurl%
\url{https://doi.org/10.5555/1795114.1795167}
\showDOI{\tempurl}


\bibitem[Ribeiro et~al\mbox{.}(2015)]%
        {Ribeiro2015MultiobjectiveSystems}
\bibfield{author}{\bibinfo{person}{Marco~Tulio Ribeiro}, \bibinfo{person}{Nivio Ziviani}, \bibinfo{person}{Edleno Silva~De Moura}, \bibinfo{person}{Itamar Hata}, \bibinfo{person}{Anisio Lacerda}, {and} \bibinfo{person}{Adriano Veloso}.} \bibinfo{year}{2015}\natexlab{}.
\newblock \showarticletitle{{Multiobjective Pareto-Efficient Approaches for Recommender Systems}}.
\newblock \bibinfo{journal}{\emph{ACM Trans. Intell. Syst. Technol.}} \bibinfo{volume}{5}, \bibinfo{number}{4} (\bibinfo{date}{12} \bibinfo{year}{2015}), \bibinfo{pages}{1--20}.
\newblock
\showISSN{2157-6904}
\urldef\tempurl%
\url{https://doi.org/10.1145/2629350}
\showDOI{\tempurl}


\bibitem[Ryu and Min(2018)]%
        {Ryu2018Multi-objectiveWeight}
\bibfield{author}{\bibinfo{person}{Namhee Ryu} {and} \bibinfo{person}{Seungjae Min}.} \bibinfo{year}{2018}\natexlab{}.
\newblock \showarticletitle{{Multi-objective Optimization with an Adaptive Weight Determination Scheme Using the Concept of Hyperplane: Multi-objective Optimization with an Adaptive Weight}}.
\newblock \bibinfo{journal}{\emph{Internat. J. Numer. Methods Engrg.}}  \bibinfo{volume}{118} (\bibinfo{date}{10} \bibinfo{year}{2018}), \bibinfo{pages}{303--319}.
\newblock
\urldef\tempurl%
\url{https://doi.org/10.1002/nme.6013}
\showDOI{\tempurl}


\bibitem[Saito and Joachims(2022)]%
        {Saito2022FairRanking}
\bibfield{author}{\bibinfo{person}{Yuta Saito} {and} \bibinfo{person}{Thorsten Joachims}.} \bibinfo{year}{2022}\natexlab{}.
\newblock \showarticletitle{{Fair Ranking as Fair Division: Impact-Based Individual Fairness in Ranking}}. In \bibinfo{booktitle}{\emph{Proceedings of the 28th ACM SIGKDD Conference on Knowledge Discovery and Data Mining (KDD '22), August 14-18, 2022, Washington, DC, USA}}, Vol.~\bibinfo{volume}{1}. \bibinfo{publisher}{ACM}, \bibinfo{address}{Washington, DC, USA}, \bibinfo{pages}{1514--1524}.
\newblock
\showISBNx{9781450393850}
\urldef\tempurl%
\url{https://doi.org/10.1145/3534678.3539353}
\showDOI{\tempurl}


\bibitem[Shannon(1948)]%
        {Shannon1948ACommunication}
\bibfield{author}{\bibinfo{person}{C~E Shannon}.} \bibinfo{year}{1948}\natexlab{}.
\newblock \showarticletitle{{A Mathematical Theory of Communication}}.
\newblock \bibinfo{journal}{\emph{The Bell System Technical Journal}}  \bibinfo{volume}{27} (\bibinfo{year}{1948}), \bibinfo{pages}{623--656}.
\newblock


\bibitem[van Veldhuizen(1999)]%
        {vanVeldhuizen1999MultiobjectiveInnovations}
\bibfield{author}{\bibinfo{person}{David~A van Veldhuizen}.} \bibinfo{year}{1999}\natexlab{}.
\newblock \emph{\bibinfo{title}{{Multiobjective evolutionary algorithms: classifications, analyses, and new innovations}}}.
\newblock \bibinfo{thesistype}{Ph.\,D. Dissertation}. \bibinfo{school}{Air Force Institute of Technology}.
\newblock
\urldef\tempurl%
\url{https://api.semanticscholar.org/CorpusID:61080988}
\showURL{%
\tempurl}


\bibitem[Voorhees(2001)]%
        {Voorhees2001EvaluationDocuments}
\bibfield{author}{\bibinfo{person}{Ellen~M Voorhees}.} \bibinfo{year}{2001}\natexlab{}.
\newblock \showarticletitle{{Evaluation by Highly Relevant Documents}}. In \bibinfo{booktitle}{\emph{Proceedings of the 24th Annual International ACM SIGIR Conference on Research and Development in Information Retrieval}} \emph{(\bibinfo{series}{SIGIR '01})}. \bibinfo{publisher}{Association for Computing Machinery}, \bibinfo{address}{New York, NY, USA}, \bibinfo{pages}{74--82}.
\newblock
\showISBNx{1581133316}
\urldef\tempurl%
\url{https://doi.org/10.1145/383952.383963}
\showDOI{\tempurl}


\bibitem[Wang et~al\mbox{.}(2016)]%
        {Wang2016AEngineering}
\bibfield{author}{\bibinfo{person}{Shuai Wang}, \bibinfo{person}{Shaukat Ali}, \bibinfo{person}{Tao Yue}, \bibinfo{person}{Yan Li}, {and} \bibinfo{person}{Marius Liaaen}.} \bibinfo{year}{2016}\natexlab{}.
\newblock \showarticletitle{{A Practical Guide to Select Quality Indicators for Assessing Pareto-Based Search Algorithms in Search-Based Software Engineering}}. In \bibinfo{booktitle}{\emph{Proceedings of the 38th International Conference on Software Engineering}} \emph{(\bibinfo{series}{ICSE '16})}. \bibinfo{publisher}{Association for Computing Machinery}, \bibinfo{address}{New York, NY, USA}, \bibinfo{pages}{631--642}.
\newblock
\showISBNx{9781450339001}
\urldef\tempurl%
\url{https://doi.org/10.1145/2884781.2884880}
\showDOI{\tempurl}


\bibitem[Wang and Wang(2022)]%
        {Wang2022ProvidingSystems}
\bibfield{author}{\bibinfo{person}{Xiuling Wang} {and} \bibinfo{person}{Wendy~Hui Wang}.} \bibinfo{year}{2022}\natexlab{}.
\newblock \showarticletitle{{Providing Item-side Individual Fairness for Deep Recommender Systems}}.
\newblock \bibinfo{journal}{\emph{ACM International Conference Proceeding Series}}  \bibinfo{volume}{22} (\bibinfo{date}{6} \bibinfo{year}{2022}), \bibinfo{pages}{117--127}.
\newblock
\showISBNx{9781450393522}
\urldef\tempurl%
\url{https://doi.org/10.1145/3531146.3533079}
\showDOI{\tempurl}


\bibitem[Wang et~al\mbox{.}(2023)]%
        {Wang2023ASystems}
\bibfield{author}{\bibinfo{person}{Yifan Wang}, \bibinfo{person}{Weizhi Ma}, \bibinfo{person}{Min Zhang}, \bibinfo{person}{Yiqun Liu}, {and} \bibinfo{person}{Shaoping Ma}.} \bibinfo{year}{2023}\natexlab{}.
\newblock \showarticletitle{{A Survey on the Fairness of Recommender Systems}}.
\newblock \bibinfo{journal}{\emph{ACM Trans. Inf. Syst.}} \bibinfo{volume}{41}, \bibinfo{number}{3} (\bibinfo{date}{2} \bibinfo{year}{2023}), \bibinfo{pages}{1--43}.
\newblock
\showISSN{1046-8188}
\urldef\tempurl%
\url{https://doi.org/10.1145/3547333}
\showDOI{\tempurl}


\bibitem[Wu et~al\mbox{.}(2022)]%
        {Wu2022JointRecommendation}
\bibfield{author}{\bibinfo{person}{Haolun Wu}, \bibinfo{person}{Bhaskar Mitra}, \bibinfo{person}{Chen Ma}, \bibinfo{person}{Fernando Diaz}, {and} \bibinfo{person}{Xue Liu}.} \bibinfo{year}{2022}\natexlab{}.
\newblock \showarticletitle{{Joint Multisided Exposure Fairness for Recommendation}}. In \bibinfo{booktitle}{\emph{SIGIR 2022 - Proceedings of the 45th International ACM SIGIR Conference on Research and Development in Information Retrieval}}. \bibinfo{publisher}{Association for Computing Machinery, Inc}, \bibinfo{address}{Madrid, Spain}, \bibinfo{pages}{703--714}.
\newblock
\showISBNx{9781450387323}
\urldef\tempurl%
\url{https://doi.org/10.1145/3477495.3532007}
\showDOI{\tempurl}


\bibitem[Xu et~al\mbox{.}(2023)]%
        {Xu2023P-MMF:System}
\bibfield{author}{\bibinfo{person}{Chen Xu}, \bibinfo{person}{Sirui Chen}, \bibinfo{person}{Jun Xu}, \bibinfo{person}{Weiran Shen}, \bibinfo{person}{Xiao Zhang}, \bibinfo{person}{Gang Wang}, {and} \bibinfo{person}{Zhenhua Dong}.} \bibinfo{year}{2023}\natexlab{}.
\newblock \showarticletitle{{P-MMF: Provider Max-min Fairness Re-ranking in Recommender System}}.
\newblock \bibinfo{journal}{\emph{ACM Web Conference 2023 - Proceedings of the World Wide Web Conference, WWW 2023}} (\bibinfo{date}{4} \bibinfo{year}{2023}), \bibinfo{pages}{3701--3711}.
\newblock
\showISBNx{9781450394161}
\urldef\tempurl%
\url{https://doi.org/10.1145/3543507.3583296}
\showDOI{\tempurl}


\bibitem[Yuan et~al\mbox{.}(2022)]%
        {Yuan2022Tenrec:Systems}
\bibfield{author}{\bibinfo{person}{Guanghu Yuan}, \bibinfo{person}{Fajie Yuan}, \bibinfo{person}{Yudong Li}, \bibinfo{person}{Beibei Kong}, \bibinfo{person}{Shujie Li}, \bibinfo{person}{Lei Chen}, \bibinfo{person}{Min Yang}, \bibinfo{person}{Chenyun YU}, \bibinfo{person}{Bo Hu}, \bibinfo{person}{Zang Li}, \bibinfo{person}{Yu Xu}, {and} \bibinfo{person}{Xiaohu Qie}.} \bibinfo{year}{2022}\natexlab{}.
\newblock \showarticletitle{{Tenrec: A Large-scale Multipurpose Benchmark Dataset for Recommender Systems}}.
\newblock \bibinfo{journal}{\emph{Advances in Neural Information Processing Systems}}  \bibinfo{volume}{35} (\bibinfo{date}{12} \bibinfo{year}{2022}), \bibinfo{pages}{11480--11493}.
\newblock
\urldef\tempurl%
\url{https://www.tencent.com/en-us/}
\showURL{%
\tempurl}


\bibitem[Zehlike et~al\mbox{.}(2017)]%
        {Zehlike2017FAIR:Algorithm}
\bibfield{author}{\bibinfo{person}{Meike Zehlike}, \bibinfo{person}{Francesco Bonchi}, \bibinfo{person}{Carlos Castillo}, \bibinfo{person}{Sara Hajian}, \bibinfo{person}{Mohamed Megahed}, {and} \bibinfo{person}{Ricardo Baeza-Yates}.} \bibinfo{year}{2017}\natexlab{}.
\newblock \showarticletitle{{FA*IR: A fair top-k ranking algorithm}}.
\newblock \bibinfo{journal}{\emph{International Conference on Information and Knowledge Management, Proceedings}}  \bibinfo{volume}{Part F131841} (\bibinfo{date}{11} \bibinfo{year}{2017}), \bibinfo{pages}{1569--1578}.
\newblock
\showISBNx{9781450349185}
\urldef\tempurl%
\url{https://doi.org/10.1145/3132847.3132938}
\showDOI{\tempurl}


\bibitem[Zehlike and Castillo(2020)]%
        {Zehlike2020ReducingApproach}
\bibfield{author}{\bibinfo{person}{Meike Zehlike} {and} \bibinfo{person}{Carlos Castillo}.} \bibinfo{year}{2020}\natexlab{}.
\newblock \showarticletitle{{Reducing Disparate Exposure in Ranking: A Learning To Rank Approach}}. In \bibinfo{booktitle}{\emph{Proceedings of The Web Conference 2020}} \emph{(\bibinfo{series}{WWW '20})}. \bibinfo{publisher}{Association for Computing Machinery}, \bibinfo{address}{New York, NY, USA}, \bibinfo{pages}{2849--2855}.
\newblock
\showISBNx{9781450370233}
\urldef\tempurl%
\url{https://doi.org/10.1145/3366424.3380048}
\showDOI{\tempurl}


\bibitem[Zehlike et~al\mbox{.}(2022)]%
        {Zehlike2022FairnessSystemsc}
\bibfield{author}{\bibinfo{person}{Meike Zehlike}, \bibinfo{person}{Ke Yang}, {and} \bibinfo{person}{Julia Stoyanovich}.} \bibinfo{year}{2022}\natexlab{}.
\newblock \showarticletitle{{Fairness in Ranking, Part II: Learning-to-Rank and Recommender Systems}}.
\newblock \bibinfo{journal}{\emph{ACM Comput. Surv.}} \bibinfo{volume}{55}, \bibinfo{number}{6} (\bibinfo{date}{12} \bibinfo{year}{2022}), \bibinfo{pages}{1--41}.
\newblock
\showISSN{0360-0300}
\urldef\tempurl%
\url{https://doi.org/10.1145/3533380}
\showDOI{\tempurl}


\bibitem[Zhao et~al\mbox{.}(2021)]%
        {Zhao2021RecBole:Algorithms}
\bibfield{author}{\bibinfo{person}{Wayne~Xin Zhao}, \bibinfo{person}{Shanlei Mu}, \bibinfo{person}{Yupeng Hou}, \bibinfo{person}{Zihan Lin}, \bibinfo{person}{Yushuo Chen}, \bibinfo{person}{Xingyu Pan}, \bibinfo{person}{Kaiyuan Li}, \bibinfo{person}{Yujie Lu}, \bibinfo{person}{Hui Wang}, \bibinfo{person}{Changxin Tian}, \bibinfo{person}{Yingqian Min}, \bibinfo{person}{Zhichao Feng}, \bibinfo{person}{Xinyan Fan}, \bibinfo{person}{Xu Chen}, \bibinfo{person}{Pengfei Wang}, \bibinfo{person}{Wendi Ji}, \bibinfo{person}{Yaliang Li}, \bibinfo{person}{Xiaoling Wang}, {and} \bibinfo{person}{Ji~Rong Wen}.} \bibinfo{year}{2021}\natexlab{}.
\newblock \showarticletitle{{RecBole: Towards a Unified, Comprehensive and Efficient Framework for Recommendation Algorithms}}. In \bibinfo{booktitle}{\emph{International Conference on Information and Knowledge Management, Proceedings}}. \bibinfo{publisher}{ACM}, \bibinfo{address}{New York, NY, USA}, \bibinfo{pages}{4653--4664}.
\newblock
\showISBNx{9781450384469}
\urldef\tempurl%
\url{https://doi.org/10.1145/3459637.3482016}
\showDOI{\tempurl}


\bibitem[Zheng and Wang(2022)]%
        {Zheng2022AOptimization}
\bibfield{author}{\bibinfo{person}{Yong Zheng} {and} \bibinfo{person}{David~(Xuejun) Wang}.} \bibinfo{year}{2022}\natexlab{}.
\newblock \showarticletitle{{A survey of recommender systems with multi-objective optimization}}.
\newblock \bibinfo{journal}{\emph{Neurocomputing}}  \bibinfo{volume}{474} (\bibinfo{year}{2022}), \bibinfo{pages}{141--153}.
\newblock
\showISSN{0925-2312}
\urldef\tempurl%
\url{https://doi.org/10.1016/j.neucom.2021.11.041}
\showDOI{\tempurl}


\bibitem[Zhu et~al\mbox{.}(2020)]%
        {Zhu2020FARM:APPs}
\bibfield{author}{\bibinfo{person}{Qiliang Zhu}, \bibinfo{person}{Qibo Sun}, \bibinfo{person}{Zengxiang Li}, {and} \bibinfo{person}{Shangguang Wang}.} \bibinfo{year}{2020}\natexlab{}.
\newblock \showarticletitle{{FARM: A Fairness-Aware Recommendation Method for High Visibility and Low Visibility Mobile APPs}}.
\newblock \bibinfo{journal}{\emph{IEEE Access}}  \bibinfo{volume}{8} (\bibinfo{year}{2020}), \bibinfo{pages}{122747--122756}.
\newblock
\showISSN{21693536}
\urldef\tempurl%
\url{https://doi.org/10.1109/ACCESS.2020.3007617}
\showDOI{\tempurl}


\end{thebibliography}

\appendix

\section{Extended dataset statistics}\label{app:stat}
Tab.~\ref{tab:data_split} presents the statistics of each dataset split. For several datasets (e.g., Amazon-lb and ML-*), the number of users in the test split is significantly less than the number of users in the train split. Tab.~\ref{tab:rel_item_stat} presents the statistics of items in the test split, per user.

\begin{table}[htbp]
    \centering
        \caption{Number of [users, items, and interactions] in the train, validation, and test split after preprocessing. 
    } 
    \label{tab:data_split}
\resizebox{\columnwidth}{!}{
\begin{tabular}{llll}
\toprule
{} &                   train &                     val &                    test \\
\midrule
Lastfm    &     [1842, 2821, 42758] &     [1831, 2448, 14248] &     [1836, 2476, 14237] \\
Amazon-lb &       [1054, 552, 8860] &        [470, 204, 1811] &        [437, 209, 1726] \\
QK-video  &     [4656, 6245, 34345] &      [3470, 4095, 8726] &      [3514, 4101, 8706] \\
Jester    &   [63724, 100, 1294511] &    [62137, 100, 427623] &    [62167, 100, 427926] \\
ML-10M    &  [49378, 6838, 4944064] &    [2695, 7828, 296914] &    [1523, 7880, 121707] \\
ML-20M    &  [89917, 8719, 9882504] &    [4987, 10742, 472243] &    [2178, 13935, 233394] \\
\bottomrule
\end{tabular}
}
\end{table}

\begin{table}[htbp]
    \caption{Statistics of items in the test split, per user, i.e., the number of relevant items per user.}
    \label{tab:rel_item_stat}
    \centering
    \resizebox{0.65\columnwidth}{!}{
\begin{tabular}{lrrrr}
\toprule
{} &    mean &  min &  median &   max \\
\midrule
Lastfm    &    7.75 &    1 &       8 &    19 \\
Amazon-lb &    3.95 &    1 &       3 &    16 \\
QK-video  &    2.48 &    1 &       2 &    16 \\
Jester    &    6.88 &    1 &       6 &    29 \\
ML-10M    &   79.91 &    1 &      46 &  1632 \\
ML-20M    &  107.16 &    1 &      53 &  2266 \\
\bottomrule
\end{tabular}
}
\end{table}

\section{Algorithms for generating Pareto Frontier}\label{app:algo}
We present the pseudocodes of the algorithms for generating the Pareto Frontier: the Oracle (Algorithm \ref{alg:oracle}) and  \textsc{Oracle2Fair} (Algorithm \ref{alg:oracle2fair}). We provide the worst-case time complexity analysis for both algorithms and discuss a possible edge case. 
We denote as $k$ the recommendation cut-off, as $m$ the number of users, as $n$ the number of items, as $H$ the maximum number of items in a user's interaction history ($H_u$) across all users, and as $R$ the maximum number of items in a user's test split ($R_u^*$) across all users. We use the binary logarithm ($\log_2{}$). For brevity, we omit lines with $O(1)$, and we state the reasoning for each line (L).

\SetKwComment{Comment}{/* }{ */}
\RestyleAlgo{ruled}
\LinesNumbered

\begin{algorithm*}[htbp]
\caption{Oracle \\ Create recommendations with the highest relevance}\label{alg:oracle}

\KwData{ \\
$I$: all items in the dataset; \\
$H_u$: items in train-val split for each user $u \in U$;  \\
$R_u^*$: items in test split (relevant items) for each user $u \in U$; \\
$k$: number of recommended items
}
\KwResult{\\
$rec$: most relevant recommendation\\\\
$result$: a list of relevance and fairness scores\\
$itemNotInRec$: items that are not in the recommendation}

\Comment{Handle users with exactly $|R_u^*|=k$}
\lForEach{$u \in U$ where $|R_u^*| = k$}{{$rec[u] \gets R_u^*$}}

\Comment{Handle users with $|R_u^*| > k$}

\For{$K=k+1$ \KwTo $max(|R_u^*|)$}{\label{ln:startgtk}
    $userWithK \gets$ get users where $|R_u^*|=K$ \\
    \ForEach{$u \in userWithK$}{
        $takenItem[u] \gets {R_u^* \cap rec}$ \\
        $weight[u] \gets sum(countInRec(takenItem[u]))$ \\
        }
    $sortUserWithK \gets$ sort $userWithK$ by the least weight \\
    $tempRec[u] \gets R_u^* \setminus takenItem[u]$ \\
    keep only max $k$ items in $tempRec[u]$ \\

    \ForEach{$u \in sortUserWithK$}{
        $rec[u] \gets tempRec[u]$ \\
        $numItemToAdd \gets k - |tempRec[u]|$ \\
        sort $takenItem[u]$ by the least item count \\
        $rec[u].append(takenItem[u][:numItemToAdd])$  
    }
}  \label{ln:endgtk}

\Comment{Handle users with $|R_u^*| < k$}
$remainUser \gets$ get users where $|R_u^*|<k$ \\ \label{ln:startltk}
\lForEach{$u \in remainUser$}{$rec[u] \gets |R_u^*|$} 
$itemNotInRec \gets I \setminus rec$ \\
\ForEach{$u \in remainUser$}{
    \While{$|rec[u]|<k$ and $itemNotInRec \neq \emptyset$}{
        \For{$item \in itemNotInRec$}{
         \If{$item \notin H_u$}{
         $rec[u].append(item)$ \\
         $itemNotInRec \gets itemNotInRec\setminus \{item\}$
         }
        }    
    }

    \If{$|rec[u]|<k$}{
        \While{$|rec[u]|<k$}{

        $candItem \gets$ least popular item in $rec$ that is not in $H_u \cup R_u^* \cup rec[u]$ \\
        $rec[u].append(candItem)$
    }
    }
} \label{ln:endltk}
$result \gets calculateScores(rec)$ 
\end{algorithm*}

\begin{algorithm*}[htbp]
\caption{\textsc{Oracle2Fair} \\After recommending maximally relevant items, iteratively change the recommendation list to increase fairness until maximum fairness is reached}\label{alg:oracle2fair}
\KwData{$H_u, R_u^*,I, k$}
\KwResult{\\
$rec$: most fair possible recommendation;\\
$result$: a list of relevance and fairness scores}

$rec, result, itemNotInRec \gets Oracle(I, H_u, R_u^*)$

\Comment{Get the most popular item in the recommendations and its frequency count}
 $newMostPop \gets mostPop \gets getMostPopItem(rec)$  \\ \label{ln:updatecountbegin}
 $newCntPop \gets cntPop \gets cnt(mostPop, rec)$ \\

$uWithMostPop \gets$ all users with $mostPop \in rec[u]$ \\
sort $uWithMostPop$ by largest index of $mostPop$ in $rec[u]$ \\ \label{ln:updatecountend}
\For{$i \in itemNotInRec$}{
\lIf{$cntPop=1$}{break}    
\If{$newMostPop \neq mostPop$}{
    $mostPop \gets newMostPop$ \\ \label{ln:mostpopstart}
    update $uWithMostPop$ following $mostPop$ \\ \label{ln:mostpopend}
}
\If{$newMostPop = mostPop$}{
    $candU \gets$ all $u$ in $uWithMostPop$ where $i \notin H_u$  \\
    \If{$\exists u \in candU$ with $i \in R_u^*$ \label{ln:recommendstart}}{
        recommend $i$ to the top $u$ from $candU$ with $i \in R_u^*$
    }
    \lElse{recommend $i$ to the top $u$ from $candU$}
    reorder $rec[u]$ so all relevant items are at the top \\
    $result.append(calculateScores(rec))$ \\ \label{ln:recommendend}
    $itemNotInRec \gets itemNotInRec \setminus \{i\}$ \\
    $newMostPop \gets getMostPopItem(rec)$ \\
    $newCntPop \gets cnt(mostPop, rec)$ 
    }
  }
\Else{
    do lines \ref{ln:updatecountbegin}--\ref{ln:updatecountend} \\ \label{ln:dolines}
    $i \gets leastPop \gets getLeastPopItem(rec)$ \\ \label{ln:getleastpop}
    $m, n \gets |U|, |I|$ \\
    \While{$cntPop > \lceil km/n \rceil$}{
    \lIf{$newMostPop \neq mostPop$}{do lines \ref{ln:mostpopstart}--\ref{ln:mostpopend}}
    \If{$newMostPop = mostPop$}{
    $candU \gets$ all $u$ in $uWithMostPop$ where $i \notin H_u \cup rec[u]$  \\
    do lines \ref{ln:recommendstart}--\ref{ln:recommendend} \\
    do lines \ref{ln:dolines}--\ref{ln:getleastpop}
    }
    }
}

\end{algorithm*}

\subsection{Time complexity of the Oracle (\Cref{alg:oracle})}
\label{app:analyse-oracle}

The line-by-line time complexity analysis of Oracle (\Cref{alg:oracle}) is as follows:

\begin{itemize}
    \item L1: $O(km)$, creating list of size $k$ for $m$ users
    \item L2--17: $O(R^2m + Rm \log{m} + Rmk\log{k})$, resulting from at most $R$ iterations of:
    \begin{itemize}
        \item L3: $O(m)$, linear search on list of size $m$
        \item L4--7: $O(mR)$, at most $m$ times looking up at most $R$ values
        \item L8: $O(m \log{m})$, sorting a list of size $m$
        \item L11--16: $O(mk \log{k})$, at most $m$ times sorting list of size $k$
    \end{itemize}
    
    \item  L18: $O(m)$, linear search on list of size $m$
    \item  L19: $O(m)$, $m$ assignment operations
    \item  L21--36: $O(k^2m^2)$, resulting from at most $m$ iterations of: 
    \begin{itemize}
        \item L22--29: $O(kH)$, at most $k$ times of linear search on list of size $H$
        \item L30--L35: $O(k^2m)$, $k$ times of counting in a list of size $km$.
    
        In most cases, the number of users $m >> H$, so the time complexity of this block is dominated by $O(k^2m)$.
    \end{itemize}
    \item L37: $O(km)$ computing relevance/fairness measures for $m$ users based on recommendation list of size $k$

\end{itemize}

Overall, the time complexity of Oracle is dominated by that of L2 and L21--36. Hence, the time complexity of Oracle is $O(R^2m + Rm \log{m} + Rmk \log{k} + k^2m^2)$.

\subsection{Time complexity of \textsc{Oracle2Fair} (\Cref{alg:oracle2fair})}

The line-by-line time complexity analysis of \textsc{Oracle2Fair} (\Cref{alg:oracle2fair}) is as follows:

\begin{itemize}
    \item L1: $O(R^2m + Rm \log{m} + Rmk \log{k} + k^2m^2)$ (derived in \Cref{app:analyse-oracle})
    \item L2--4: $O(km)$, $m$ times of counting and linear search on list of size $k$
    \item L5: $O(m \log{m})$, sorting a list of size $m$ (e.g., with Tim Sort)
    \item L6--24: $O(max(Hmn, kmn))$, resulting from at most $n$ iterations of: 
    \begin{itemize}
        \item L10: $O(km)$, $m$ times of linear search on list of size $k$
        \item L13: $O(Hm)$, $m$ times linear search on list of size at most $H$
        \item L14: $O(km)$
        \item L18: $O(k \log{k})$, sorting a list of size $k$
        \item L21--L22: $O(km)$; the term $O (k \log{k})$ is dominated by $O(km)$, as typically the cut-off $k$ is much smaller than the number of users $m$ (and hence $\log{k} < m$).
    \end{itemize}

     \item L26: $O(km + m \log{m})$, combining L2--4 and L5
     \item L27: $O(km)$, $m$ times counting on list of size $k$
     \item L29--34: $O(max(Hkm^2,k^2m^2)+km^2\log{m})$, resulting from at most $km$ iterations (derived below) of:
     \begin{itemize}
         \item L30: $O(km)$, from L10
         \item L32: $O(Hm)$, $m$ times linear search on list of size at most $H$
         \item L33: $O(km)$, from L14 and L18
         \item L34: $O(km+m \log{m})$, from L26--27
     \end{itemize}
\end{itemize}

We estimate the worst-case complexity of the number of iterations of the code block in L29--34, by assuming that the initial recommendation is the unfairest. The unfairest recommendation happens when the same $k$ unique items are recommended to all $m$ users. 
This code block aims to reduce the max frequency count of all items to $\left\lceil km/n \right\rceil$. Hence, the number of iterations is $\sum^{k} (m-\left\lceil km/n \right\rceil) \leq k(m-\frac{km}{n}-1) = km-\frac{k^2m}{n}-k \leq km$.

All in all, the time complexity of \textsc{Oracle2Fair} is dominated by that of L1, L6--24, and L29--34. Hence, the overall time complexity of \textsc{Oracle2Fair} is $O(R^2m + Rm \log{m} + Rmk \log{k} +  k^2m^2 + km^2\log{m} + kmn )$ if $k\geq H$, or $O(R^2m + Rm \log{m} + Rmk \log{k} +  Hkm^2 + km^2\log{m} + Hmn )$ otherwise. To simplify the time complexity, further assumptions need to be made for one or more variables.

\subsection{Possible edge cases}
There might be edge cases, for example, datasets where the train/val set of each user contains almost all items in the dataset (each user has rated/clicked most items in the dataset). In this case, if we do not want to re-recommend items in the train/val set to users, some items may have item counts more than
$\left\lceil km/n \right\rceil$ at the end of the process, and \Cref{alg:oracle2fair} might not halt. However, such datasets are rare in recommender systems.

\section{Modifications to the GS algorithm}\label{app:gs}
The original GS algorithm \cite{Wang2022ProvidingSystems} increases individual item fairness within clusters of similar items. The item similarity is determined based on the item embedding. As our experiments and the \textsc{Fair} measures do not deal with the additional constraint of item similarity, we consider all items as similar. Therefore, we only have a single cluster of items.

On top of that, we also modify GS to increase computational efficiency. In the original GS algorithm, for each pair of candidate items for replacement $i$ and candidate items to be replaced $i'$, the algorithm finds all users that have $i$ in the original recommendation list. The algorithm then computes the loss in relevance (computed using predicted relevance value) if item $i$ is replaced by $i'$. Until this point, our modified algorithm does the same. The difference is that we save each $i, i', u$, and the loss associated, while the original algorithm only saves the information for the one user $u^*$, whose recommendation list will suffer the least loss when we replace $i$ with $i'$. The original GS then proceeds to make the replacement, update the pool of candidate items for replacement and to be replaced, and go through the entire process again. Initially, we found that with the GS algorithm, around 20\% of the initial recommendations are replaced during the process, meaning that for Amazon-lb, there are at least $437 \times 10 \times 0.2 \geq 800$ iterations of the process (Tab.~\ref{tab:data_split}). The number of iterations is much bigger for ML-10M, which has more than three times the  recommendation slots as Amazon-lb, and hence it is very costly to use the GS algorithm as is. 

Our modified GS utilises the saved information earlier. After going through all pairs of $(i,i')$, we sort the saved list from the smallest to the largest loss, and (attempt to) perform the replacement using the first $P$ pairs, where $P$ is 25\% of the number of recommendation slots. During the replacement process, if the item that is supposed to be replaced no longer exists in the user's recommendation list, we simply skip the replacement.

\section{Extended results}\label{app:actual-scores}
\begin{figure*}[tbp]
    \centering
    \includegraphics[width=\columnwidth]{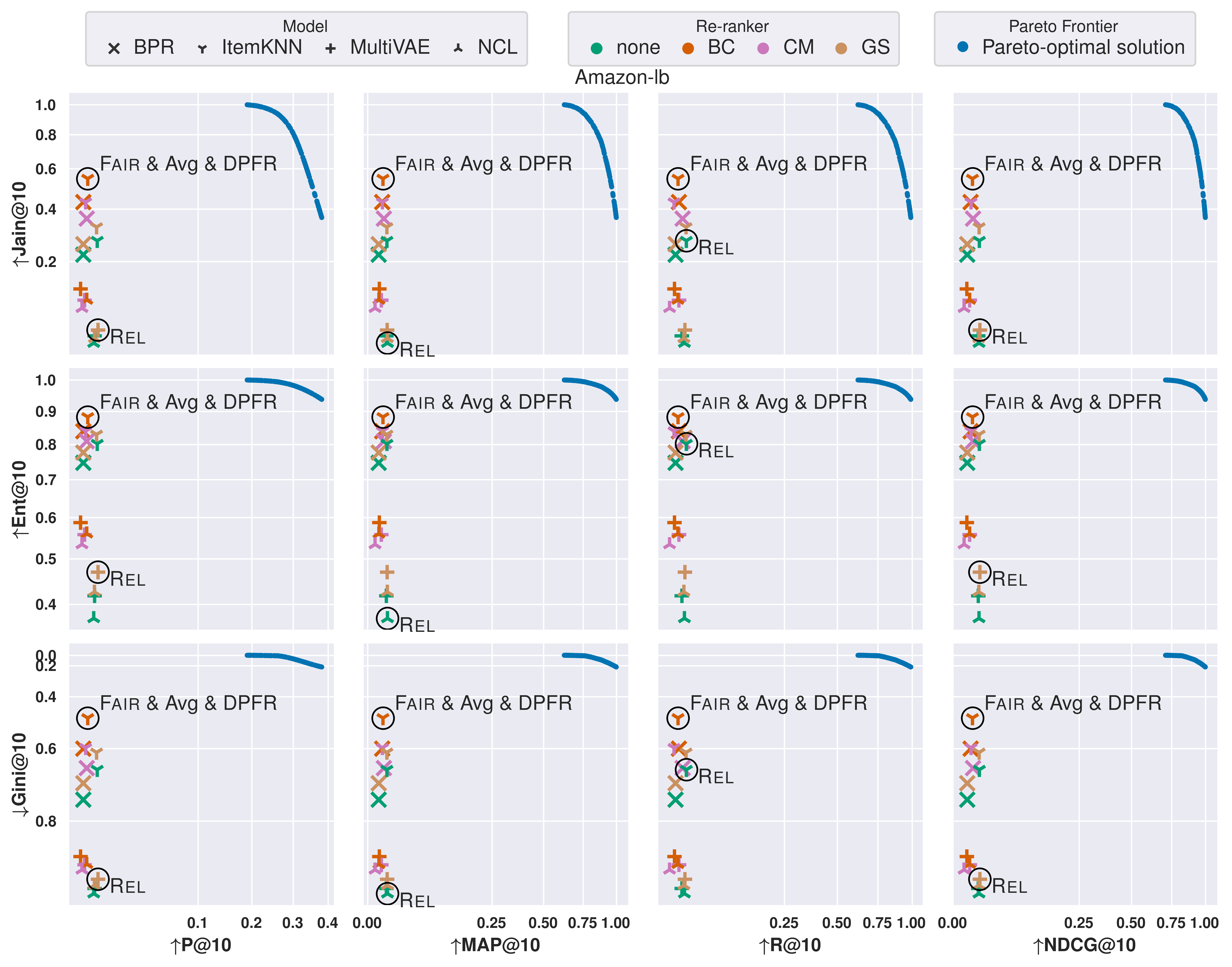}
    \includegraphics[width=\columnwidth]{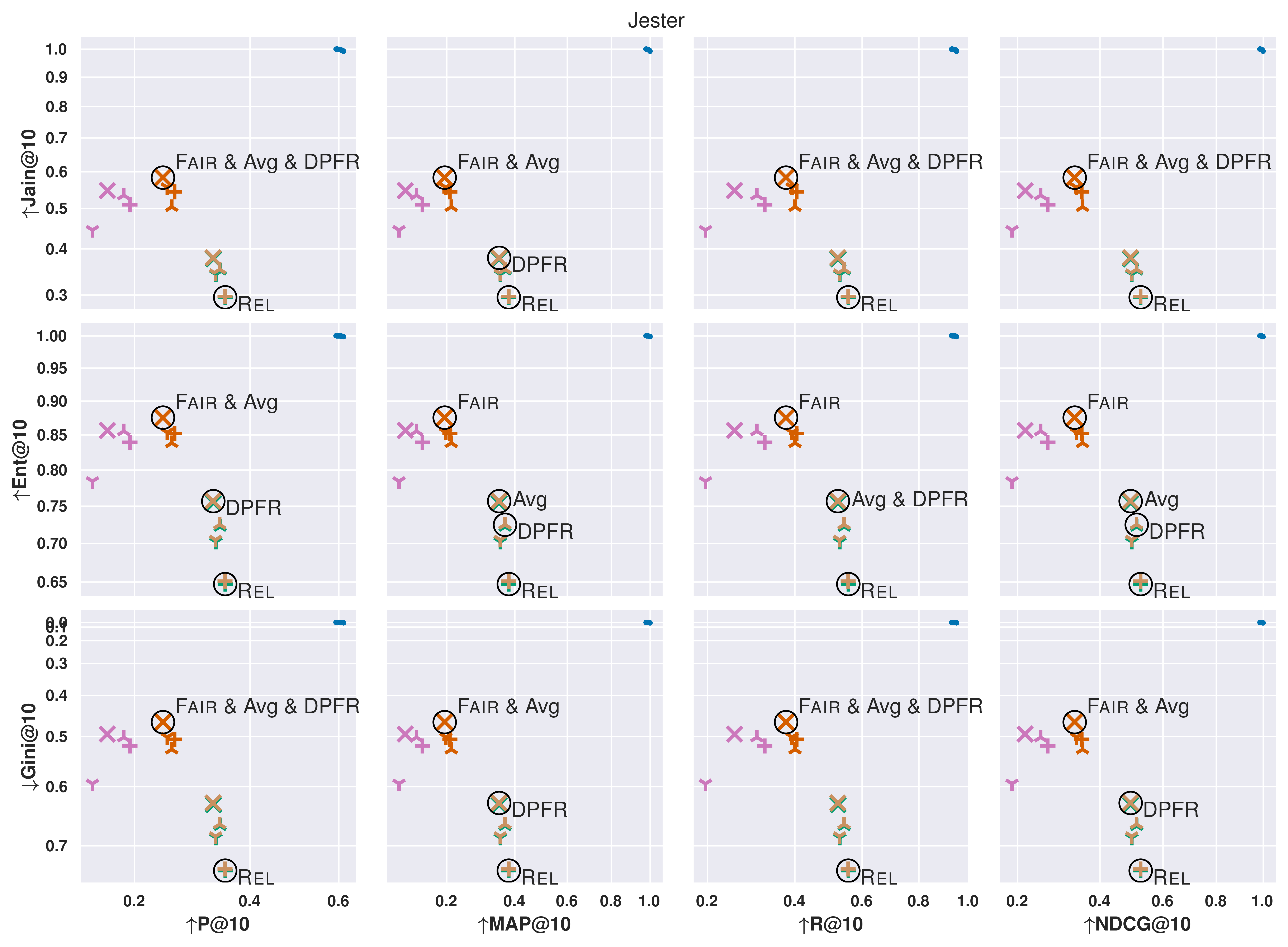}
    \includegraphics[width=\columnwidth]{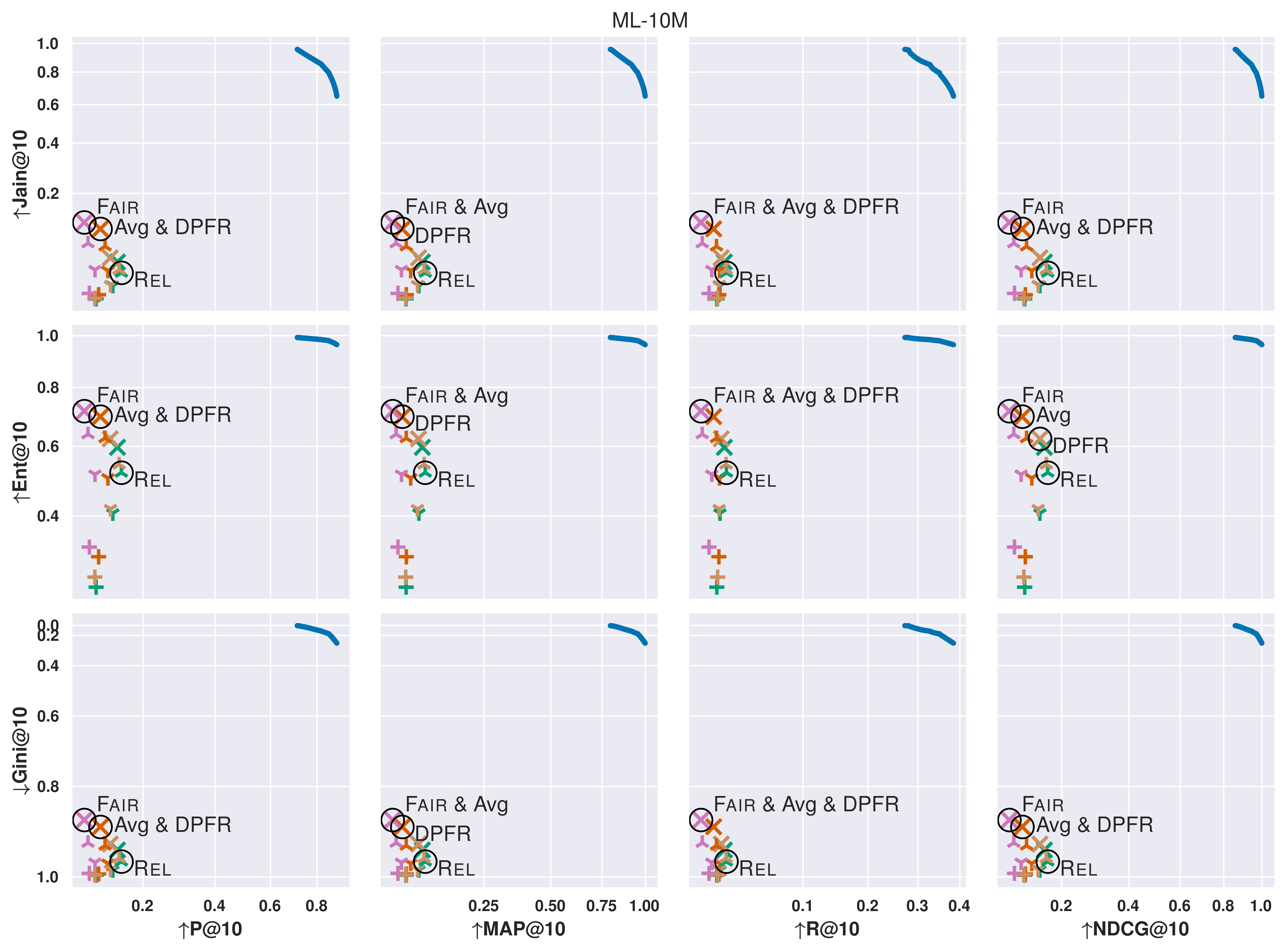}
    \includegraphics[width=\columnwidth]{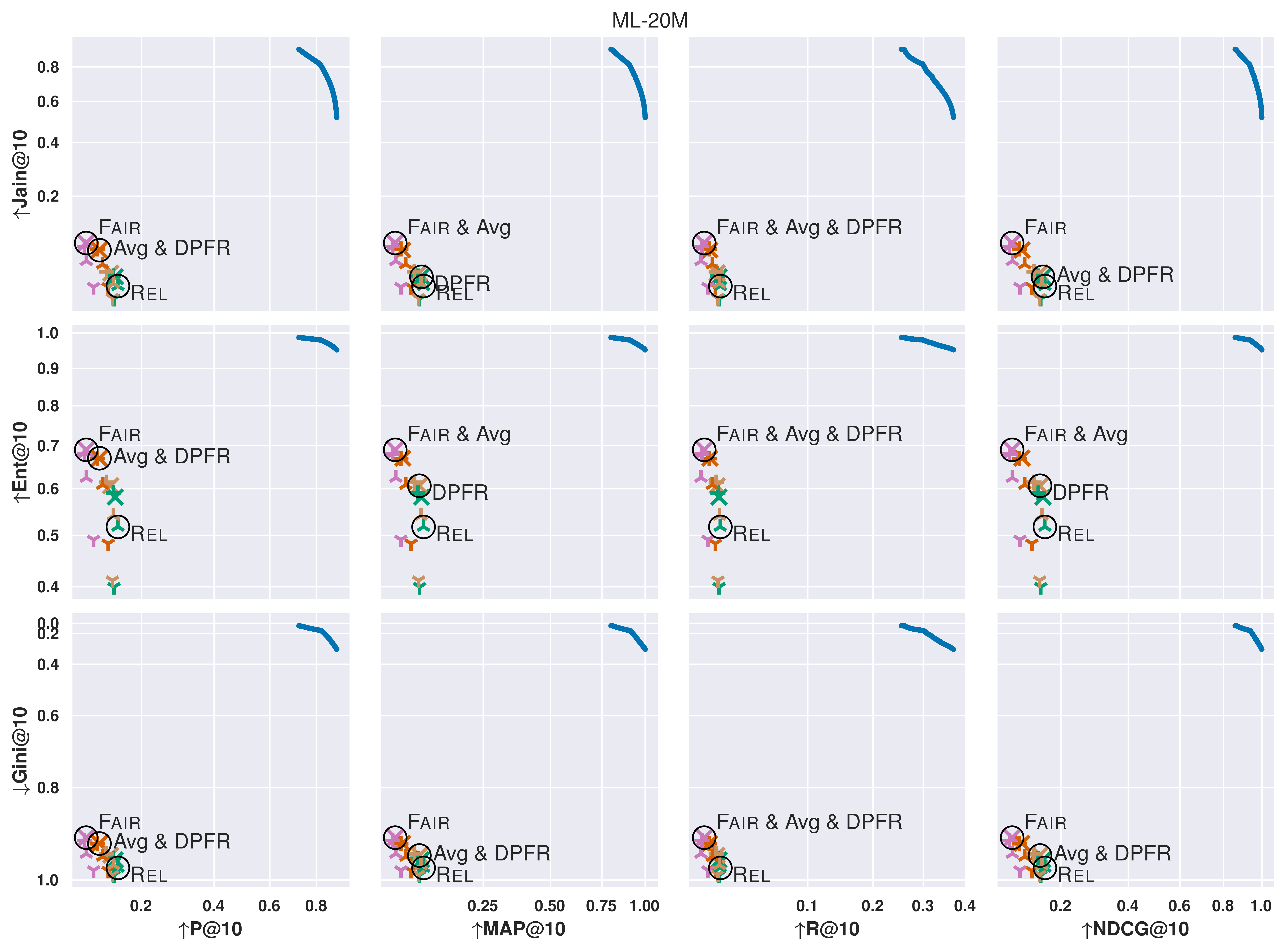}
    
    \caption{Pareto Frontier of fairness and relevance (in blue), together with recommender model scores for Amazon-lb, Jester, and ML-*. \textsc{Fair} measures are on the $y$-axis and \textsc{Rel} measures are on the $x$-axis. We implement exponential-like scales to enhance the visibility of the model plots. The \textsc{Rel, Fair}, Avg, and DPFR denote the best model based on each evaluation approach.}
    \label{fig:app-pairplot}

\end{figure*}

We present the actual scores of the recommender models in Tab.~\ref{tab:base-rerank-all-1}--\ref{tab:base-rerank-all-2}. In Tab.~\ref{tab:app-gradient}, we present the gradient values of the PF, used in determining which pair of measures are suitable for DPFR. In Fig.~\ref{fig:app-pairplot} we present the Pareto Frontier (PF) of fairness and relevance together with recommender model scores in Tab.~\ref{tab:base-rerank-all-1}--\ref{tab:base-rerank-all-2} for Amazon-lb, Jester, and ML-*. In \Cref{tab:dpfr-scores,tab:dpfr-scores2} we present the DPFR scores for all datasets. 
In Tab.~\ref{tab:corr_est} we present the Kendall's $\tau$ correlation scores of the DPFR from estimated PF and the PF.

\begin{table*}[tbp]
\caption{Relevance (\textsc{Rel}), fairness (\textsc{Fair}), and joint fairness and relevance (\textsc{Fair+Rel}) scores at $k=10$ of the recommender models for Lastfm, Amazon-lb, and QK-video, without and with re-ranking the the top $k'=25$ items using Borda Count (BC), COMBMNZ (CM), and Greedy Substitution (GS). The most relevant or most fair score per measure is in bold. $\uparrow$ means the higher the better, $\downarrow$ the lower the better.}
\label{tab:base-rerank-all-1}
\resizebox{0.90\textwidth}{!}{
\begin{tabular}{lll*{4}{r}|*{4}{r}|*{4}{r}|*{4}{r}}
\toprule
 &  & model & \multicolumn{4}{c|}{ItemKNN} & \multicolumn{4}{c|}{BPR} & \multicolumn{4}{c|}{MultiVAE} & \multicolumn{4}{c}{NCL} \\ 
\midrule
 &  & reranking & - & BC & CM & GS & - & BC & CM & GS & - & BC & CM & GS & - & BC & CM & GS \\
\midrule
\multirow[c]{16}{*}{\rotatebox[origin=r]{90}{Lastfm}} & \multirow[c]{6}{*}{\rotatebox[origin=r]{90}{\textsc{Rel}}} & $\uparrow$ $\text{HR}$ & 0.765 & 0.742 & 0.581 & 0.750 & 0.773 & 0.729 & 0.587 & 0.751 & 0.778 & 0.693 & 0.523 & 0.734 & \bfseries 0.793 & 0.726 & 0.571 & 0.765 \\
 &  & $\uparrow$ $\text{MRR}$ & 0.484 & 0.333 & 0.270 & 0.481 & 0.492 & 0.323 & 0.280 & 0.488 & 0.476 & 0.285 & 0.232 & 0.470 & \bfseries 0.503 & 0.311 & 0.260 & 0.499 \\
 &  & $\uparrow$ $\text{P}$ & 0.172 & 0.147 & 0.089 & 0.167 & 0.178 & 0.140 & 0.092 & 0.169 & 0.176 & 0.129 & 0.076 & 0.161 & \bfseries 0.184 & 0.141 & 0.087 & 0.173 \\
 &  & $\uparrow$ $\text{MAP}$ & 0.137 & 0.085 & 0.053 & 0.135 & 0.141 & 0.080 & 0.058 & 0.138 & 0.138 & 0.070 & 0.045 & 0.132 & \bfseries 0.148 & 0.079 & 0.050 & 0.144 \\
 &  & $\uparrow$ $\text{R}$ & 0.218 & 0.186 & 0.114 & 0.211 & 0.224 & 0.180 & 0.119 & 0.211 & 0.224 & 0.163 & 0.098 & 0.205 & \bfseries 0.234 & 0.180 & 0.110 & 0.220 \\
 &  & $\uparrow$ $\text{NDCG}$ & 0.245 & 0.181 & 0.119 & 0.241 & 0.252 & 0.173 & 0.126 & 0.244 & 0.247 & 0.155 & 0.102 & 0.235 & \bfseries 0.261 & 0.170 & 0.115 & 0.252 \\
\cline{2-19}
 & \multirow[c]{5}{*}{\rotatebox[origin=r]{90}{\textsc{Fair}}} & $\uparrow$ $\text{Jain}$ & 0.042 & 0.101 & 0.094 & 0.046 & 0.058 & 0.151 & 0.140 & 0.067 & 0.097 & \bfseries 0.236 & 0.222 & 0.115 & 0.082 & 0.216 & 0.215 & 0.095 \\
 &  & $\uparrow$ $\text{QF}$ & 0.474 & 0.642 & \bfseries 0.679 & 0.533 & 0.362 & 0.491 & 0.528 & 0.402 & 0.517 & 0.658 & 0.678 & 0.554 & 0.453 & 0.622 & 0.657 & 0.502 \\
 &  & $\uparrow$ $\text{Ent}$ & 0.589 & 0.727 & 0.735 & 0.622 & 0.610 & 0.736 & 0.740 & 0.646 & 0.707 & 0.820 & \bfseries 0.826 & 0.740 & 0.671 & 0.801 & 0.810 & 0.705 \\
 &  & $\uparrow$ $\text{FSat}$ & 0.129 & 0.197 & 0.216 & 0.152 & 0.147 & 0.211 & 0.228 & 0.177 & 0.202 & 0.293 & \bfseries 0.321 & 0.249 & 0.178 & 0.269 & 0.286 & 0.221 \\
 &  & $\downarrow$ $\text{Gini}$ & 0.904 & 0.810 & 0.790 & 0.879 & 0.910 & 0.827 & 0.818 & 0.887 & 0.839 & 0.715 & \bfseries 0.696 & 0.803 & 0.872 & 0.748 & 0.728 & 0.840 \\
\cline{2-19}
 & \multirow[c]{5}{*}{\rotatebox[origin=r]{90}{\textsc{Fair+Rel}}} & $\uparrow$ $\text{IBO}$ & 0.209 & 0.270 & 0.256 & 0.227 & 0.208 & 0.263 & 0.253 & 0.228 & 0.261 & \bfseries 0.314 & 0.278 & 0.281 & 0.242 & 0.308 & 0.292 & 0.265 \\
 &  & $\downarrow$ $\text{MME}$ & 0.001 & 0.001 & 0.001 & 0.001 & 0.001 & 0.001 & 0.001 & 0.001 & 0.001 & 0.000 & 0.001 & 0.001 & 0.001 & \bfseries 0.000 & 0.001 & 0.001 \\
 &  & $\downarrow$ $\text{IAA}$ & 0.004 & 0.004 & 0.004 & 0.004 & 0.004 & 0.004 & 0.004 & 0.004 & 0.004 & 0.004 & 0.004 & 0.004 & \bfseries 0.004 & 0.004 & 0.004 & 0.004 \\
 &  & $\downarrow$ $\text{II-F}$ & 0.001 & 0.001 & 0.002 & 0.001 & 0.001 & 0.001 & 0.002 & 0.001 & 0.001 & 0.001 & 0.002 & 0.001 & \bfseries 0.001 & 0.001 & 0.002 & 0.001 \\
 &  & $\downarrow$ $\text{AI-F}$ & 0.000 & 0.000 & 0.000 & 0.000 & 0.000 & 0.000 & 0.000 & 0.000 & 0.000 & 0.000 & 0.000 & 0.000 & 0.000 & \bfseries 0.000 & 0.000 & 0.000 \\
\cline{1-19}
\multirow[c]{16}{*}{\rotatebox[origin=r]{90}{Amazon-lb}} & \multirow[c]{6}{*}{\rotatebox[origin=r]{90}{\textsc{Rel}}} & $\uparrow$ $\text{HR}$ & \bfseries 0.046 & 0.021 & 0.016 & 0.043 & 0.011 & 0.014 & 0.021 & 0.011 & 0.039 & 0.007 & 0.014 & \bfseries 0.046 & 0.034 & 0.021 & 0.011 & 0.034 \\
 &  & $\uparrow$ $\text{MRR}$ & 0.020 & 0.011 & 0.011 & 0.020 & 0.003 & 0.005 & 0.007 & 0.003 & 0.023 & 0.003 & 0.004 & \bfseries 0.024 & 0.022 & 0.006 & 0.003 & 0.022 \\
 &  & $\uparrow$ $\text{P}$ & 0.005 & 0.002 & 0.002 & 0.005 & 0.001 & 0.001 & 0.002 & 0.001 & 0.004 & 0.001 & 0.002 & \bfseries 0.005 & 0.004 & 0.002 & 0.001 & 0.004 \\
 &  & $\uparrow$ $\text{MAP}$ & 0.006 & 0.004 & 0.004 & 0.006 & 0.002 & 0.003 & 0.004 & 0.002 & 0.006 & 0.002 & 0.003 & 0.006 & \bfseries 0.006 & 0.002 & 0.001 & 0.006 \\
 &  & $\uparrow$ $\text{R}$ & \bfseries 0.013 & 0.007 & 0.005 & 0.013 & 0.005 & 0.008 & 0.010 & 0.005 & 0.010 & 0.005 & 0.008 & 0.012 & 0.012 & 0.007 & 0.003 & 0.011 \\
 &  & $\uparrow$ $\text{NDCG}$ & 0.011 & 0.006 & 0.005 & 0.011 & 0.003 & 0.005 & 0.006 & 0.003 & 0.010 & 0.003 & 0.004 & \bfseries 0.011 & 0.011 & 0.004 & 0.002 & 0.011 \\
\cline{2-19}
 & \multirow[c]{5}{*}{\rotatebox[origin=r]{90}{\textsc{Fair}}} & $\uparrow$ $\text{Jain}$ & 0.271 & \bfseries 0.547 & 0.431 & 0.324 & 0.223 & 0.432 & 0.359 & 0.259 & 0.035 & 0.123 & 0.097 & 0.043 & 0.026 & 0.098 & 0.080 & 0.031 \\
 &  & $\uparrow$ $\text{QF}$ & 0.650 & \bfseries 0.679 & 0.612 & 0.663 & 0.549 & 0.630 & 0.594 & 0.571 & 0.222 & 0.294 & 0.286 & 0.254 & 0.229 & 0.315 & 0.310 & 0.265 \\
 &  & $\uparrow$ $\text{Ent}$ & 0.802 & \bfseries 0.882 & 0.839 & 0.829 & 0.747 & 0.839 & 0.809 & 0.776 & 0.418 & 0.587 & 0.558 & 0.469 & 0.371 & 0.560 & 0.534 & 0.426 \\
 &  & $\uparrow$ $\text{FSat}$ & 0.370 & \bfseries 0.538 & 0.438 & 0.435 & 0.314 & 0.410 & 0.376 & 0.364 & 0.114 & 0.159 & 0.152 & 0.138 & 0.091 & 0.146 & 0.138 & 0.115 \\
 &  & $\downarrow$ $\text{Gini}$ & 0.665 & \bfseries 0.492 & 0.598 & 0.613 & 0.747 & 0.601 & 0.660 & 0.703 & 0.949 & 0.882 & 0.899 & 0.930 & 0.959 & 0.898 & 0.910 & 0.943 \\
\cline{2-19}
 & \multirow[c]{5}{*}{\rotatebox[origin=r]{90}{\textsc{Fair+Rel}}} & $\uparrow$ $\text{IBO}$ & 0.062 & 0.038 & 0.029 & \bfseries 0.067 & 0.019 & 0.029 & 0.038 & 0.019 & 0.029 & 0.019 & 0.029 & 0.033 & 0.038 & 0.033 & 0.024 & 0.033 \\
 &  & $\downarrow$ $\text{MME}$ & 0.001 & 0.001 & 0.001 & 0.001 & 0.001 & \bfseries 0.001 & 0.001 & 0.001 & 0.003 & 0.001 & 0.001 & 0.003 & 0.004 & 0.001 & 0.001 & 0.004 \\
 &  & $\downarrow$ $\text{IAA}$ & 0.011 & 0.011 & 0.011 & \bfseries 0.011 & 0.011 & 0.011 & 0.011 & 0.011 & 0.011 & 0.011 & 0.011 & 0.011 & 0.011 & 0.011 & 0.011 & 0.011 \\
 &  & $\downarrow$ $\text{II-F}$ & 0.006 & 0.006 & 0.006 & \bfseries 0.006 & 0.006 & 0.006 & 0.006 & 0.006 & 0.006 & 0.006 & 0.006 & 0.006 & 0.006 & 0.006 & 0.006 & 0.006 \\
 &  & $\downarrow$ $\text{AI-F}$ & 0.000 & \bfseries 0.000 & 0.000 & 0.000 & 0.000 & 0.000 & 0.000 & 0.000 & 0.001 & 0.000 & 0.000 & 0.001 & 0.002 & 0.000 & 0.000 & 0.002 \\
\cline{1-19}
\multirow[c]{16}{*}{\rotatebox[origin=r]{90}{QK-video}} & \multirow[c]{6}{*}{\rotatebox[origin=r]{90}{\textsc{Rel}}} & $\uparrow$ $\text{HR}$ & 0.040 & 0.046 & 0.047 & 0.038 & 0.099 & 0.063 & 0.045 & 0.089 & 0.109 & 0.089 & 0.061 & 0.103 & \bfseries 0.130 & 0.102 & 0.077 & 0.124 \\
 &  & $\uparrow$ $\text{MRR}$ & 0.013 & 0.014 & 0.013 & 0.013 & 0.039 & 0.018 & 0.015 & 0.038 & 0.039 & 0.028 & 0.021 & 0.038 & \bfseries 0.048 & 0.030 & 0.024 & 0.047 \\
 &  & $\uparrow$ $\text{P}$ & 0.004 & 0.005 & 0.005 & 0.004 & 0.011 & 0.007 & 0.005 & 0.010 & 0.012 & 0.009 & 0.006 & 0.011 & \bfseries 0.014 & 0.011 & 0.008 & 0.013 \\
 &  & $\uparrow$ $\text{MAP}$ & 0.005 & 0.005 & 0.005 & 0.005 & 0.017 & 0.008 & 0.006 & 0.016 & 0.018 & 0.012 & 0.009 & 0.017 & \bfseries 0.022 & 0.013 & 0.010 & 0.021 \\
 &  & $\uparrow$ $\text{R}$ & 0.014 & 0.018 & 0.019 & 0.014 & 0.043 & 0.028 & 0.019 & 0.039 & 0.051 & 0.039 & 0.027 & 0.047 & \bfseries 0.061 & 0.045 & 0.033 & 0.058 \\
 &  & $\uparrow$ $\text{NDCG}$ & 0.009 & 0.011 & 0.010 & 0.009 & 0.029 & 0.015 & 0.011 & 0.027 & 0.031 & 0.022 & 0.016 & 0.030 & \bfseries 0.038 & 0.025 & 0.019 & 0.037 \\
\cline{2-19}
 & \multirow[c]{5}{*}{\rotatebox[origin=r]{90}{\textsc{Fair}}} & $\uparrow$ $\text{Jain}$ & 0.483 & \bfseries 0.815 & 0.589 & 0.567 & 0.081 & 0.333 & 0.379 & 0.101 & 0.012 & 0.038 & 0.032 & 0.014 & 0.020 & 0.076 & 0.071 & 0.023 \\
 &  & $\uparrow$ $\text{QF}$ & 0.901 & \bfseries 0.956 & 0.790 & 0.924 & 0.625 & 0.809 & 0.823 & 0.678 & 0.100 & 0.155 & 0.163 & 0.127 & 0.201 & 0.331 & 0.365 & 0.253 \\
 &  & $\uparrow$ $\text{Ent}$ & 0.933 & \bfseries 0.979 & 0.937 & 0.950 & 0.755 & 0.888 & 0.903 & 0.792 & 0.420 & 0.557 & 0.547 & 0.458 & 0.507 & 0.667 & 0.674 & 0.549 \\
 &  & $\uparrow$ $\text{FSat}$ & 0.443 & \bfseries 0.659 & 0.547 & 0.522 & 0.212 & 0.346 & 0.382 & 0.259 & 0.052 & 0.089 & 0.090 & 0.070 & 0.077 & 0.140 & 0.150 & 0.104 \\
 &  & $\downarrow$ $\text{Gini}$ & 0.472 & \bfseries 0.235 & 0.442 & 0.397 & 0.807 & 0.613 & 0.570 & 0.761 & 0.982 & 0.957 & 0.959 & 0.976 & 0.966 & 0.909 & 0.902 & 0.952 \\
\cline{2-19}
 & \multirow[c]{5}{*}{\rotatebox[origin=r]{90}{\textsc{Fair+Rel}}} & $\uparrow$ $\text{IBO}$ & 0.033 & 0.038 & 0.038 & 0.035 & 0.054 & 0.050 & 0.036 & 0.052 & 0.031 & 0.042 & 0.036 & 0.033 & 0.043 & \bfseries 0.060 & 0.054 & 0.047 \\
 &  & $\downarrow$ $\text{MME}$ & 0.000 & \bfseries 0.000 & 0.000 & 0.000 & 0.000 & 0.000 & 0.000 & 0.000 & 0.000 & 0.000 & 0.000 & 0.000 & 0.000 & 0.000 & 0.000 & 0.000 \\
 &  & $\downarrow$ $\text{IAA}$ & 0.001 & 0.001 & 0.001 & 0.001 & 0.001 & 0.001 & 0.001 & 0.001 & 0.001 & 0.001 & 0.001 & 0.001 & \bfseries 0.001 & 0.001 & 0.001 & 0.001 \\
 &  & $\downarrow$ $\text{II-F}$ & 0.001 & 0.001 & 0.001 & 0.001 & 0.001 & 0.001 & 0.001 & 0.001 & 0.001 & 0.001 & 0.001 & 0.001 & \bfseries 0.001 & 0.001 & 0.001 & 0.001 \\
 &  & $\downarrow$ $\text{AI-F}$ & 0.000 & \bfseries 0.000 & 0.000 & 0.000 & 0.000 & 0.000 & 0.000 & 0.000 & 0.000 & 0.000 & 0.000 & 0.000 & 0.000 & 0.000 & 0.000 & 0.000 \\
\bottomrule
\end{tabular}}
\end{table*}

\begin{table*}[tbp]
\caption{Relevance (\textsc{Rel}), fairness (\textsc{Fair}), and joint fairness and relevance (\textsc{Fair+Rel}) scores at $k=10$ of the recommender models for Jester and ML-*, without and with re-ranking the the top $k'=25$ items using Borda Count (BC), COMBMNZ (CM), and Greedy Substitution (GS) evaluated at $k=10$. The most relevant or most fair score per measure is in bold. $\uparrow$ means the higher the better, $\downarrow$ the lower the better.}
\label{tab:base-rerank-all-2}
\resizebox{0.90\textwidth}{!}{
\begin{tabular}{lll*{4}{r}|*{4}{r}|*{4}{r}|*{4}{r}}
\toprule
 &  & model & \multicolumn{4}{c|}{ItemKNN} & \multicolumn{4}{c|}{BPR} & \multicolumn{4}{c|}{MultiVAE} & \multicolumn{4}{c}{NCL} \\ 
\midrule
 &  & reranking & - & BC & CM & GS & - & BC & CM & GS & - & BC & CM & GS & - & BC & CM & GS \\
\midrule
\multirow[c]{16}{*}{\rotatebox[origin=r]{90}{Jester}} & \multirow[c]{6}{*}{\rotatebox[origin=r]{90}{\textsc{Rel}}} & $\uparrow$ $\text{HR}$ & 0.933 & 0.888 & 0.652 & 0.932 & 0.929 & 0.876 & 0.742 & 0.928 & \bfseries 0.944 & 0.899 & 0.818 & 0.944 & 0.939 & 0.893 & 0.804 & 0.939 \\
 &  & $\uparrow$ $\text{MRR}$ & 0.632 & 0.443 & 0.307 & 0.632 & 0.635 & 0.455 & 0.322 & 0.635 & \bfseries 0.661 & 0.465 & 0.370 & 0.661 & 0.651 & 0.479 & 0.349 & 0.651 \\
 &  & $\uparrow$ $\text{P}$ & 0.334 & 0.250 & 0.144 & 0.333 & 0.330 & 0.243 & 0.163 & 0.329 & \bfseries 0.351 & 0.262 & 0.194 & 0.351 & 0.342 & 0.257 & 0.185 & 0.341 \\
 &  & $\uparrow$ $\text{MAP}$ & 0.352 & 0.198 & 0.101 & 0.352 & 0.348 & 0.195 & 0.112 & 0.348 & \bfseries 0.379 & 0.208 & 0.145 & 0.379 & 0.367 & 0.211 & 0.133 & 0.367 \\
 &  & $\uparrow$ $\text{R}$ & 0.529 & 0.393 & 0.197 & 0.529 & 0.524 & 0.377 & 0.255 & 0.523 & \bfseries 0.555 & 0.405 & 0.324 & 0.555 & 0.543 & 0.400 & 0.305 & 0.542 \\
 &  & $\uparrow$ $\text{NDCG}$ & 0.496 & 0.336 & 0.189 & 0.496 & 0.493 & 0.331 & 0.216 & 0.492 & \bfseries 0.525 & 0.350 & 0.265 & 0.524 & 0.512 & 0.352 & 0.249 & 0.511 \\
\cline{2-19}
 & \multirow[c]{5}{*}{\rotatebox[origin=r]{90}{\textsc{Fair}}} & $\uparrow$ $\text{Jain}$ & 0.343 & 0.556 & 0.445 & 0.345 & 0.377 & \bfseries 0.583 & 0.547 & 0.380 & 0.295 & 0.544 & 0.509 & 0.297 & 0.351 & 0.504 & 0.534 & 0.354 \\
 &  & $\uparrow$ $\text{QF}^*$ & \bfseries 1.000 & \bfseries 1.000 & \bfseries 1.000 & \bfseries 1.000 & \bfseries 1.000 & \bfseries 1.000 & \bfseries 1.000 & \bfseries 1.000 & 0.967 & \bfseries 1.000 & \bfseries 1.000 & \bfseries 1.000 & \bfseries 1.000 & \bfseries 1.000 & \bfseries 1.000 & \bfseries 1.000 \\
 &  & $\uparrow$ $\text{Ent}$ & 0.702 & 0.854 & 0.784 & 0.705 & 0.754 & \bfseries 0.875 & 0.857 & 0.757 & 0.648 & 0.852 & 0.839 & 0.651 & 0.722 & 0.838 & 0.855 & 0.725 \\
 &  & $\uparrow$ $\text{FSat}$ & 0.267 & \bfseries 0.378 & 0.289 & 0.267 & 0.244 & 0.344 & 0.333 & 0.244 & 0.256 & 0.344 & 0.300 & 0.256 & 0.222 & 0.344 & 0.311 & 0.222 \\
 &  & $\downarrow$ $\text{Gini}$ & 0.687 & 0.502 & 0.595 & 0.685 & 0.632 & \bfseries 0.467 & 0.495 & 0.629 & 0.738 & 0.506 & 0.520 & 0.735 & 0.668 & 0.528 & 0.502 & 0.665 \\
\cline{2-19}
 & \multirow[c]{5}{*}{\rotatebox[origin=r]{90}{\textsc{Fair+Rel}}} & $\uparrow$ $\text{IBO}$ & 0.600 & \bfseries 0.930 & 0.740 & 0.600 & 0.840 & 0.910 & 0.780 & 0.840 & 0.500 & 0.870 & 0.810 & 0.500 & 0.740 & 0.920 & 0.780 & 0.740 \\
 &  & $\downarrow$ $\text{MME}$ & 0.003 & 0.003 & 0.006 & 0.003 & 0.004 & \bfseries 0.002 & 0.005 & 0.004 & 0.008 & 0.003 & 0.004 & 0.008 & 0.004 & 0.003 & 0.006 & 0.004 \\
 &  & $\downarrow$ $\text{IAA}$ & 0.081 & 0.093 & 0.104 & 0.081 & 0.081 & 0.094 & 0.103 & 0.081 & 0.078 & 0.092 & 0.100 & \bfseries 0.078 & 0.079 & 0.092 & 0.101 & 0.079 \\
 &  & $\downarrow$ $\text{II-F}$ & 0.028 & 0.035 & 0.040 & 0.028 & 0.029 & 0.035 & 0.040 & 0.029 & 0.027 & 0.035 & 0.038 & \bfseries 0.027 & 0.028 & 0.034 & 0.038 & 0.028 \\
 &  & $\downarrow$ $\text{AI-F}$ & 0.002 & 0.002 & 0.003 & 0.002 & 0.002 & \bfseries 0.001 & 0.002 & 0.002 & 0.003 & 0.002 & 0.002 & 0.003 & 0.002 & 0.002 & 0.002 & 0.002 \\
\cline{1-19}
\multirow[c]{16}{*}{\rotatebox[origin=r]{90}{ML-10M}} & \multirow[c]{6}{*}{\rotatebox[origin=r]{90}{\textsc{Rel}}} & $\uparrow$ $\text{HR}$ & 0.487 & 0.480 & 0.443 & 0.481 & 0.512 & 0.462 & 0.386 & 0.485 & 0.417 & 0.438 & 0.387 & 0.410 & \bfseries 0.521 & 0.473 & 0.402 & 0.513 \\
 &  & $\uparrow$ $\text{MRR}$ & 0.282 & 0.242 & 0.225 & 0.279 & 0.299 & 0.208 & 0.185 & 0.295 & 0.237 & 0.231 & 0.191 & 0.235 & \bfseries 0.302 & 0.216 & 0.203 & 0.301 \\
 &  & $\uparrow$ $\text{P}$ & 0.137 & 0.128 & 0.105 & 0.133 & 0.146 & 0.114 & 0.088 & 0.132 & 0.107 & 0.111 & 0.096 & 0.105 & \bfseries 0.154 & 0.123 & 0.094 & 0.149 \\
 &  & $\uparrow$ $\text{MAP}$ & 0.089 & 0.074 & 0.060 & 0.086 & 0.095 & 0.061 & 0.047 & 0.088 & 0.067 & 0.067 & 0.054 & 0.066 & \bfseries 0.101 & 0.067 & 0.052 & 0.099 \\
 &  & $\uparrow$ $\text{R}$ & 0.022 & 0.022 & 0.018 & 0.022 & 0.025 & 0.019 & 0.012 & 0.023 & 0.020 & 0.021 & 0.016 & 0.021 & \bfseries 0.026 & 0.020 & 0.013 & 0.026 \\
 &  & $\uparrow$ $\text{NDCG}$ & 0.150 & 0.133 & 0.113 & 0.147 & 0.160 & 0.115 & 0.092 & 0.150 & 0.119 & 0.121 & 0.100 & 0.118 & \bfseries 0.167 & 0.123 & 0.100 & 0.164 \\
\cline{2-19}
 & \multirow[c]{5}{*}{\rotatebox[origin=r]{90}{\textsc{Fair}}} & $\uparrow$ $\text{Jain}$ & 0.011 & 0.026 & 0.027 & 0.012 & 0.037 & 0.100 & \bfseries 0.115 & 0.044 & 0.003 & 0.005 & 0.006 & 0.004 & 0.024 & 0.063 & 0.069 & 0.027 \\
 &  & $\uparrow$ $\text{QF}$ & 0.044 & 0.062 & 0.068 & 0.047 & 0.145 & 0.199 & \bfseries 0.216 & 0.160 & 0.014 & 0.021 & 0.025 & 0.016 & 0.086 & 0.123 & 0.132 & 0.094 \\
 &  & $\uparrow$ $\text{Ent}$ & 0.407 & 0.503 & 0.514 & 0.418 & 0.596 & 0.697 & \bfseries 0.716 & 0.624 & 0.238 & 0.302 & 0.324 & 0.258 & 0.519 & 0.625 & 0.638 & 0.544 \\
 &  & $\uparrow$ $\text{FSat}$ & 0.044 & 0.062 & 0.068 & 0.047 & 0.145 & 0.199 & \bfseries 0.216 & 0.160 & 0.014 & 0.021 & 0.025 & 0.016 & 0.086 & 0.123 & 0.132 & 0.094 \\
 &  & $\downarrow$ $\text{Gini}$ & 0.987 & 0.973 & 0.971 & 0.985 & 0.945 & 0.895 & \bfseries 0.879 & 0.932 & 0.997 & 0.994 & 0.993 & 0.996 & 0.969 & 0.936 & 0.930 & 0.963 \\
\cline{2-19}
 & \multirow[c]{5}{*}{\rotatebox[origin=r]{90}{\textsc{Fair+Rel}}} & $\uparrow$ $\text{IBO}$ & 0.031 & 0.043 & 0.046 & 0.034 & 0.069 & 0.089 & \bfseries 0.091 & 0.076 & 0.012 & 0.016 & 0.018 & 0.014 & 0.054 & 0.073 & 0.074 & 0.058 \\
 &  & $\downarrow$ $\text{MME}$ & 0.001 & 0.001 & 0.001 & 0.001 & 0.001 & \bfseries 0.001 & 0.001 & 0.001 & 0.003 & 0.002 & 0.001 & 0.002 & 0.001 & 0.001 & 0.001 & 0.001 \\
 &  & $\downarrow$ $\text{IAA}$ & 0.008 & 0.009 & 0.009 & 0.008 & 0.008 & 0.009 & 0.009 & 0.008 & 0.009 & 0.009 & 0.009 & 0.009 & \bfseries 0.008 & 0.009 & 0.009 & 0.008 \\
 &  & $\downarrow$ $\text{II-F}$ & 0.000 & 0.000 & 0.000 & 0.000 & 0.000 & 0.000 & 0.000 & 0.000 & 0.000 & 0.000 & 0.000 & 0.000 & \bfseries 0.000 & 0.000 & 0.000 & 0.000 \\
 &  & $\downarrow$ $\text{AI-F}$ & 0.000 & 0.000 & 0.000 & 0.000 & 0.000 & \bfseries 0.000 & 0.000 & 0.000 & 0.000 & 0.000 & 0.000 & 0.000 & 0.000 & 0.000 & 0.000 & 0.000 \\
\cline{1-19}
\multirow[c]{16}{*}{\rotatebox[origin=r]{90}{ML-20M}} & \multirow[c]{6}{*}{\rotatebox[origin=r]{90}{\textsc{Rel}}} & $\uparrow$ $\text{HR}$ & 0.488 & 0.473 & 0.420 & 0.483 & \bfseries 0.505 & 0.444 & 0.392 & 0.483 & 0.489 & 0.432 & 0.391 & 0.465 & \bfseries 0.505 & 0.453 & 0.388 & 0.493 \\
 &  & $\uparrow$ $\text{MRR}$ & 0.280 & 0.237 & 0.213 & 0.278 & 0.293 & 0.205 & 0.190 & 0.290 & 0.259 & 0.193 & 0.180 & 0.256 & \bfseries 0.293 & 0.206 & 0.193 & 0.292 \\
 &  & $\uparrow$ $\text{P}$ & 0.142 & 0.131 & 0.106 & 0.139 & 0.145 & 0.116 & 0.094 & 0.136 & 0.141 & 0.112 & 0.091 & 0.128 & \bfseries 0.150 & 0.121 & 0.094 & 0.141 \\
 &  & $\uparrow$ $\text{MAP}$ & 0.092 & 0.077 & 0.061 & 0.090 & 0.096 & 0.063 & 0.052 & 0.092 & 0.089 & 0.060 & 0.049 & 0.082 & \bfseries 0.100 & 0.068 & 0.053 & 0.095 \\
 &  & $\uparrow$ $\text{R}$ & 0.019 & 0.017 & 0.014 & 0.019 & 0.019 & 0.014 & 0.012 & 0.018 & 0.019 & 0.014 & 0.011 & 0.018 & \bfseries 0.020 & 0.016 & 0.011 & 0.020 \\
 &  & $\uparrow$ $\text{NDCG}$ & 0.154 & 0.135 & 0.112 & 0.151 & 0.158 & 0.116 & 0.098 & 0.152 & 0.148 & 0.111 & 0.093 & 0.139 & \bfseries 0.163 & 0.121 & 0.099 & 0.157 \\
\cline{2-19}
 & \multirow[c]{5}{*}{\rotatebox[origin=r]{90}{\textsc{Fair}}} & $\uparrow$ $\text{Jain}$ & 0.008 & 0.017 & 0.018 & 0.009 & 0.028 & 0.068 & \bfseries 0.081 & 0.033 & 0.029 & 0.070 & 0.074 & 0.034 & 0.018 & 0.044 & 0.049 & 0.021 \\
 &  & $\uparrow$ $\text{QF}$ & 0.035 & 0.047 & 0.051 & 0.037 & 0.114 & 0.154 & \bfseries 0.165 & 0.125 & 0.117 & 0.146 & 0.154 & 0.126 & 0.074 & 0.103 & 0.112 & 0.082 \\
 &  & $\uparrow$ $\text{Ent}$ & 0.399 & 0.483 & 0.491 & 0.411 & 0.581 & 0.670 & \bfseries 0.690 & 0.606 & 0.591 & 0.669 & 0.680 & 0.615 & 0.517 & 0.608 & 0.624 & 0.541 \\
 &  & $\uparrow$ $\text{FSat}$ & 0.035 & 0.047 & 0.051 & 0.037 & 0.114 & 0.154 & \bfseries 0.165 & 0.125 & 0.117 & 0.146 & 0.154 & 0.126 & 0.074 & 0.103 & 0.112 & 0.082 \\
 &  & $\downarrow$ $\text{Gini}$ & 0.991 & 0.982 & 0.981 & 0.990 & 0.960 & 0.926 & \bfseries 0.914 & 0.951 & 0.957 & 0.927 & 0.920 & 0.948 & 0.976 & 0.953 & 0.947 & 0.971 \\
\cline{2-19}
 & \multirow[c]{5}{*}{\rotatebox[origin=r]{90}{\textsc{Fair+Rel}}} & $\uparrow$ $\text{IBO}$ & 0.021 & 0.031 & 0.033 & 0.022 & 0.049 & 0.064 & \bfseries 0.067 & 0.054 & 0.052 & 0.064 & 0.065 & 0.056 & 0.039 & 0.051 & 0.054 & 0.042 \\
 &  & $\downarrow$ $\text{MME}$ & 0.001 & 0.001 & 0.001 & 0.001 & 0.001 & 0.000 & \bfseries 0.000 & 0.001 & 0.001 & 0.000 & 0.000 & 0.001 & 0.001 & 0.001 & 0.001 & 0.001 \\
 &  & $\downarrow$ $\text{IAA}$ & 0.007 & 0.007 & 0.007 & 0.007 & 0.007 & 0.007 & 0.007 & 0.007 & 0.007 & 0.007 & 0.007 & 0.007 & \bfseries 0.007 & 0.007 & 0.007 & 0.007 \\
 &  & $\downarrow$ $\text{II-F}$ & 0.000 & 0.000 & 0.000 & 0.000 & 0.000 & 0.000 & 0.000 & 0.000 & 0.000 & 0.000 & 0.000 & 0.000 & \bfseries 0.000 & 0.000 & 0.000 & 0.000 \\
 &  & $\downarrow$ $\text{AI-F}$ & 0.000 & 0.000 & 0.000 & 0.000 & 0.000 & 0.000 & \bfseries 0.000 & 0.000 & 0.000 & 0.000 & 0.000 & 0.000 & 0.000 & 0.000 & 0.000 & 0.000 \\
\bottomrule
\multicolumn{19}{l}{\small *QF $=1$ means that all items in the dataset appear in the recommendation across all users.} \\
\multicolumn{19}{l}{\small $^{\dag}$The scores of QF are the same as FSat for ML-*, as QF is computed based on the percentage of items in the dataset that are recommended, which in this dataset} \\
\multicolumn{19}{l}{\small is equivalent to FSat: the percentage of items in the dataset that are recommended at least $\left \lfloor \frac{km}{n}\right \rfloor=1$ time.}
\end{tabular}}
\end{table*}

\begin{table*}[tbp]
\centering
\caption{The gradient values of the PF, based on the extreme points (starting and ending points). We consider a gradient to be `good' if it is not zero or undefined (-).}\label{tab:app-gradient} 
\begin{tabular}{lrrrrrrrl}
\toprule
{} & Lastfm & Amazon-lb & QK-video & Jester &  ML-10M &  ML-20M &  \# good &    conclusion \\
\midrule
HR-Ent    &   -97.57 &      -1.86 &     -0.31 &      - &  -14.74 &   -6.95 &        5 &  inconsistent \\
HR-FSat   & -1439.17 &     -19.92 &      0.00 &      - &  -30.48 &  -18.97 &        4 &  inconsistent \\
HR-Gini   &   561.63 &       6.23 &      3.71 &      - &  117.19 &   43.44 &        5 &  inconsistent \\
HR-Jain   &  -979.86 &     -18.77 &     -5.80 &      - & -157.73 &  -78.22 &        5 &  inconsistent \\
HR-QF     &     0.00 &       0.00 &      0.00 &      - &  -30.48 &  -18.97 &        2 &  inconsistent \\
MAP-Ent   &    -0.17 &      -0.17 &     -0.03 &  -0.07 &   -0.14 &   -0.18 &        6 &   always good \\
MAP-FSat  &    -2.46 &      -1.81 &      0.00 & -44.47 &   -0.29 &   -0.48 &        5 &  inconsistent \\
MAP-Gini  &     0.96 &       0.56 &      0.34 &   1.42 &    1.12 &    1.10 &        6 &   always good \\
MAP-Jain  &    -1.68 &      -1.70 &     -0.54 &  -0.37 &   -1.51 &   -1.98 &        6 &   always good \\
MAP-QF    &     0.00 &       0.00 &      0.00 &    0.0 &   -0.29 &   -0.48 &        2 &  inconsistent \\
MRR-Ent   &   -97.57 &      -1.86 &     -0.31 &      - &  -14.74 &   -6.95 &        5 &  inconsistent \\
MRR-FSat  & -1439.17 &     -19.92 &      0.00 &      - &  -30.48 &  -18.97 &        4 &  inconsistent \\
MRR-Gini  &   561.63 &       6.23 &      3.71 &      - &  117.19 &   43.44 &        5 &  inconsistent \\
MRR-Jain  &  -979.86 &     -18.77 &     -5.80 &      - & -157.73 &  -78.22 &        5 &  inconsistent \\
MRR-QF    &     0.00 &       0.00 &      0.00 &      - &  -30.48 &  -18.97 &        2 &  inconsistent \\
NDCG-Ent  &    -0.24 &      -0.22 &     -0.04 &   -0.1 &   -0.20 &   -0.25 &        6 &   always good \\
NDCG-FSat &    -3.50 &      -2.32 &      0.00 & -68.56 &   -0.42 &   -0.68 &        5 &  inconsistent \\
NDCG-Gini &     1.37 &       0.73 &      0.47 &    2.2 &    1.62 &    1.55 &        6 &   always good \\
NDCG-Jain &    -2.38 &      -2.19 &     -0.73 &  -0.57 &   -2.18 &   -2.79 &        6 &   always good \\
NDCG-QF   &     0.00 &       0.00 &      0.00 &    0.0 &   -0.42 &   -0.68 &        2 &  inconsistent \\
P-Ent     &    -0.20 &      -0.33 &     -0.07 &  -0.08 &   -0.16 &   -0.20 &        6 &   always good \\
P-FSat    &    -2.95 &      -3.53 &      0.00 & -51.41 &   -0.33 &   -0.55 &        5 &  inconsistent \\
P-Gini    &     1.15 &       1.10 &      0.89 &   1.65 &    1.26 &    1.26 &        6 &   always good \\
P-Jain    &    -2.01 &      -3.33 &     -1.40 &  -0.43 &   -1.70 &   -2.27 &        6 &   always good \\
P-QF      &     0.00 &       0.00 &      0.00 &    0.0 &   -0.33 &   -0.55 &        2 &  inconsistent \\
R-Ent     &    -0.17 &      -0.17 &     -0.03 &  -0.07 &   -0.26 &   -0.30 &        6 &   always good \\
R-FSat    &    -2.57 &      -1.83 &      0.00 & -48.04 &   -0.53 &   -0.82 &        5 &  inconsistent \\
R-Gini    &     1.00 &       0.57 &      0.34 &   1.54 &    2.04 &    1.88 &        6 &   always good \\
R-Jain    &    -1.75 &      -1.73 &     -0.54 &   -0.4 &   -2.75 &   -3.39 &        6 &   always good \\
R-QF      &     0.00 &       0.00 &      0.00 &    0.0 &   -0.53 &   -0.82 &        2 &  inconsistent \\
\bottomrule
\end{tabular}
\end{table*}

\begin{table*}[tbp]
\caption{$\downarrow$DPFR scores at $k=10$ of the recommender models for Lastfm, Amazon-lb, and QK-video, without and with re-ranking the the top $k'=25$ items using Borda Count (BC), COMBMNZ (CM), and Greedy Substitution (GS). The best score per measure pair is in bold.}
\label{tab:dpfr-scores}
\resizebox{0.90\textwidth}{!}{
\begin{tabular}{ll*{4}{r}|*{4}{r}|*{4}{r}|*{4}{r}}
\toprule
 &  model & \multicolumn{4}{c|}{ItemKNN} & \multicolumn{4}{c|}{BPR} & \multicolumn{4}{c|}{MultiVAE} & \multicolumn{4}{c}{NCL} \\ 
\midrule
 &  reranking & - & BC & CM & GS & - & BC & CM & GS & - & BC & CM & GS & - & BC & CM & GS \\
\midrule
\multirow[c]{12}{*}{\rotatebox[origin=r]{90}{Lastfm}} & P-Jain & 0.853 & 0.819 & 0.861 & 0.853 & 0.837 & 0.784 & 0.824 & 0.834 & 0.805 & \bfseries 0.728 & 0.776 & 0.799 & 0.813 & 0.735 & 0.773 & 0.809 \\
 & P-Ent & 0.585 & 0.524 & 0.571 & 0.567 & 0.567 & 0.525 & 0.566 & 0.551 & 0.510 & 0.501 & 0.550 & 0.505 & 0.524 & \bfseries 0.497 & 0.544 & 0.513 \\
 & P-Gini & 0.869 & 0.802 & 0.821 & 0.850 & 0.871 & 0.820 & 0.841 & 0.856 & 0.811 & \bfseries 0.737 & 0.759 & 0.789 & 0.835 & 0.756 & 0.775 & 0.813 \\
 & MAP-Jain & 1.044 & 1.042 & 1.072 & 1.042 & 1.029 & 1.015 & 1.040 & 1.026 & 1.005 & \bfseries 0.974 & 1.004 & 0.998 & 1.008 & 0.979 & 1.003 & 1.003 \\
 & MAP-Ent & 0.811 & 0.802 & 0.830 & 0.797 & 0.797 & 0.803 & 0.824 & 0.784 & 0.759 & 0.791 & 0.815 & \bfseries 0.753 & 0.764 & 0.787 & 0.813 & 0.755 \\
 & MAP-Gini & 1.038 & 1.009 & 1.022 & 1.021 & 1.039 & 1.025 & 1.036 & 1.025 & 0.991 & \bfseries 0.961 & 0.972 & 0.970 & 1.007 & 0.975 & 0.986 & 0.987 \\
 & R-Jain & 0.968 & 0.947 & 1.004 & 0.969 & 0.952 & 0.917 & 0.970 & 0.954 & 0.923 & \bfseries 0.875 & 0.936 & 0.923 & 0.927 & \bfseries 0.875 & 0.930 & 0.928 \\
 & R-Ent & 0.720 & 0.683 & 0.747 & 0.708 & 0.703 & 0.686 & 0.741 & 0.696 & \bfseries 0.656 & 0.675 & 0.738 & 0.660 & 0.664 & 0.665 & 0.729 & 0.661 \\
 & R-Gini & 0.968 & 0.918 & 0.955 & 0.953 & 0.969 & 0.935 & 0.971 & 0.959 & 0.914 & \bfseries 0.868 & 0.907 & 0.900 & 0.932 & 0.878 & 0.918 & 0.917 \\
 & NDCG-Jain & 0.989 & 0.995 & 1.046 & 0.989 & 0.973 & 0.968 & 1.013 & 0.972 & 0.948 & 0.933 & 0.985 & 0.945 & 0.949 & \bfseries 0.932 & 0.978 & 0.947 \\
 & NDCG-Ent & 0.760 & 0.758 & 0.813 & 0.747 & 0.743 & 0.762 & 0.805 & 0.733 & 0.703 & 0.756 & 0.807 & 0.702 & 0.706 & 0.746 & 0.797 & \bfseries 0.700 \\
 & NDCG-Gini & 1.000 & 0.976 & 1.009 & 0.984 & 1.000 & 0.994 & 1.023 & 0.987 & 0.949 & 0.934 & 0.966 & \bfseries 0.932 & 0.964 & 0.943 & 0.974 & 0.947 \\
\cline{1-18}
\multirow[c]{12}{*}{\rotatebox[origin=r]{90}{Amazon-lb}} & P-Jain & 0.529 & \bfseries 0.352 & 0.415 & 0.489 & 0.571 & 0.415 & 0.464 & 0.542 & 0.733 & 0.656 & 0.679 & 0.726 & 0.742 & 0.678 & 0.694 & 0.736 \\
 & P-Ent & 0.340 & \bfseries 0.307 & 0.324 & 0.327 & 0.376 & 0.325 & 0.339 & 0.358 & 0.636 & 0.493 & 0.517 & 0.590 & 0.678 & 0.515 & 0.537 & 0.629 \\
 & P-Gini & 0.633 & \bfseries 0.486 & 0.575 & 0.587 & 0.708 & 0.578 & 0.630 & 0.669 & 0.896 & 0.834 & 0.850 & 0.878 & 0.906 & 0.849 & 0.860 & 0.890 \\
 & MAP-Jain & 0.998 & \bfseries 0.904 & 0.936 & 0.974 & 1.025 & 0.936 & 0.962 & 1.007 & 1.129 & 1.078 & 1.093 & 1.123 & 1.134 & 1.093 & 1.105 & 1.130 \\
 & MAP-Ent & 0.830 & \bfseries 0.818 & 0.825 & 0.825 & 0.848 & 0.825 & 0.830 & 0.840 & 0.990 & 0.906 & 0.919 & 0.961 & 1.017 & 0.919 & 0.933 & 0.985 \\
 & MAP-Gini & 0.993 & \bfseries 0.906 & 0.958 & 0.963 & 1.045 & 0.959 & 0.991 & 1.018 & 1.180 & 1.135 & 1.147 & 1.166 & 1.187 & 1.146 & 1.156 & 1.175 \\
 & R-Jain & 0.984 & \bfseries 0.893 & 0.927 & 0.960 & 1.015 & 0.924 & 0.948 & 0.997 & 1.119 & 1.069 & 1.082 & 1.112 & 1.123 & 1.082 & 1.096 & 1.119 \\
 & R-Ent & 0.817 & \bfseries 0.809 & 0.818 & 0.811 & 0.839 & 0.815 & 0.818 & 0.831 & 0.981 & 0.898 & 0.909 & 0.950 & 1.007 & 0.909 & 0.925 & 0.975 \\
 & R-Gini & 0.981 & \bfseries 0.897 & 0.951 & 0.951 & 1.037 & 0.949 & 0.980 & 1.010 & 1.172 & 1.127 & 1.138 & 1.157 & 1.178 & 1.137 & 1.149 & 1.166 \\
 & NDCG-Jain & 1.022 & \bfseries 0.937 & 0.966 & 1.000 & 1.051 & 0.967 & 0.990 & 1.034 & 1.148 & 1.102 & 1.116 & 1.142 & 1.153 & 1.116 & 1.127 & 1.149 \\
 & NDCG-Ent & 0.868 & \bfseries 0.859 & 0.866 & 0.863 & 0.889 & 0.867 & 0.871 & 0.881 & 1.023 & 0.945 & 0.957 & 0.994 & 1.049 & 0.956 & 0.970 & 1.018 \\
 & NDCG-Gini & 1.028 & \bfseries 0.947 & 0.997 & 1.000 & 1.083 & 0.999 & 1.029 & 1.056 & 1.211 & 1.170 & 1.181 & 1.197 & 1.218 & 1.180 & 1.190 & 1.206 \\
\cline{1-18}
\multirow[c]{12}{*}{\rotatebox[origin=r]{90}{QK-video}} & P-Jain & 0.563 & \bfseries 0.297 & 0.468 & 0.487 & 0.941 & 0.700 & 0.658 & 0.922 & 1.007 & 0.983 & 0.989 & 1.005 & 0.999 & 0.945 & 0.950 & 0.996 \\
 & P-Ent & 0.245 & \bfseries 0.236 & 0.243 & 0.241 & 0.335 & 0.258 & 0.254 & 0.310 & 0.623 & 0.499 & 0.510 & 0.588 & 0.542 & 0.403 & 0.400 & 0.504 \\
 & P-Gini & 0.521 & \bfseries 0.327 & 0.495 & 0.456 & 0.833 & 0.649 & 0.611 & 0.789 & 1.002 & 0.978 & 0.981 & 0.996 & 0.985 & 0.931 & 0.925 & 0.972 \\
 & MAP-Jain & 1.106 & \bfseries 0.996 & 1.061 & 1.070 & 1.332 & 1.180 & 1.157 & 1.318 & 1.379 & 1.365 & 1.371 & 1.378 & 1.371 & 1.338 & 1.343 & 1.369 \\
 & MAP-Ent & 0.979 & \bfseries 0.977 & 0.979 & 0.978 & 0.995 & 0.980 & 0.980 & 0.987 & 1.125 & 1.066 & 1.073 & 1.106 & 1.079 & 1.024 & 1.025 & 1.061 \\
 & MAP-Gini & 1.082 & \bfseries 1.003 & 1.070 & 1.052 & 1.254 & 1.147 & 1.127 & 1.225 & 1.372 & 1.358 & 1.362 & 1.367 & 1.357 & 1.324 & 1.322 & 1.347 \\
 & R-Jain & 1.097 & \bfseries 0.982 & 1.047 & 1.060 & 1.312 & 1.162 & 1.145 & 1.301 & 1.355 & 1.345 & 1.357 & 1.356 & 1.343 & 1.314 & 1.326 & 1.342 \\
 & R-Ent & 0.969 & 0.962 & 0.964 & 0.968 & 0.969 & \bfseries 0.959 & 0.966 & 0.964 & 1.096 & 1.040 & 1.056 & 1.079 & 1.043 & 0.993 & 1.002 & 1.027 \\
 & R-Gini & 1.073 & \bfseries 0.989 & 1.056 & 1.043 & 1.233 & 1.129 & 1.114 & 1.207 & 1.348 & 1.338 & 1.348 & 1.346 & 1.329 & 1.300 & 1.304 & 1.321 \\
 & NDCG-Jain & 1.106 & \bfseries 0.996 & 1.060 & 1.070 & 1.326 & 1.178 & 1.156 & 1.314 & 1.373 & 1.361 & 1.369 & 1.372 & 1.362 & 1.332 & 1.340 & 1.361 \\
 & NDCG-Ent & 0.980 & \bfseries 0.976 & 0.978 & 0.979 & 0.988 & 0.977 & 0.980 & 0.982 & 1.118 & 1.061 & 1.071 & 1.099 & 1.069 & 1.017 & 1.021 & 1.051 \\
 & NDCG-Gini & 1.083 & \bfseries 1.002 & 1.069 & 1.053 & 1.248 & 1.145 & 1.127 & 1.221 & 1.366 & 1.354 & 1.360 & 1.362 & 1.349 & 1.319 & 1.318 & 1.340 \\
\bottomrule

\end{tabular}}
\end{table*}

\begin{table*}[tbp]
\caption{$\downarrow$DPFR scores at $k=10$ of the recommender models for Jester and ML-*, without and with re-ranking the the top $k'=25$ items using Borda Count (BC), COMBMNZ (CM), and Greedy Substitution (GS). The best score per measure pair is in bold.}
\label{tab:dpfr-scores2}
\resizebox{0.90\textwidth}{!}{
\begin{tabular}{ll*{4}{r}|*{4}{r}|*{4}{r}|*{4}{r}}
\toprule
 &  model & \multicolumn{4}{c|}{ItemKNN} & \multicolumn{4}{c|}{BPR} & \multicolumn{4}{c|}{MultiVAE} & \multicolumn{4}{c}{NCL} \\ 
\midrule
 &  reranking & - & BC & CM & GS & - & BC & CM & GS & - & BC & CM & GS & - & BC & CM & GS \\
\midrule
\multirow[c]{12}{*}{\rotatebox[origin=r]{90}{Jester}} & P-Jain & 0.709 & 0.566 & 0.719 & 0.707 & 0.679 & \bfseries 0.550 & 0.631 & 0.677 & 0.747 & 0.569 & 0.638 & 0.745 & 0.698 & 0.604 & 0.625 & 0.696 \\
 & P-Ent & 0.401 & 0.382 & 0.507 & 0.400 & 0.367 & 0.381 & 0.462 & \bfseries 0.366 & 0.433 & 0.372 & 0.440 & 0.430 & 0.381 & 0.381 & 0.442 & 0.380 \\
 & P-Gini & 0.724 & 0.602 & 0.739 & 0.722 & 0.675 & \bfseries 0.578 & 0.651 & 0.673 & 0.766 & 0.598 & 0.651 & 0.763 & 0.704 & 0.619 & 0.642 & 0.701 \\
 & MAP-Jain & 0.915 & 0.908 & 1.048 & 0.914 & 0.894 & 0.897 & 0.987 & \bfseries 0.892 & 0.932 & 0.905 & 0.976 & 0.930 & 0.898 & 0.923 & 0.974 & 0.897 \\
 & MAP-Ent & 0.704 & 0.805 & 0.914 & 0.703 & 0.687 & 0.804 & 0.889 & 0.686 & 0.704 & 0.795 & 0.859 & 0.703 & 0.681 & 0.794 & 0.868 & \bfseries 0.680 \\
 & MAP-Gini & 0.927 & 0.930 & 1.062 & 0.926 & 0.891 & 0.915 & 1.001 & \bfseries 0.889 & 0.947 & 0.924 & 0.985 & 0.945 & 0.903 & 0.933 & 0.986 & 0.901 \\
 & R-Jain & 0.774 & 0.704 & 0.927 & 0.773 & 0.748 & \bfseries 0.700 & 0.821 & 0.746 & 0.802 & 0.702 & 0.787 & 0.800 & 0.759 & 0.732 & 0.787 & 0.758 \\
 & R-Ent & 0.507 & 0.566 & 0.773 & 0.506 & 0.483 & 0.576 & 0.699 & \bfseries 0.482 & 0.521 & 0.555 & 0.636 & 0.519 & 0.484 & 0.563 & 0.651 & 0.483 \\
 & R-Gini & 0.788 & 0.733 & 0.943 & 0.787 & 0.745 & \bfseries 0.723 & 0.837 & 0.743 & 0.819 & 0.727 & 0.798 & 0.817 & 0.765 & 0.745 & 0.801 & 0.763 \\
 & NDCG-Jain & 0.823 & 0.793 & 0.977 & 0.822 & 0.798 & \bfseries 0.782 & 0.900 & 0.797 & 0.845 & 0.788 & 0.878 & 0.844 & 0.807 & 0.810 & 0.878 & 0.805 \\
 & NDCG-Ent & 0.579 & 0.673 & 0.833 & 0.578 & 0.558 & 0.673 & 0.791 & 0.557 & 0.586 & 0.660 & 0.745 & 0.584 & 0.556 & 0.661 & 0.758 & \bfseries 0.555 \\
 & NDCG-Gini & 0.837 & 0.818 & 0.992 & 0.835 & 0.795 & 0.802 & 0.914 & \bfseries 0.793 & 0.862 & 0.810 & 0.887 & 0.860 & 0.812 & 0.821 & 0.890 & 0.810 \\
\cline{1-18}
\multirow[c]{12}{*}{\rotatebox[origin=r]{90}{ML-10M}} & P-Jain & 1.072 & 1.067 & 1.082 & 1.075 & 1.047 & \bfseries 1.023 & 1.031 & 1.051 & 1.098 & 1.094 & 1.104 & 1.099 & 1.052 & 1.044 & 1.059 & 1.053 \\
 & P-Ent & 0.884 & 0.831 & 0.844 & 0.880 & 0.765 & \bfseries 0.749 & 0.766 & 0.764 & 1.023 & 0.974 & 0.970 & 1.010 & 0.801 & 0.771 & 0.791 & 0.791 \\
 & P-Gini & 1.080 & 1.075 & 1.088 & 1.081 & 1.042 & \bfseries 1.025 & 1.032 & 1.041 & 1.106 & 1.102 & 1.111 & 1.107 & 1.055 & 1.050 & 1.065 & 1.053 \\
 & MAP-Jain & 1.172 & 1.172 & 1.182 & 1.173 & 1.149 & \bfseries 1.134 & 1.135 & 1.150 & 1.193 & 1.192 & 1.201 & 1.194 & 1.155 & 1.153 & 1.161 & 1.154 \\
 & MAP-Ent & 0.995 & 0.955 & 0.962 & 0.991 & 0.893 & \bfseries 0.886 & 0.893 & 0.888 & 1.119 & 1.076 & 1.073 & 1.106 & 0.924 & 0.906 & 0.915 & 0.914 \\
 & MAP-Gini & 1.173 & 1.173 & 1.182 & 1.174 & 1.139 & \bfseries 1.130 & \bfseries 1.130 & 1.135 & 1.195 & 1.193 & 1.202 & 1.195 & 1.152 & 1.153 & 1.160 & 1.149 \\
 & R-Jain & 0.860 & 0.846 & 0.847 & 0.859 & 0.835 & 0.780 & \bfseries 0.768 & 0.829 & 0.868 & 0.866 & 0.867 & 0.867 & 0.847 & 0.813 & 0.810 & 0.844 \\
 & R-Ent & 0.653 & 0.570 & 0.563 & 0.643 & 0.492 & 0.423 & \bfseries 0.415 & 0.472 & 0.807 & 0.748 & 0.730 & 0.788 & 0.554 & 0.473 & 0.468 & 0.534 \\
 & R-Gini & 0.889 & 0.877 & 0.876 & 0.888 & 0.850 & 0.805 & \bfseries 0.793 & 0.838 & 0.900 & 0.897 & 0.898 & 0.899 & 0.872 & 0.843 & 0.840 & 0.866 \\
 & NDCG-Jain & 1.140 & 1.142 & 1.156 & 1.142 & 1.115 & \bfseries 1.107 & 1.114 & 1.118 & 1.168 & 1.166 & 1.180 & 1.169 & 1.119 & 1.125 & 1.138 & 1.119 \\
 & NDCG-Ent & 0.972 & 0.932 & 0.944 & 0.968 & 0.863 & 0.866 & 0.881 & \bfseries 0.860 & 1.103 & 1.060 & 1.062 & 1.091 & 0.894 & 0.884 & 0.901 & 0.885 \\
 & NDCG-Gini & 1.151 & 1.153 & 1.166 & 1.153 & 1.114 & \bfseries 1.112 & 1.119 & \bfseries 1.112 & 1.180 & 1.177 & 1.190 & 1.180 & 1.127 & 1.134 & 1.147 & 1.124 \\
\cline{1-18}
\multirow[c]{12}{*}{\rotatebox[origin=r]{90}{ML-20M}} & P-Jain & 1.014 & 1.015 & 1.033 & 1.016 & 0.998 & \bfseries 0.992 & 1.000 & 1.001 & 1.000 & 0.993 & 1.006 & 1.006 & 1.002 & 1.004 & 1.020 & 1.006 \\
 & P-Ent & 0.885 & 0.842 & 0.858 & 0.880 & 0.776 & \bfseries 0.761 & 0.773 & 0.771 & 0.775 & 0.765 & 0.780 & 0.774 & 0.807 & 0.783 & 0.800 & 0.801 \\
 & P-Gini & 1.056 & 1.057 & 1.073 & 1.057 & 1.031 & \bfseries 1.026 & 1.033 & 1.030 & 1.031 & 1.029 & 1.039 & 1.033 & 1.040 & 1.042 & 1.057 & 1.042 \\
 & MAP-Jain & 1.118 & 1.124 & 1.137 & 1.119 & \bfseries 1.103 & \bfseries 1.103 & 1.104 & \bfseries 1.103 & 1.107 & 1.105 & 1.111 & 1.109 & 1.106 & 1.114 & 1.123 & 1.108 \\
 & MAP-Ent & 0.996 & 0.963 & 0.973 & 0.991 & 0.900 & 0.894 & 0.898 & \bfseries 0.892 & 0.902 & 0.897 & 0.905 & 0.897 & 0.926 & 0.913 & 0.921 & 0.919 \\
 & MAP-Gini & 1.151 & 1.156 & 1.167 & 1.151 & 1.127 & 1.129 & 1.129 & \bfseries 1.124 & 1.130 & 1.131 & 1.136 & 1.129 & 1.135 & 1.143 & 1.151 & 1.135 \\
 & R-Jain & 0.775 & 0.768 & 0.769 & 0.775 & 0.757 & 0.723 & \bfseries 0.712 & 0.753 & 0.756 & 0.721 & 0.718 & 0.752 & 0.766 & 0.744 & 0.741 & 0.764 \\
 & R-Ent & 0.645 & 0.573 & 0.568 & 0.634 & 0.490 & 0.426 & \bfseries 0.413 & 0.471 & 0.483 & 0.426 & 0.421 & 0.465 & 0.542 & 0.471 & 0.462 & 0.523 \\
 & R-Gini & 0.836 & 0.829 & 0.829 & 0.836 & 0.808 & 0.778 & \bfseries 0.768 & 0.800 & 0.805 & 0.779 & 0.774 & 0.797 & 0.822 & 0.803 & 0.799 & 0.818 \\
 & NDCG-Jain & 1.081 & 1.089 & 1.107 & 1.083 & \bfseries 1.065 & 1.072 & 1.078 & 1.066 & 1.072 & 1.075 & 1.086 & 1.075 & 1.068 & 1.083 & 1.097 & 1.071 \\
 & NDCG-Ent & 0.971 & 0.939 & 0.955 & 0.966 & 0.871 & 0.873 & 0.883 & \bfseries 0.865 & 0.876 & 0.878 & 0.891 & 0.873 & 0.898 & 0.892 & 0.906 & 0.892 \\
 & NDCG-Gini & 1.121 & 1.129 & 1.144 & 1.122 & 1.097 & 1.105 & 1.110 & \bfseries 1.095 & 1.102 & 1.109 & 1.118 & 1.102 & 1.105 & 1.119 & 1.132 & 1.105 \\
\bottomrule

\end{tabular}}
\end{table*}

\begin{table*}[tbp]
    \centering
    \caption{Range of agreement $\tau$ between estimated PF and PF across 12 measure pairs, using the estimated PF with 3--12 points.}
    \label{tab:corr_est}
    \resizebox{0.72\textwidth}{!}{
\begin{tabular}{lrrrrrr}
\toprule
{\#pts} &      Lastfm &   Amazon-lb &    QK-video &      Jester &      ML-10M &      ML-20M \\
\midrule
3  &  0.78--1.00 &  0.98--1.00 &  1.00--1.00 &  1.00--1.00 &  0.97--1.00 &  0.75--1.00 \\
4  &  0.88--1.00 &  0.98--1.00 &  0.98--1.00 &  0.98--1.00 &  0.98--1.00 &  0.93--1.00 \\
5  &  0.78--1.00 &  0.98--1.00 &  1.00--1.00 &  1.00--1.00 &  0.97--1.00 &  0.92--1.00 \\
6  &  0.90--1.00 &  0.97--1.00 &  1.00--1.00 &  0.98--1.00 &  0.95--1.00 &  0.92--1.00 \\
7  &  0.88--1.00 &  1.00--1.00 &  1.00--1.00 &  1.00--1.00 &  0.98--1.00 &  0.93--1.00 \\
8  &  0.90--1.00 &  0.98--1.00 &  1.00--1.00 &  0.98--1.00 &  1.00--1.00 &  0.95--1.00 \\
9  &  0.98--1.00 &  1.00--1.00 &  1.00--1.00 &  1.00--1.00 &  0.97--1.00 &  0.98--1.00 \\
10 &  0.88--1.00 &  1.00--1.00 &  1.00--1.00 &  0.98--1.00 &  1.00--1.00 &  0.95--1.00 \\
11 &  0.92--1.00 &  1.00--1.00 &  1.00--1.00 &  1.00--1.00 &  0.98--1.00 &  0.97--1.00 \\
12 &  0.95--1.00 &  1.00--1.00 &  1.00--1.00 &  0.98--1.00 &  0.98--1.00 &  0.97--1.00 \\
\bottomrule
\end{tabular}}
\end{table*}

\section{Further discussions}
\label{app:discussion}

\subsection{The impact of replacing frequently recommended items}
In this work, we replace frequently recommended items in two cases: during the Pareto Frontier generation (PF) with the {\sc Oracle2Fair} algorithm (\Cref{ss:generation}) and as part of the fair rerankers (\Cref{s:experiments}). In both cases, none of the replacements significantly affect the overall recommendation performance:

\noindent\textbf{1. Replacement in the {\sc Oracle2Fair} algorithm.} The {\sc Oracle2Fair} algorithm is used to generate the recommendation lists whose scores make up the PF (not the scores from the model-generated recommendation lists, that we compare to a point in the PF). The replacements are done on lists separate from the model's recommendation lists. Therefore, the replacements in the {\sc Oracle2Fair} algorithm do not affect the recommendation performance based on relevance and fairness measures. 

\noindent\textbf{2. Replacement in the fair rerankers.} We look at recommendation performance based on NDCG (relevance) and Gini (fairness). In all six datasets, when comparing the best non-reranked model to its reranked versions (e.g., for Lastfm, it is NCL vs NCL-BC, NCL-CM, and NCL-GS), the decrease in NDCG is not more than 0.26, and the decrease is even below 0.15 in all datasets excluding Jester, (\Cref{app:actual-scores}, \Cref{tab:base-rerank-all-1,tab:base-rerank-all-2}). For Jester, the 0.26 decrease in NDCG is exchanged for an improvement in Gini by 0.218. Hence, we believe that the impact of item replacement is reasonable.

\subsection{Using rank-based fairness measures}

Suppose we have two \textsc{Fair} measures, e.g., set-based Gini (Gini) and rank-based Gini (Gini-w). In \cite{Rampisela2024EvaluationStudy}, the absolute scores of Gini and Gini-w do not differ considerably and the two measures correlate strongly with Kendall's $\tau \geq 0.9$. Hence, generating the PF with rank-based fairness measures such as Gini-w is not expected to result in significantly different conclusions from the set-based version.

\end{document}